\documentclass[aps,prd,twocolumn,superscriptaddress,nofootinbib]{revtex4-1}

\usepackage{graphicx,color}
\usepackage{latexsym,amsmath,amssymb,graphicx,booktabs}
\usepackage{hyperref}
\usepackage{float}
\definecolor{MyBlue}{rgb}{0.15,0.15,0.70}
\hypersetup{
colorlinks=true,
citecolor=MyBlue,
linkcolor=MyBlue,
urlcolor=MyBlue
}

\newcommand{\dgw}{d_L^{\,\rm gw}}
\newcommand{\dem}{d_L^{\,\rm em}}

\renewcommand\({\left(}
\renewcommand\){\right)}

\def\lsim{\raise 0.4ex\hbox{$<$}\kern -0.8em\lower 0.62
ex\hbox{$\sim$}}

\def\gsim{\raise 0.4ex\hbox{$>$}\kern -0.7em\lower 0.62
ex\hbox{$\sim$}}

\def\lbar{{\hbox{$\lambda$}\kern -0.7em\raise 0.6ex
\hbox{$-$}}}

\newcommand\eq[1]{eq.~(\ref{#1})}
\newcommand\eqs[2]{eqs.~(\ref{#1}) and (\ref{#2})}
\newcommand\Eq[1]{Equation~(\ref{#1})}

\newcommand\p{\partial}

\newcommand\ee{\end{equation}}
\newcommand\be{\begin{equation}}
\def\bea{\begin{array}}
\def\eea{\end{array}}\def\ea{\end{array}}
\newcommand\ees{\end{eqnarray}}
\newcommand\bees{\begin{eqnarray}}

\def\dslash{\hspace{-1mm}\not{\hbox{\kern-2pt $\partial$}}}
\def\Dslash{\not{\hbox{\kern-2pt $D$}}}
\def\pslash{\not{\hbox{\kern-2.1pt $p$}}}
\def\kslash{\not{\hbox{\kern-2.3pt $k$}}}
\def\qslash{\not{\hbox{\kern-2.3pt $q$}}}

\newcommand{\vk}{{\bf k}}

\def\p1{{\bf p}_1}
\def\p2{{\bf p}_2}
\def\k1{{\bf k}_1}
\def\k2{{\bf k}_2}

\newcommand{\dddM}{\kern 0.2em \raise 1.9ex\hbox{$...$}\kern -1.0em \hbox{$M$}}
\newcommand{\dddQ}{\kern 0.2em \raise 1.9ex\hbox{$...$}\kern -1.0em \hbox{$Q$}}
\newcommand{\dddI}{\kern 0.2em \raise 1.9ex\hbox{$...$}\kern -1.0em\hbox{$I$}}
\newcommand{\dddJ}{\kern 0.2em \raise 1.9ex\hbox{$...$}\kern-1.0em
\hbox{$J$}}
\newcommand{\dddcalJ}{\kern 0.2em \raise 1.9ex\hbox{$...$}\kern-1.0em
\hbox{${\cal J}$}}

\newcommand{\dddO}{\kern 0.2em \raise 1.9ex\hbox{$...$}\kern -1.0em
\hbox{${\cal O}$}}
\def\dddz{\raise 1.5ex\hbox{$...$}\kern -0.8em \hbox{$z$}}
\def\dddd{\raise 1.8ex\hbox{$...$}\kern -0.8em \hbox{$d$}}
\def\dddbd{\raise 1.8ex\hbox{$...$}\kern -0.8em \hbox{${\bf d}$}}
\def\ddbd{\raise 1.8ex\hbox{$..$}\kern -0.8em \hbox{${\bf d}$}}
\def\dddx{\raise 1.6ex\hbox{$...$}\kern -0.8em \hbox{$x$}}

\newcommand{\ode}{\Omega_{\rm DE}}
\newcommand{\oma}{\Omega_{M}}
\newcommand{\ora}{\Omega_{R}}

\newcommand{\rde}{\rho_{\rm DE}}
\newcommand{\wde}{w_{\rm DE}}

\graphicspath{ {./Graphs/} }

\begin{document}


\title{Gaussian processes reconstruction of  modified gravitational wave propagation}


\author{Enis Belgacem}
\email{enis.belgacem@unige.ch}
\affiliation{D\'epartement de Physique Th\'eorique and Center for Astroparticle Physics, Universit\'e de Gen\`eve, 24 quai Ansermet, CH--1211 Gen\`eve 4, Switzerland}

\author{Stefano Foffa}
\email{stefano.foffa@unige.ch}
\affiliation{D\'epartement de Physique Th\'eorique and Center for Astroparticle Physics, Universit\'e de Gen\`eve, 24 quai Ansermet, CH--1211 Gen\`eve 4, Switzerland}

\author{Michele Maggiore}
\email{michele.maggiore@unige.ch}
\affiliation{D\'epartement de Physique Th\'eorique and Center for Astroparticle Physics, Universit\'e de Gen\`eve, 24 quai Ansermet, CH--1211 Gen\`eve 4, Switzerland}

\author{Tao Yang}
\email{tao.yang@apctp.org}
\affiliation{Asia Pacific Center for Theoretical Physics, Pohang, 37673, Korea}



\begin{abstract}
Recent work has shown that  modified   gravitational wave (GW)  propagation can be a powerful probe of dark energy and modified gravity, specific to GW observations. We use the technique of Gaussian processes,  that allows the reconstruction of a function from the data without assuming any parametrization,  to measurements of the GW luminosity distance from simulated joint GW-GRB  detections, combined with  measurements of the electromagnetic luminosity distance by simulated  DES data. For the GW events we consider both a second-generation  LIGO/Virgo/Kagra (HVLKI) network, and  a third-generation detector such as the Einstein Telescope. We find that the  HVLKI network at target sensitivity, with $O(15)$ neutron star binaries with electromagnetic counterpart, could already detect deviations from GR at a level predicted by some modified gravity models, and a third-generation detector such as ET would have a remarkable discovery potential.   We discuss the complementarity of the Gaussian processes  technique  to the  $(\Xi_0,n)$ parametrization of modified GW propagation.
\end{abstract}


\maketitle

\section{Introduction}

In cosmology there are several  possible extensions  of the flat-space notion of distance. In many contexts, the relevant quantity is the  luminosity distance $d_L$, which  is defined by ${\cal F}={\cal L}/(4\pi d_L^2)$, where ${\cal F}$ is  the energy flux  received by the observer and ${\cal L}$ is the intrinsic luminosity  of the source. 
The relation between  $d_L$ and the source redshift $z$ carries important cosmological information. In a generic cosmological model with dark energy (DE) density $\rde(z)$, it is given by
\be\label{dLem}
d_L(z)=\frac{1+z}{H_0}\int_0^z\, 
\frac{d\tilde{z}}{E(\tilde{z})}\, ,
\ee
where 
\be\label{E(z)}
E(z)=\sqrt{\ora (1+z)^4+\oma (1+z)^3+\ode(z) }
\ee
and we used standard notation: $H_0$ is the Hubble parameter; $\Omega_{M,R}=\rho_{M,R}(t_0)/\rho_0$ are the  densities  of matter and radiation at the present time $t_0$, normalized to closure energy density $\rho_0$, and  $\ode(z)= \rho_{\rm DE}(z)/\rho_0$  (we also assumed flatness, for simplicity). In particular, in  $\Lambda$CDM, $\ode(z)=\Omega_{\Lambda}$ is a constant, while in a generic theory with DE equation of state $\wde(z)$, 
\be\label{4rdewdeproofs}
\ode(z)  =\Omega_{\rm DE}\exp\left\{ 3\int_{0}^z\, \frac{d\tilde{z}}{1+\tilde{z}}\, [1+\wde(\tilde{z})]\right\}\, ,
\ee
where $\Omega_{\rm DE}\equiv \ode(z=0)$. At $z\ll 1$ \eq{dLem} reduces to Hubble's law $d_L\simeq H_0^{-1}z$, so the $d_L-z$ relation is only sensitive to $H_0$. In contrast, a determination of both $d_L$ and $z$ for sources at higher redshift can give information also on dark energy, by measuring  $\ode$ and $\wde(z)$. 
Type Ia supernovae (SNe) are the classic example of use of the $d_L-z$ relation  and provided the first evidence for the accelerated expansion of the Universe~\cite{Riess:1998cb,Perlmutter:1998np}. In the context of standard General Relativity (GR), the coalescence of compact binaries provides another way of measuring $d_L$, as was realized long ago~\cite{Schutz:1986gp} (see e.g. ~\cite{Holz:2005df,Dalal:2006qt,MacLeod:2007jd,Nissanke:2009kt,Cutler:2009qv,Sathyaprakash:2009xt,Zhao:2010sz,DelPozzo:2011yh,Nishizawa:2011eq,Taylor:2012db,Camera:2013xfa,Tamanini:2016zlh,Caprini:2016qxs,Cai:2016sby,DelPozzo:2017kme,Belgacem:2017ihm,Belgacem:2018lbp,Mendonca:2019yfo} for a sample of the many developments of the idea), so these sources are referred to as ``standard sirens", the GW analogue of standard candles. Together with a determination of the redshift (which is not provided by the gravitational signal and must be obtained from an electromagnetic counterpart
or with statistical techniques), this allows us to use GW observations as probes of cosmology.
Recently, thanks to the observation of the electromagnetic counterpart to the binary neutron star (NS) coalescence GW170817~\cite{TheLIGOScientific:2017qsa,GBM:2017lvd,Monitor:2017mdv}, a measurement of $H_0$ with standard sirens has been performed for the first time~\cite{Abbott:2017xzu}. Given the  low redshift of  GW170817, $z\simeq 0.01$,  no direct information on dark energy could be obtained from it. However, in the near future, the second-generation (2G) network made by the two advanced LIGO detectors, advanced Virgo, Kagra and LIGO India (HLVKI), at target sensitivity, will detect NS-NS binaries up to $z\sim 0.1$, out of which several could have an electromagnetic counterpart. Third-generation detectors such as the Einstein Telescope (ET)~\cite{Punturo:2010zz} or Cosmic Explorer~\cite{Dwyer:2014fpa}, that could be operating in the mid 2030s, would extend the reach to fully cosmological distances. ET would detect binary black hole (BH) mergers up to $z\sim 20$, and NS-NS mergers up to $z\simeq 2-3$, with rates of order $10^5-10^6$ events per year~\cite{Sathyaprakash:2009xt,Sathyaprakash:2019nnu,Belgacem:2019tbw}. Depending on the network of telescopes and GRB satellites that will be operating at that time, in a few years ET could  collect  hundreds of NS-NS coalescences with observed electromagnetic counterpart~\cite{Stratta:2017bwq,Belgacem:2019tbw}.

Recently it has been appreciated that, in theories where GR is modified on cosmological scales, standard sirens do not measure the same luminosity distance as electromagnetic probes~\cite{Saltas:2014dha,Gleyzes:2014rba,Lombriser:2015sxa,Nishizawa:2017nef,Arai:2017hxj,Belgacem:2017ihm,Amendola:2017ovw,Linder:2018jil,Belgacem:2018lbp,Lagos:2019kds,Nishizawa:2019rra}. One must then distinguish between an `electromagnetic luminosity distance' $\dem$, given by the standard expression (\ref{dLem},\ref{E(z)}),  and a `GW luminosity distance' $\dgw$. To see how the difference emerges  recall that, in GR, the propagation of GWs in 
Friedmann-Robertson-Walker (FRW) metric is governed by 
\be\label{4eqtensorsect}
\tilde{h}''_A+2{\cal H}\tilde{h}'_A+k^2\tilde{h}_A=0\, ,
\ee
where $\tilde{h}_A(\eta, \vk)$ are  the Fourier modes of the GW amplitude with polarization labeled by $A=+,\times$; the prime denotes the derivative with respect to cosmic time $\eta$ and ${\cal H}=a'/a$ is the Hubble parameter in conformal time, where $a(\eta)$ is the scale factor
(we use units $c=1$). In  modified gravity  this equation, in general, is also modified. However,   a change of the numerical coefficient of the $k^2\tilde{h}_A$ is now basically excluded  because it would affect the  speed of GWs. This is now bounded   at the level  $|c_{\rm gw}-c|/c< O(10^{-15})$ by the observation of  GW170817/GRB~170817A ~\cite{Monitor:2017mdv}. A consequence of \eq{4eqtensorsect} is that, in the propagation across cosmological distances, the amplitude of GWs decreases as $1/a(\eta)$. In the amplitude emitted from coalescing binaries this factors combines with other factors coming from the transformation of masses and GW frequency from the source frame to the detector frame, to produce  the usual  dependence of the GW amplitude from the luminosity distance,
$\tilde{h}_A(\eta, \vk)\propto 1/d_L(z)$
(see e.g.  Section 4.1.4 of \cite{Maggiore:1900zz} for a textbook derivation). 

However, in general  models of modified gravity that pass the speed-of-gravity test still modify the `friction term'  $2{\cal H}\tilde{h}'_A$, so that \eq{4eqtensorsect} is replaced by
\be\label{prophmodgrav}
\tilde{h}''_A  +2 {\cal H}[1-\delta(\eta)] \tilde{h}'_A+k^2\tilde{h}_A=0\, ,
\ee
for some function $\delta(\eta)$. This behavior has been first found explicitly in scalar-tensor theories of the Horndeski class~\cite{Saltas:2014dha,Lombriser:2015sxa,Arai:2017hxj,Amendola:2017ovw,Linder:2018jil} and in nonlocal infrared modifications of gravity~\cite{Belgacem:2017ihm,Belgacem:2018lbp}. The analysis in \cite{Belgacem:2019pkk} showed that also scalar-tensor theories of the DHOST type, as well as bigravity, display this phenomenon, so an equation of propagation of the form (\ref{prophmodgrav}) is completely generic and appears in all the best studied models of modified gravity.
In some instances  the effect is related to  the time variation of the Newton constant (or, equivalently, of the Planck mass) that takes place in these models. This however, is not generic, and there are models where the function $\delta(\eta)$ in \eq{prophmodgrav} is not related to a time-varying Planck mass; see the discussion in sect.~III of \cite{Belgacem:2018lbp}. Modified GW propagation also takes place in a different context, namely in theories with extra dimensions, where it is a consequence  of the  loss of gravitons to a higher-dimensional bulk. It was in fact first discovered in this context, within the DGP model~\cite{Deffayet:2007kf} (see also \cite{Pardo:2018ipy} for recent discussion). In this case, modified GW propagation is in general no longer governed by an equation of the form (\ref{prophmodgrav}), and rather expresses the flux conservation in the higher-dimensional theory.

When the propagation equation is governed by (\ref{prophmodgrav}), it is possible to show that, after propagation from the source to the observer,  in the GW amplitude the standard factor  $1/\dem$ obtained from a GR computation is replaced by  $1/\dgw$, where~\cite{Belgacem:2017ihm,Belgacem:2018lbp} 
\be\label{eq:dLgwdLem}
d_L^{\,\rm gw}(z)=d_L^{\,\rm em}(z)\exp\left\{-\int_0^z \,\frac{dz'}{1+z'}\,\delta(z')\right\}\, ,
\ee
(see also Sect.~19.6.3  of \cite{Maggiore:2018zz} for a  textbook discussion). GW measurements can therefore access the quantity $\delta(z)$, or equivalently $\dgw(z)/\dem(z)$, 
which is a smoking gun of modified gravity and of the dark-energy sector of alternatives to $\Lambda$CDM.

There are several reasons why this observable is particularly interesting:

\begin{itemize}

\item  A generic 
modified gravity theories affects both the  cosmological evolution at  the background level, and the cosmological perturbations. The modification of the background evolution are  encoded in the DE equation of state $\wde(z)$, as in \eq{4rdewdeproofs}. At the perturbation level, modified gravity in general affects both  scalar and  tensor  perturbations. The modification in the background evolution and in the scalar perturbation sector are probed with standard electromagnetic observations (in particular, scalar perturbations are probed by the growth of structures and by lensing). Tensor perturbations, i.e. GWs propagating over a FRW background, can however be probed only by GW observations. This is a new observational window, that we are beginning to open  now, thanks to the first spectacular detections by advanced LIGO and advanced Virgo, and that will be fully exploited with 3G detectors such as the Einstein Telescope; apart from the result on the speed of GWs, this is still virgin territory, where surprises might await for us. The fact that this territory can only be explored by GW observation makes modified GW propagation  a privileged observable for GW experiments. 

\item As pointed out in~\cite{Belgacem:2017ihm,Belgacem:2018lbp}, a  second reason for the importance of this observable is that, in a  modified gravity model where the deviation of 
$d_L^{\,\rm gw}(z)/d_L^{\,\rm em}(z)$ from 1 is of the same order as the deviation of $\wde(z)$ from the $\Lambda$CDM value $-1$, observationally the effect of  $d_L^{\,\rm gw}(z)/d_L^{\,\rm em}(z)$ dominates on the effect of $\wde(z)$.
This is due to the fact that the parameters $H_0$ and $\oma$ that appear in \eqs{dLem}{E(z)} are not fixed, but must be determined within each model by comparison with the data. If we change $\wde$, with respect to the $\Lambda$CDM value $\wde=-1$, the values of $H_0$ and $\oma$ will also change. Since a fit to CMB and BAO data basically amounts to requiring that the model reproduces some fixed distance scales at large redshifts, $H_0$ and $\oma$ will change in such a way to partially compensate  the change in luminosity distance induced by $\wde$ at large $z$. Thus, a change in $\wde$ by, say, $10\%$ with respect to the $\Lambda$CDM value $\wde=-1$, at the large redshift where we are no longer in the Hubble law regime will result in a relative change of $\dem$  by only, say, $1\%$. In contrast, the effect from modified GW propagation encoded in \eq{eq:dLgwdLem} is an extra effect, which is not compensated by a corresponding change in the value of the cosmological parameters.

\item Last but not least, the deviations from GR in the tensor sector can be much larger than for the background evolution or for the scalar sector. Indeed,  at the background level, assuming a constant DE equation of state  $\wde(z)=w_0$,  {\em Planck} 2015  combined with other datasets gives $w_0=-1.006\pm 0.045$~\cite{Planck_2015_CP}, which indicates that  the deviation from $\Lambda$CDM is bounded at the $4.5\%$ level.  Using the more realistic $(w_0,w_a)$ parametrization [see \eq{w0wa} below], one finds $w_0=-0.961\pm 0.077$ and $w_a=-0.28^{+0.31}_{-0.27}$, so in this case deviations  for $w_0$ from $-1$ are bounded at the $7.7\%$ level.
Similarly, the DES Y1 results~\cite{Abbott:2018xao} put bounds at the level of $7\%$ on   the deviation  of  scalar perturbations from  $\Lambda$CDM. One could then naturally guess that a modified gravity model that fits current data, and therefore does not deviate by more than $(5-7)\%$ from $\Lambda$CDM in the background evolution and in scalar perturbations, will not deviate much more in the tensor sector, either. Surprisingly, this is not true. In \cite{Belgacem:2019lwx} were discussed the cosmological predictions of a modified gravity model (a variant of the nonlocal infrared modification of gravity proposed in \cite{Maggiore:2013mea}, with initial conditions set during inflation; see \cite{Maggiore:2016gpx} for review) that is very close to $\Lambda$CDM in the background evolution and scalar perturbations, and fit all cosmological data very well, and still in the tensor sector, at the redshifts available to 3G detectors, can differ from GR by as much as $60\%$, see Fig.~\ref{fig:dgw_su_dem}. Independently of the virtues of this specific model, this gives an explicit example of the fact that modified GW propagation provides a window on dark energy and modified gravity that could reserve great surprises.

\end{itemize}

\section{Gaussian processes versus explicit parametrizations}
\label{sec:param}

\begin{figure}[t]
 \includegraphics[width=0.40\textwidth]{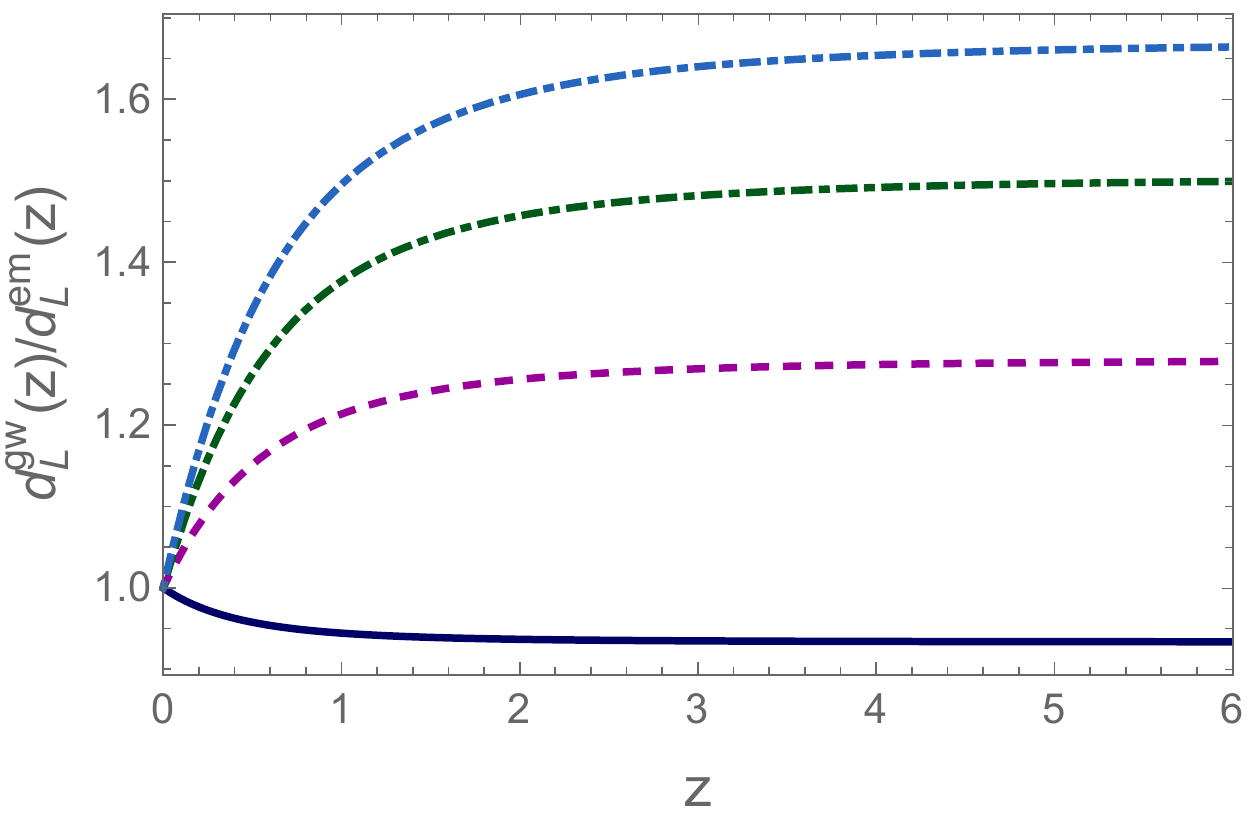}
 \caption{The function $d_L^{\,\rm gw}(z)/d_L^{\,\rm em}(z)$   for the nonlocal RT model with initial conditions during radiation dominance, i.e.  $\Delta N=0$ (blue solid line), and for initial conditions set during inflation at $\Delta N$ efolds before the end of inflation, for $\Delta N=34$ (magenta, dashed), $\Delta N=50$ (green, dot-dashed) and $\Delta N=64$ (cyan, dot-dashed). From ref.~\cite{Belgacem:2019lwx}.}
  \label{fig:dgw_su_dem}
\end{figure}

In order to see the effect of modified GW propagation in  observational data, the simplest approach is to use a parametrization of the redshift dependence, either of $\delta(z)$, or of $\dgw(z)/\dem(z)$. This is similar to what is done for the equation of state of DE, which is usually parametrized in terms of the two parameters $(w_0,w_a)$, as~\cite{Chevallier:2000qy,Linder:2002et}
\be\label{w0wa}
w_{\rm DE}(z)= w_0+z (1+z)^{-1} w_a\, .
\ee 
It turns out that there is a very natural parametrization for $\dgw(z)/\dem(z)$  in terms of two parameters $(\Xi_0,n)$, given by~\cite{Belgacem:2018lbp}
\be\label{eq:fit}
\frac{d_L^{\,\rm gw}(z)}{d_L^{\,\rm em}(z)}=\Xi_0 +\frac{1-\Xi_0}{(1+z)^n}\, .
\ee
This expression reproduces the fact that,  as the redshift of the source goes to zero,  there is no  effect from modified GW propagation, so $d_L^{\,\rm gw}/d_L^{\,\rm em}$ must go to one. At large $z$, with the parametrization (\ref{eq:fit}), $d_L^{\,\rm gw}/d_L^{\,\rm em}$ approaches  a constant value $\Xi_0$. This is motivated by the fact that,
in  typical modified gravity models, DE is a recent phenomenon,  so the modifications to GR, and therefore the function $\delta(z)$,  go to zero at large redshift. Then, the integral that determines
$d_L^{\,\rm gw}(z)/d_L^{\,\rm em}(z)$  in \eq{eq:dLgwdLem} saturates to constant at  large $z$.

Indeed, it has been found that \eq{eq:fit} accurately reproduces the behavior of $d_L^{\,\rm gw}/d_L^{\,\rm em}$ predicted from most modified gravity models. Fig.~\ref{fig:dgw_su_dem} shows the result in a nonlocal modification of gravity (the so-called  `RT' model)  proposed in \cite{Maggiore:2013mea}, as a function of the only free parameter of the model, $\Delta N$, that corresponds to the number of e-folds before the end of inflation at the epoch when initial conditions $O(1)$ are set on some auxiliary field of the theory (see 
\cite{Belgacem:2019lwx} for details,  refs.~\cite{Maggiore:2016gpx,Belgacem:2017cqo} for reviews of the model, and   refs.~\cite{Foffa:2013vma,Dirian:2014ara,Dirian:2014bma,Dirian:2016puz,Dirian:2017pwp,Belgacem:2017cqo,Belgacem:2018wtb} for further related work). For all  values $\Delta N$,
the predictions  for
$d_L^{\,\rm gw}(z)/d_L^{\,\rm em}(z)$ are very accurately reproduced by \eq{eq:fit}, at a level that the fit would be basically indistinguishable from the numerical prediction, on the scale of Fig.~\ref{fig:dgw_su_dem}; for instance, the values of
$(\Xi_0,n)$ from a fit to these curves are $\Xi_0=\{0.93,1.28,1.50,1.67\}$ and $n=\{2.59,2.07,1.99,1.94\}$ for $\Delta N=\{0,34,50,64\}$, respectively~\cite{Belgacem:2019lwx}. Note in particular, that, in the last case, the deviation of $\Xi_0$ from the GR value $\Xi_0=1$ is larger than $60\%$!

\Eq{eq:dLgwdLem} can be inverted for $\delta(z)$, giving 
\be\label{deltalogd}
\delta(z)=-(1+z)\frac{d}{dz}\log\( \frac{d_L^{\,\rm gw}(z)}{d_L^{\,\rm em}(z)} \)
\, .
\ee
The parametrization (\ref{eq:fit}) of $d_L^{\,\rm gw}(z)/d_L^{\,\rm em}(z)$ then implies a corresponding parametrization   of $\delta(z)$,~\cite{Belgacem:2018lbp}
\be\label{paramdeltaz}
\delta(z)=\frac{n  (1-\Xi_0)}{1-\Xi_0+ \Xi_0 (1+z)^n}
\, .
\ee
A detailed study in \cite{Belgacem:2019pkk} has revealed that the parametrizations (\ref{eq:fit})
and (\ref{paramdeltaz}) fit remarkably well the predictions 
of a large class of models (various Horndeski theories with different choices of the functions that characterize them, or several variant of nonlocal gravity), with two notable exceptions. One is bigravity, where the coupling between the two metrics gives rise to a series of oscillations in $d_L^{\,\rm gw}(z)/d_L^{\,\rm em}(z)$. The second are DHOST theories, where the parametrization (\ref{paramdeltaz}) misses a bump in $\delta(z)$. Nevertheless, in the latter case \eq{eq:fit} still reproduces reasonably well the behavior of $d_L^{\,\rm gw}(z)/d_L^{\,\rm em}(z)$. This is due to the fact that the effect of the bump in $\delta(z)$ is smoothed out by the integration in \eq{eq:dLgwdLem}, so that $d_L^{\,\rm gw}(z)/d_L^{\,\rm em}(z)$ maintains the monotonic behavior described by \eq{eq:fit}. This shows that the parametrization (\ref{eq:fit}) for $d_L^{\,\rm gw}(z)/d_L^{\,\rm em}(z)$ is more solid than the corresponding parametrization for $\delta(z)$. From some point of view this is good news, since it shows that \eq{eq:fit} is a good starting point for searching for modified GW propagation in the data, given that $d_L^{\,\rm gw}(z)$ and $d_L^{\,\rm em}(z)$ are the directly observable quantities. On the other hand, this might not be sufficient to reconstruct detailed features of $\delta(z)$, which might carry distinct imprints of the cosmological model. 

This fact, together with the existence of at least one example, bigravity, where \eq{eq:fit} is not appropriate, stimulates the development of a complementary approach,  not based on the use of a specific parametrization.
A natural possibility is provided by the method of Gaussian processes (GP).
GP is a regression method based on  ``Supervised Machine Learning".\footnote{\url{http://www.gaussianprocess.org}} It is a nonparametric technique which can reconstruct the distribution of a function from the training of the dataset, without assuming any parametrization for it. The GP method is very suitable for  reconstructing the functions  $d_L^{\,\rm gw}(z)$ and $d_L^{\,\rm em}(z)$ and their derivatives directly from the data. Having the distributions and covariances of these distance functions from GP, the reconstruction of $\delta(z)$ can be obtained. Many applications of GP in cosmology can be found in~\cite{Seikel:2012uu,Seikel:2012cs,Yahya:2013xma,Busti:2014dua,Busti:2015aqa,Cai:2015zoa,Cai:2015pia,Cai:2016vmn,Cai:2016sby,Cai:2017yww}. In particular, in \cite{Cai:2016sby} this technique has been applied to the reconstruction of the dark energy equation of state to simulated data from the Einstein Telescope, in combination with other cosmological informations.
In the present paper we use the Python package \textit{Gaussian process in Python (GaPP)}, which is developed by~\citet{Seikel:2012uu}, to derive our results.  The principle, algorithm and the code details can also be found in~\cite{Seikel:2012uu}. 


\section{Mock datasets}
\label{sec:data}

\begin{figure}
 \centering
 \includegraphics[width=0.40\textwidth]{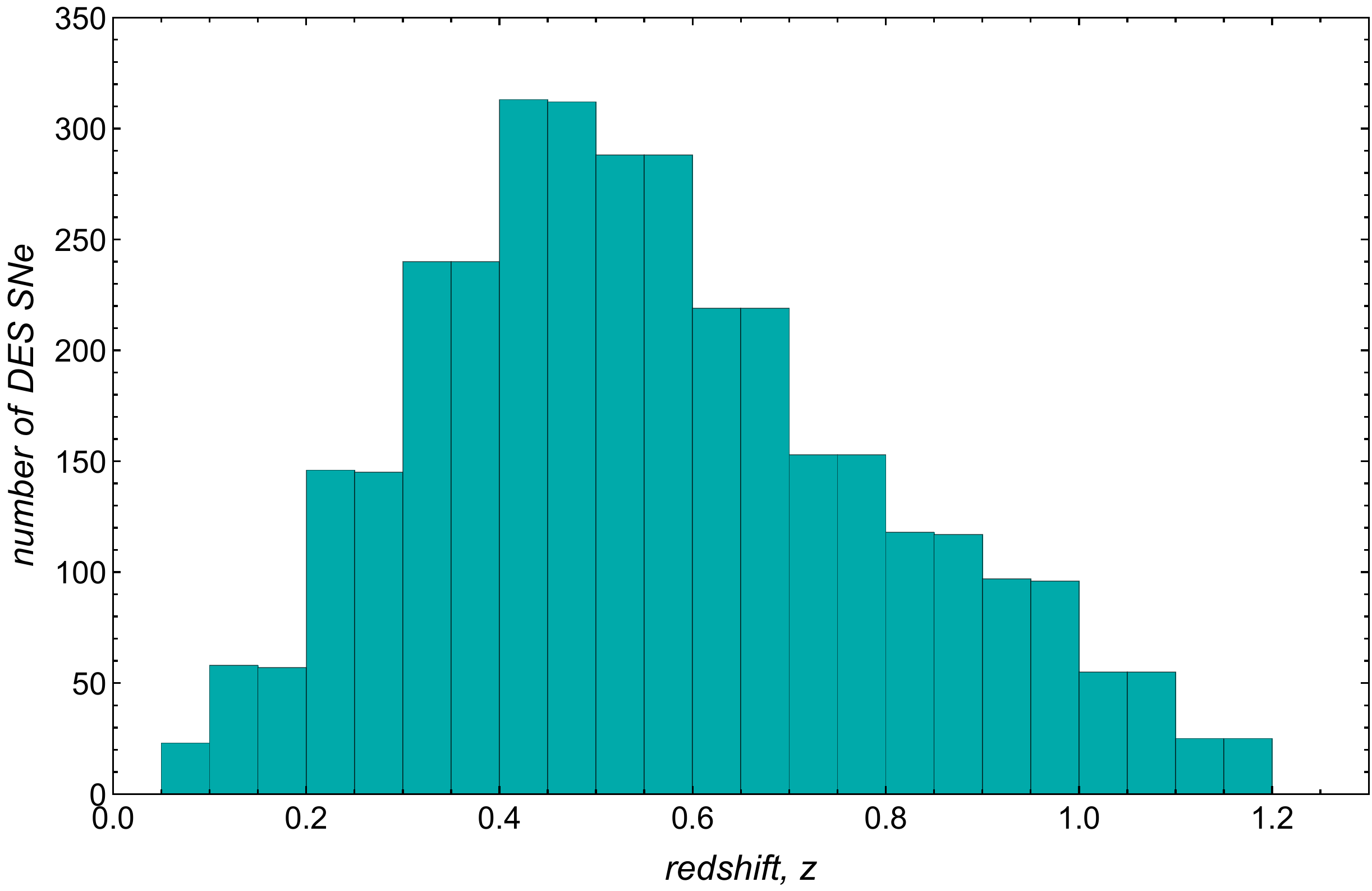}
 \caption{The redshift distribution of mock DES supernovae.}
  \label{fig:DES_redshift_dist}
\end{figure}

\begin{figure}
 \centering
 \includegraphics[width=0.45\textwidth]{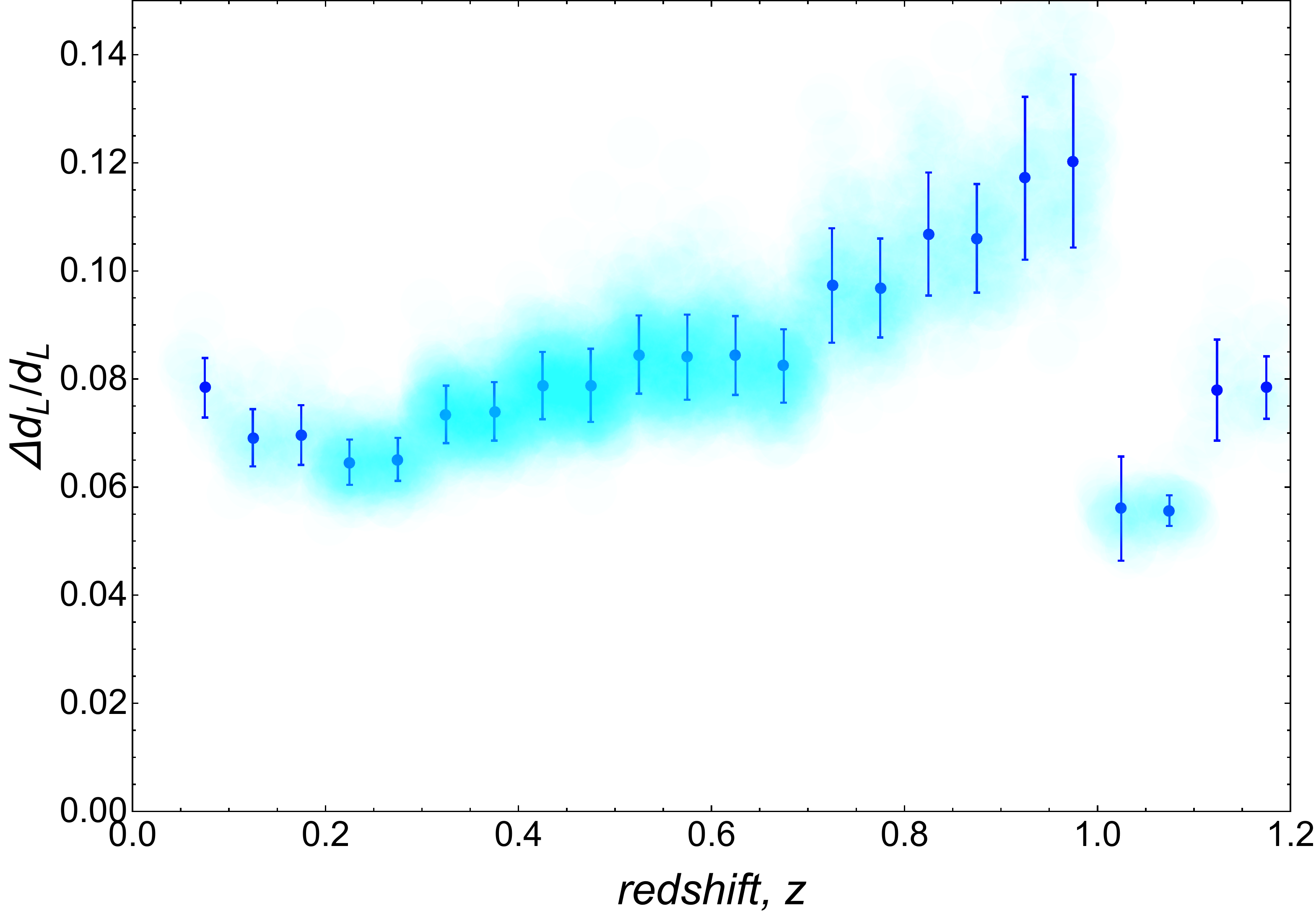}
 \caption{The relative error $\Delta d_L/d_L$ for  the  catalog of DES supernovae shown in
Fig.~\ref{fig:DES_redshift_dist} and Table~\ref{tab:relerr_DES}. The  cyan shaded area corresponds to the single supernovae, while the coordinates of the blue points are given by the mean values of the redshift and of  $\Delta d_L/d_L$ in each redshift bin, with the bins chosen as  in Fig.~\ref{fig:DES_redshift_dist}. The blue error bars are the standard deviations of 
$\Delta d_L/d_L$ in each redshift bin. }
  \label{fig:DES_rel_err}
\end{figure}

To perform our analysis, we consider simulated measurements of the electromagnetic luminosity distance from DES supernovae. The generation of the mock data follows Section III~B of \cite{Cai:2015zoa}, where a redshift range from $z_{\rm min}^{(\rm DES)}$=0.05 to $z_{\rm max}^{(\rm DES)}$=1.2 is considered and the errors on  luminosity distance are estimated by using Table 14 of \cite{Bernstein:2011zf}.  We assume a fiducial $\Lambda$CDM cosmology with Hubble parameter $H_0=67.64  \, \rm{km} \, \rm{s}^{-1} \rm{Mpc}^{-1}$ and present fraction of matter energy density $\Omega_M=0.3087$, corresponding to the mean values obtained from the CMB+SNe+BAO dataset described in Section 3 of~\cite{Belgacem:2019tbw}. For each SN at given $z$, the `measured' value of the luminosity distance is then obtained by randomly scattering the corresponding value of $d_L(z)$ in the fiducial model, according to a Gaussian distribution centered around it and with a standard deviation given by the estimated DES error on luminosity distance.
Fig.~\ref{fig:DES_redshift_dist} shows the redshift distribution of the resulting catalog, containing 3443 SNe Ia whose light curves are obtained in a time period of five years. Table \ref{tab:relerr_DES} gives a simplified description of the mock data, using the same redshift bins as in the histogram of Fig.~\ref{fig:DES_redshift_dist}.  The table shows the mean value and the standard deviation of the relative error on luminosity distance $\Delta d_L/d_L$ for each redshift bin and the same quantities are plotted in Fig.~\ref{fig:DES_rel_err}. The drop in the relative error on luminosity distance beyond redshift $z=1$ is due to selection effects at high redshift and is explained in Section 5.1 of \cite{Bernstein:2011zf}. 

\begin{table}[]
\centering
\begin{tabular}{|c|c|c|c|c|}
\hline
redshift  &       number of          & mean & mean  & standard deviation \\
    bin        &  DES SNe &  redshift                             &                  $\Delta d_L/d_L$                  & of $\Delta d_L/d_L$ \\ \hline
(0.05 , 0.1) & 23 & 0.07500 & 0.07837&0.00552 \\
(0.1 , 0.15) & 58 &0.12478 &0.06913&0.00530 \\
(0.15 , 0.2) & 57 & 0.17478 &0.06963&0.00553 \\
(0.2 , 0.25) & 146 & 0.22491 &0.06461&0.00418 \\
(0.25 , 0.3) & 145 &0.27491 &0.06514&0.00398 \\
(0.3 , 0.35) & 240 &0.32490 &0.07345&0.00533 \\
(0.35 , 0.4) & 240 & 0.37490 &0.07401&0.00542 \\
(0.4 , 0.45) & 313 &0.42496 & 0.07878&0.00623 \\
(0.45 , 0.5) & 312 & 0.47496 &0.07882&0.00676 \\
(0.5 , 0.55) & 288 &0.52491 & 0.08453&0.00725 \\
(0.55 , 0.6) & 288 &0.57491 &0.08404&0.00788 \\
(0.6 , 0.65) & 219 &0.62489 &0.08434&0.00729 \\
(0.65 , 0.7) & 219 &0.67489 &0.08239&0.00679 \\
(0.7 , 0.75)& 153 &0.72484&0.09729&0.01058 \\
(0.75 , 0.8) & 153 & 0.77484 & 0.09683&0.00916 \\
(0.8 , 0.85) & 118 & 0.82490 &0.10683&0.01139 \\
(0.85 , 0.9) & 117 & 0.87490 &0.10603&0.01003 \\
(0.9 , 0.95) & 97 & 0.92487 &0.11714&0.01506 \\
(0.95 , 1) & 96 & 0.97487 & 0.12033&0.01603 \\
(1 , 1.05) & 55 & 1.02455 &0.05602&0.00964 \\
(1.05 , 1.1) & 55 & 1.07455 &0.05565&0.00283 \\
(1.1 , 1.15) & 25 & 1.12400 & 0.07794&0.00936 \\
(1.15 , 1.2) & 26 & 1.17500 &0.07842&0.00579 \\
\hline
\end{tabular}
\caption{The mean value and the variance of the relative error on luminosity distance   $\Delta d_L/d_L$, averaging over the events   in the given redshift bin,  for  the  catalog of DES supernovae shown in
Fig.~\ref{fig:DES_redshift_dist}.
\label{tab:relerr_DES}}
\end{table}

\begin{figure}
 \centering
 \includegraphics[width=0.45\textwidth]{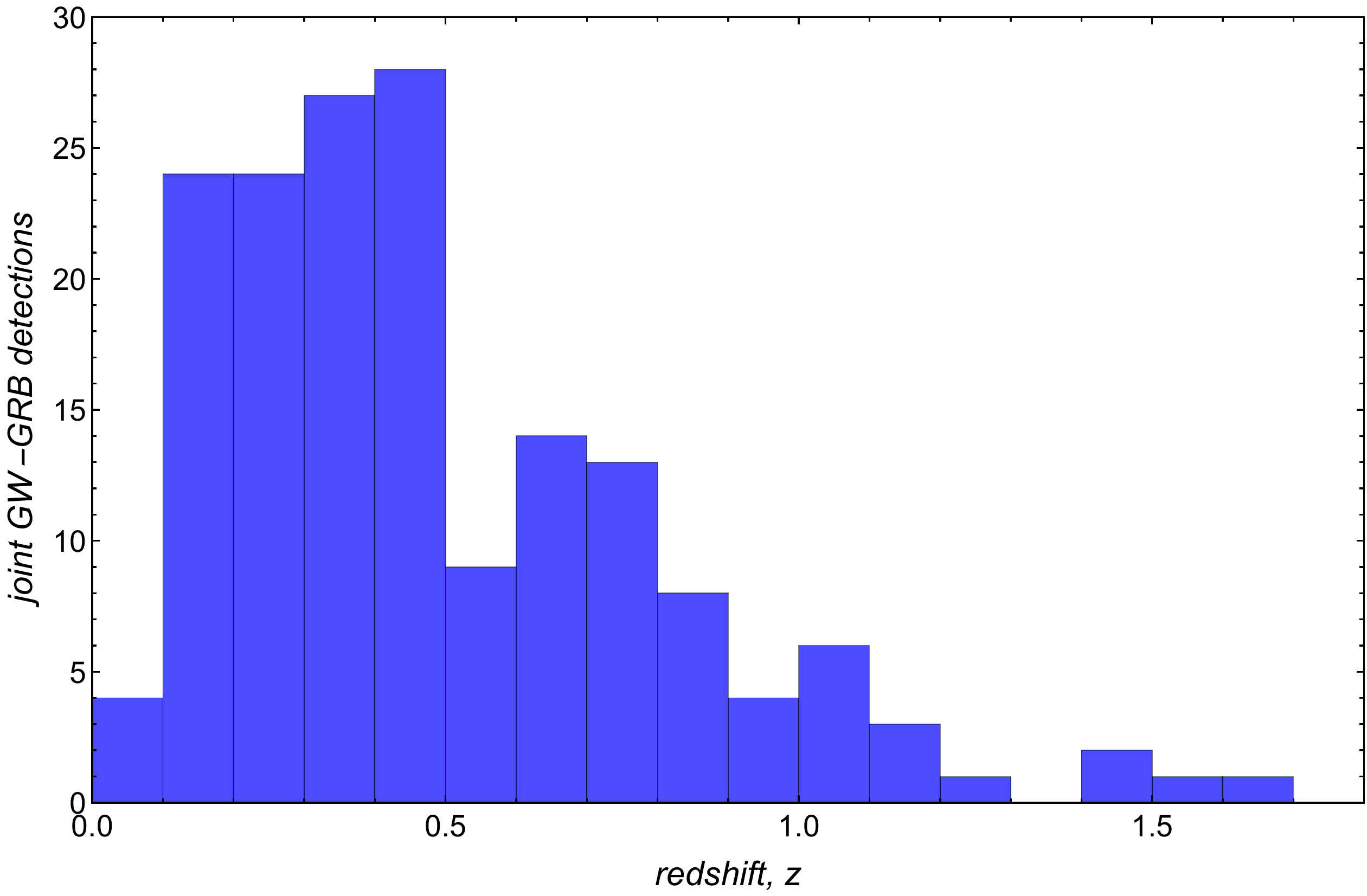}
 \caption{The redshift distribution of joint GW-GRB mock sources, with GW detections at the Einstein Telescope and gamma-ray burst detections at   THESEUS (data from Table~27 of ref.~\cite{Belgacem:2019tbw}).
 }
  \label{fig:ET_redshift_dist}
\end{figure}

\begin{figure}
 \centering
 \includegraphics[width=0.45\textwidth]{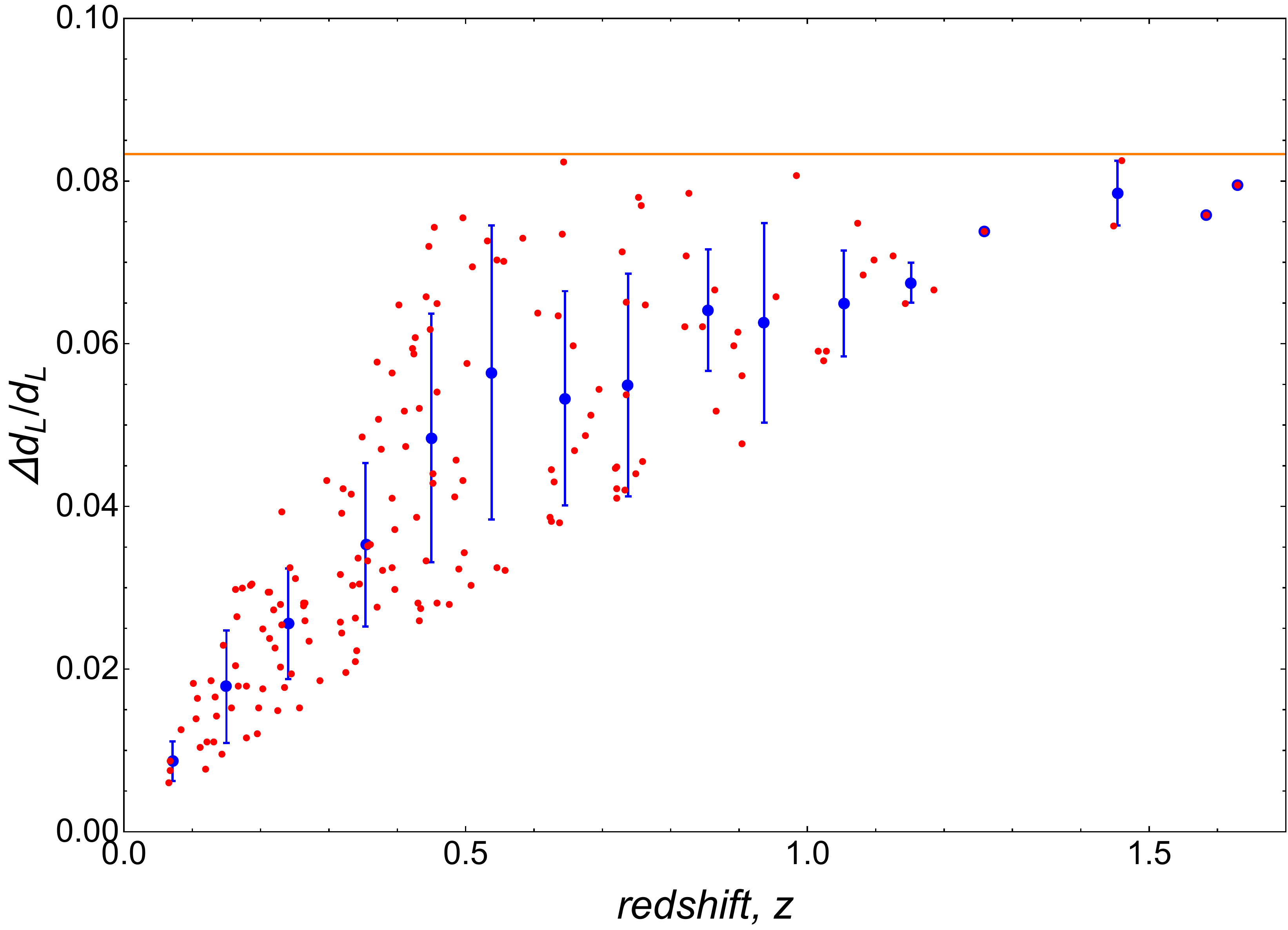}
 \caption{The ET instrumental contribution to the relative error $\Delta d_L/d_L$ for  the  catalog of joint GW-GRB detections at the Einstein Telescope and THESEUS shown in Fig.~\ref{fig:ET_redshift_dist} and described in Table~\ref{tab:relerr_ET}. The  red points correspond to the individual events, while the coordinates of the blue points are given by the mean values of the redshift and of  $\Delta d_L/d_L$ in each redshift bin, with the bins chosen as  in Fig.~\ref{fig:ET_redshift_dist}. The blue error bars are the standard deviations of $\Delta d_L/d_L$ in each redshift bin. The orange horizontal line at $\Delta d_L/d_L$=1/12 corresponds to the SNR=12 threshold for detection.}
  \label{fig:ET_rel_err}
\end{figure}

For the reconstruction of the GW luminosity distance, we consider two cases: the second-generation network made by advanced LIGO, advanced Virgo, Kagra and LIGO India (HLVKI) at target sensitivity, or  a third-generation detector such as the Einstein Telescope. In both cases, we consider GW events from binary neutron star mergers with a redshift determined through the joint observation of a gamma-ray burst (GRB), using the mock catalogs of joint GW-GRB events generated in~\cite{Belgacem:2019tbw}. The generation of those catalogs uses a state-of-the-art treatment of the merger  of binary neutron stars, that takes into account both the star formation rate and models of the time delay between the formation of the binary and the merger. The binary formation is assumed to follow the cosmic star formation rate modeled as in~\cite{Vangioni:2014axa}, while for the time delay distribution we use a  power law with a minimum allowed time to coalescence of 20 Myr. Following Section 2.1 of~\cite{Belgacem:2019tbw}, the overall normalization of the merger rate is fixed by matching with the local rate estimated from the O1 LIGO and O2 LIGO/Virgo observation runs~\cite{LIGOScientific:2018mvr}. We assume a Gaussian distribution for the neutron star masses and we consider 10 yr of data and an 80$\%$ duty cycle for each of the detectors included in the analysis, with a SNR=12 threshold for the total signal-to-noise ratio of a detection (the reader is referred to Section 2.1 of~\cite{Belgacem:2019tbw} for further details about the SNR calculation). The instrumental contribution to the error on luminosity distance has been estimated as $\Delta d_L/d_L$=1/SNR. In principle, a further contribution to  $\Delta d_L/d_L$ comes from weak lensing. In~\cite{Sathyaprakash:2009xt,Zhao:2010sz} it was modeled as $(\Delta d_L(z)/d_L(z))_{\rm lensing}=0.05 z$, while the more recent study in~\cite{Bertacca:2017vod} gives a significantly smaller effect. In any case, even with the pessimistic estimate  $(\Delta d_L(z)/d_L(z))_{\rm lensing}=0.05 z$, the lensing contribution is negligible for GW detections at HLVKI and it is also subdominant for sources detected at the Einstein Telescope with $z<1.5$, which constitute 99\% of all the events with a detected GRB  counterpart in the final ET catalog that we will consider.  On the other hand, for low-redshift sources the contribution due to the peculiar Hubble flow is important and, if not corrected for,  gives an error on the redshift that can be modeled as due to an unknown  peculiar velocity of order $\pm 200\, \rm{km/s}$~\cite{Chen:2017rfc}. Here we assume, conservatively, that  this error is not corrected for, and we then  propagate it to determine its contribution to the error on luminosity distance (see however \cite{Mukherjee:2019qmm} for strategies for correcting the errors due to  peculiar velocities). The gravitational wave luminosity distance for the events of the mock catalog is drawn from a Gaussian distribution centered around the value from a fiducial cosmology and with standard deviation given by the sum in quadrature of the errors on luminosity distance described before. In this work we consider two different fiducial cosmologies for GW detections:
\begin{itemize}
\item the $\Lambda$CDM model with $H_0=67.64  \, \rm{km} \, \rm{s}^{-1} \rm{Mpc}^{-1}$ and $\Omega_M=0.3087$ already used for the DES mock catalog;
\item the RT nonlocal model with initial conditions set at $\Delta N=64$ e-folds before the end of inflation, introduced in Section~\ref{sec:param}.
\end{itemize}
In the following, we will simply refer to those choices as the ``$\Lambda$CDM" and the ``RT" fiducial cosmologies, without repeating again the specifications.
In the case of the RT fiducial, the modifications in the cosmological background with respect to $\Lambda$CDM are so small that the electromagnetic luminosity distances in the two models differ by just a few parts per thousand. That contribution is utterly negligible when compared to the modified gravitational wave propagation effect in~\eq{eq:dLgwdLem} for the  RT model, which amounts to a deviation from GR by more than $60\%$. As a consequence the fiducial values of the gravitational wave luminosity distance can be simply obtained by multiplicating those from $\Lambda$CDM by the factor $d_L^{\,\rm gw}(z)/d_L^{\,\rm em}(z)$ in~\eq{eq:fit} with $\Xi_0=1.67$ and $n=1.94$. For the error on luminosity distance, we reasonably assume the relative error to be the same as in the case of the $\Lambda$CDM fiducial cosmology, so we also multiply the $\Lambda$CDM absolute error by the same factor in~\eq{eq:fit}. Beside the intrinsic interest of the model, the use of the RT nonlocal model as a fiducial cosmological model can be seen as a case study for exploring the results of Gaussian processes reconstruction when the reference theory has a non-trivial GW propagation equation, despite being fully compatible with electromagnetic observations.
Once the mock catalog of GW detections is built, we extract the events whose  GRB emission is actually detected. A full explanation for this final stage in the construction of mock data can be found in Section 2.2 of~\cite{Belgacem:2019tbw}, where the criterion for retaining a GW event in the GW-GRB catalog requires the peak flux of the GRB emission to be above the flux limit of the satellite considered for detection, which is Fermi-GBM for GW events at the HLVKI 2G network, and the proposed THESEUS mission~\cite{Amati:2017npy} for GW events detected at the Einstein Telescope. In the ET/THESEUS catalog we only consider a fraction $1/3$ of the events selected by the procedure above, since only the central part of the XGIS spectrometer on board THESEUS will be capable of arcmin localization of sources (this corresponds to the ``realistic" assumption for the FOV of THESEUS in Section 2.2.2 of~\cite{Belgacem:2019tbw}).
Table~\ref{tab:cat2G} shows a realization (containing 15 sources) for the catalog of joint GW-GRB events at the second-generation network HLVKI, assuming  $\Lambda$CDM as fiducial cosmology. For the case of Einstein Telescope we present the redshift distribution of the sources in Fig.~\ref{fig:ET_redshift_dist} and the corresponding description of the instrumental error on luminosity distance in Table~\ref{tab:relerr_ET}, with the same meaning of columns as in Table~\ref{tab:relerr_DES}. Fig.~\ref{fig:ET_rel_err} is a plot for the error on luminosity distance of the 169 events in the catalog realization considered in Table~\ref{tab:relerr_ET}.

\begin{table}
\begin{tabular}{|c|c|c|}
\hline
$z$ & $d_L$ (Mpc) & $\Delta d_L$ (Mpc) \\ \hline
0.029271 & 134.815 & 4.000 \\
0.035195 & 157.475 & 5.636 \\
0.060585 & 283.567 & 18.706 \\
0.066283 & 316.373 & 14.509 \\
0.071053 & 327.381 & 20.085 \\
0.071730 & 342.952 & 16.957 \\
0.076180 & 341.595 & 22.360 \\
0.081819 & 418.469 & 30.238 \\
0.088698 & 396.734 & 25.757 \\
0.091869 & 402.590 & 34.170 \\
0.094237 & 406.423 & 31.472 \\
0.095288 & 432.996 & 36.423 \\
0.099956 & 491.071 & 31.721 \\
0.102531 & 461.627 & 36.858 \\
0.114869 & 626.939 & 43.010 \\
\hline
\end{tabular}
\caption{The events in a realization of joint GW-GRB mock sources, with gravitational wave and gamma-ray burst detections at the HLVKI network and Fermi-GBM respectively (from ref.~\cite{Belgacem:2019tbw}).
\label{tab:cat2G}}
\end{table}

\begin{table}
\centering
\begin{tabular}{|c|c|c|c|c|}
\hline
redshift  &       number of         & mean & mean  & standard deviation \\
    bin        &  GW-GRB &  redshift                             &                  $\Delta d_L/d_L$                  & of $\Delta d_L/d_L$ \\ \hline
(0 , 0.1) & 4 &0.07108 & 0.00868&0.00244 \\
(0.1 , 0.2) & 24 &0.15001 &0.01784&0.00692 \\
(0.2 , 0.3) & 24 &0.24043 &0.02558&0.00680 \\
(0.3 , 0.4) & 27 &0.35355 & 0.03529&0.01004 \\
(0.4 , 0.5) & 28 &0.44966 & 0.04843&0.01528 \\
(0.5 , 0.6) & 9 & 0.53785 & 0.05646&0.01807 \\
(0.6 , 0.7) & 14 & 0.64540 &0.05329&0.01318 \\
(0.7 , 0.8) & 13 &0.73793 &0.05493&0.01368 \\
(0.8 , 0.9) & 8 &0.85497 &0.06413&0.00746 \\
(0.9 , 1.0) & 4 & 0.93702 & 0.06257&0.01228 \\
(1.0 , 1.1) & 6 & 1.05334 & 0.06494&0.00651 \\
(1.1 , 1.2) & 3 &1.15162 &0.06749&0.00246 \\
(1.2 , 1.3) & 1 &1.25943 & 0.07373&0 \\
(1.3 , 1.4)& -- &-- & --&-- \\
(1.4 , 1.5) &2 & 1.45375 &0.07851&0.00398 \\
(1.5 , 1.6) & 1 & 1.58407 &0.07577&0 \\
(1.6 , 1.7) & 1 & 1.62843 & 0.07947&0 \\
\hline
\end{tabular}
\caption{The mean value and the variance of the ET instrumental contribution to   $\Delta d_L/d_L$, averaging over the events   in the given redshift bin,  for  the specific realization of the catalog of joint GW-GRB detections shown in
Fig.~\ref{fig:ET_redshift_dist} (from ref.~\cite{Belgacem:2019tbw}).
\label{tab:relerr_ET}}
\end{table}

\section{Results}

\begin{figure*}
\includegraphics[width=0.8\textwidth]{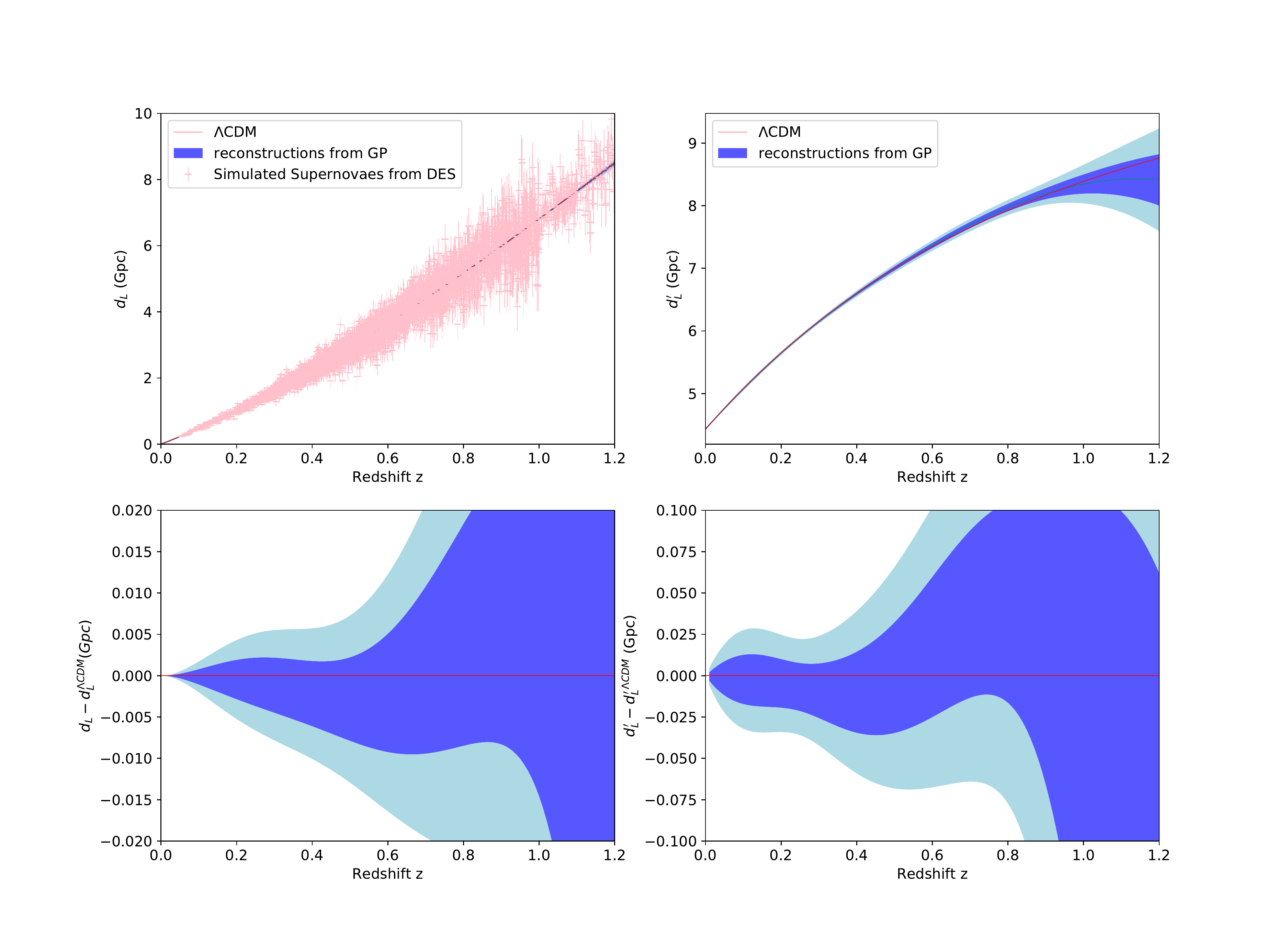}
\caption{Reconstruction of the EM luminosity distance from the simulated DES catalog. In all  panels, the blue and light blue regions correspond to $68\%$ and $95\%$ confidence levels respectively, while the red curve is the fiducial to be reconstructed. Upper left panel: $\dem(z)$ and the mock DES supernovae used. Lower left: the difference between the reconstructed $\dem(z)$ and the fiducial one. Upper right: the derivative $d_L'^{\rm em}(z)$. Lower right: the difference between the reconstructed $d_L'^{\rm em}(z))$ and its fiducial curve.}
\label{fig:dL_DES}
\end{figure*}

We use the Gaussian processes method to reconstruct the functions $\dem(z)$ and $\dgw(z)$, as well as their derivatives with respect to redshift $d_L'^{\rm em}(z)$ and $d_L'^{\rm gw}(z)$, with the mock datasets of electromagnetic and gravitational-wave observations described in Section~\ref{sec:data}. Fig.~\ref{fig:dL_DES} shows the result of the reconstruction for the electromagnetic luminosity distance and its derivative with respect to redshift, by using the DES mock dataset\footnote{We recall from Section~\ref{sec:data} that, contrary to $\dgw(z)$, the difference in $\dem(z)$ between $\Lambda$CDM and the RT model is very small (a few parts per thousand). Therefore we can safely neglect it and consider the same mock catalog of supernovae for both  fiducial cosmological models.}. For the GW luminosity distance, we show the results in two separate subsections for  each mock dataset used (HLVKI or ET). Given a reconstruction of the functions $\dgw(z)$ and $\dem(z)$,  we can then reconstruct the luminosity distance ratio 
\begin{equation}
D(z)\equiv\dgw(z)/\dem(z)
\label{eq:Dz}
\end{equation}
as well as the function $\delta(z)$ that, from \eq{deltalogd},   is given by
\begin{equation}
\delta(z)=(1+z)\left(\frac{d_L'^{\rm em}(z)}{d_L^{\rm em}(z)}-\frac{d_L'^{\rm gw}(z)}{d_L^{\rm gw}(z)}\right) \,.
\label{eq:deltaz}
\end{equation}
For each dataset used (HLVKI or ET) we will show the result both with $\Lambda$CDM as fiducial, and with RT as fiducial.
More precisely, when we take the RT model (with $\Delta N=64$) as fiducial, we will assume that its prediction is exactly given by \eq{eq:fit} with $\Xi_0=1.67$ and $n=1.94$.
Actually, for the RT model with large $\Delta N$, \eq{eq:fit}   fits extremely well the prediction of the model
for $\dgw(z)/\dem(z)$  (obtained from the numerical integration of the relevant equations, that involve the numerical evolution of the auxiliary fields of the model, see \cite{Dirian:2016puz}). In contrast, at very small $z$ the numerical result for $\delta(z)$ differs somewhat from that obtained by the parametrization (\ref{paramdeltaz}); the correct numerical result for $\delta(z)$ is shown in Fig.~2 of \cite{Belgacem:2019lwx}. Again, the difference is due to the fact that small details in  $\delta(z)$ gets smoothed out when performing  the integration in \eq{eq:dLgwdLem}, as we already discussed above.
Here, in order to illustrate the methodology, we simply assume that the result for $\dgw(z)/\dem(z)$ is {\em exactly}  given by \eq{eq:fit}, so that the result for $\delta(z)$ would also be given  by the corresponding equation (\ref{paramdeltaz}), which is obtained  by applying \eq{deltalogd} to \eq{eq:fit}.

\subsection{Results for the HLVKI network}

\begin{figure*}
\includegraphics[width=0.9\textwidth]{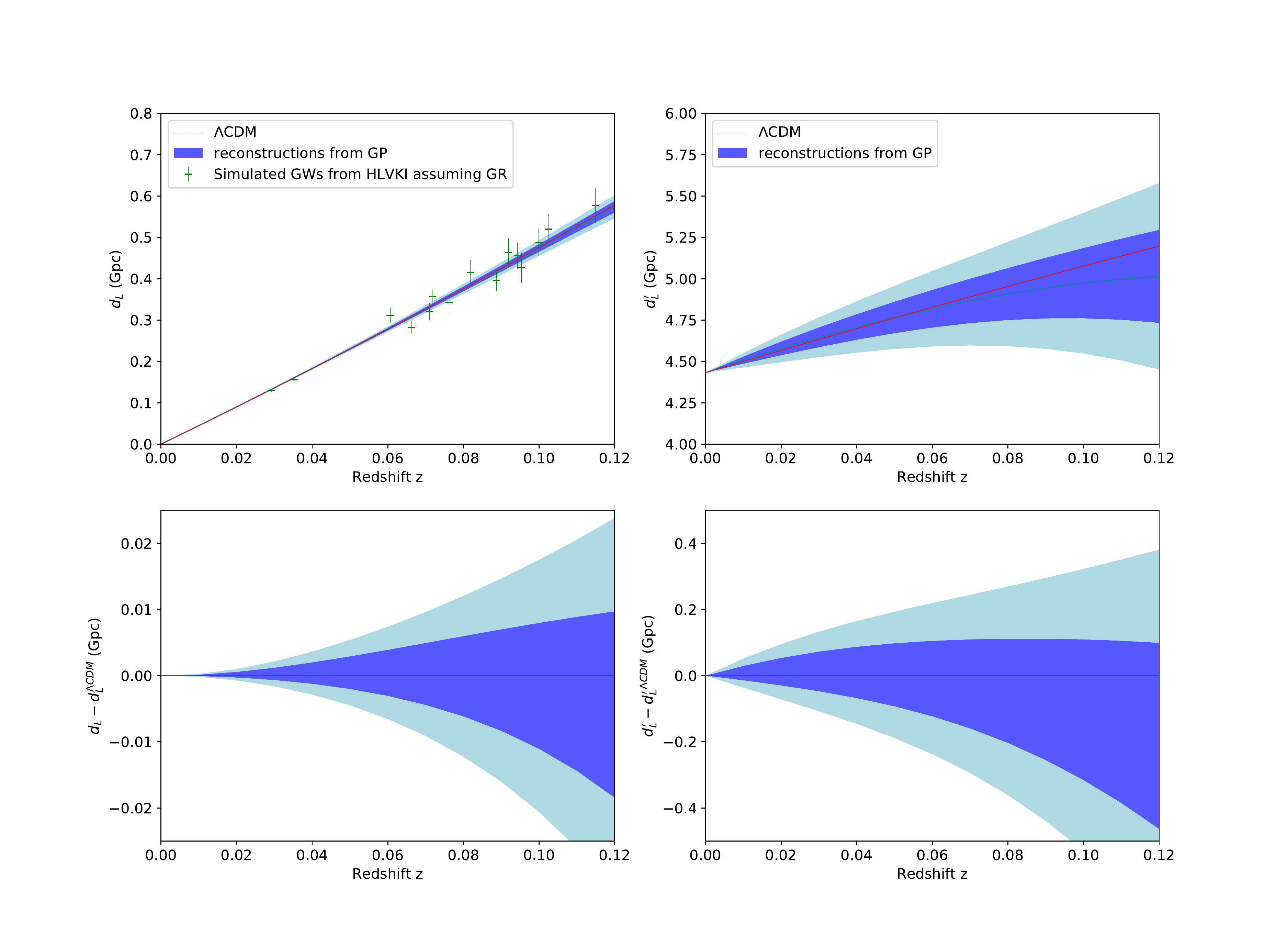}
\caption{Reconstruction of the GW luminosity distance from mock detections at HLVKI with electromagnetic counterpart, for the $\Lambda$CDM fiducial cosmology. The panels are organized as in Fig.~\ref{fig:dL_DES} and the green points in the upper left panel are the mock data in the HLVKI catalog.
Note the difference in vertical scale between the two lower panels.}
\label{fig:dL_HLVKI}
\end{figure*}

\begin{figure*}
\includegraphics[width=0.9\textwidth]{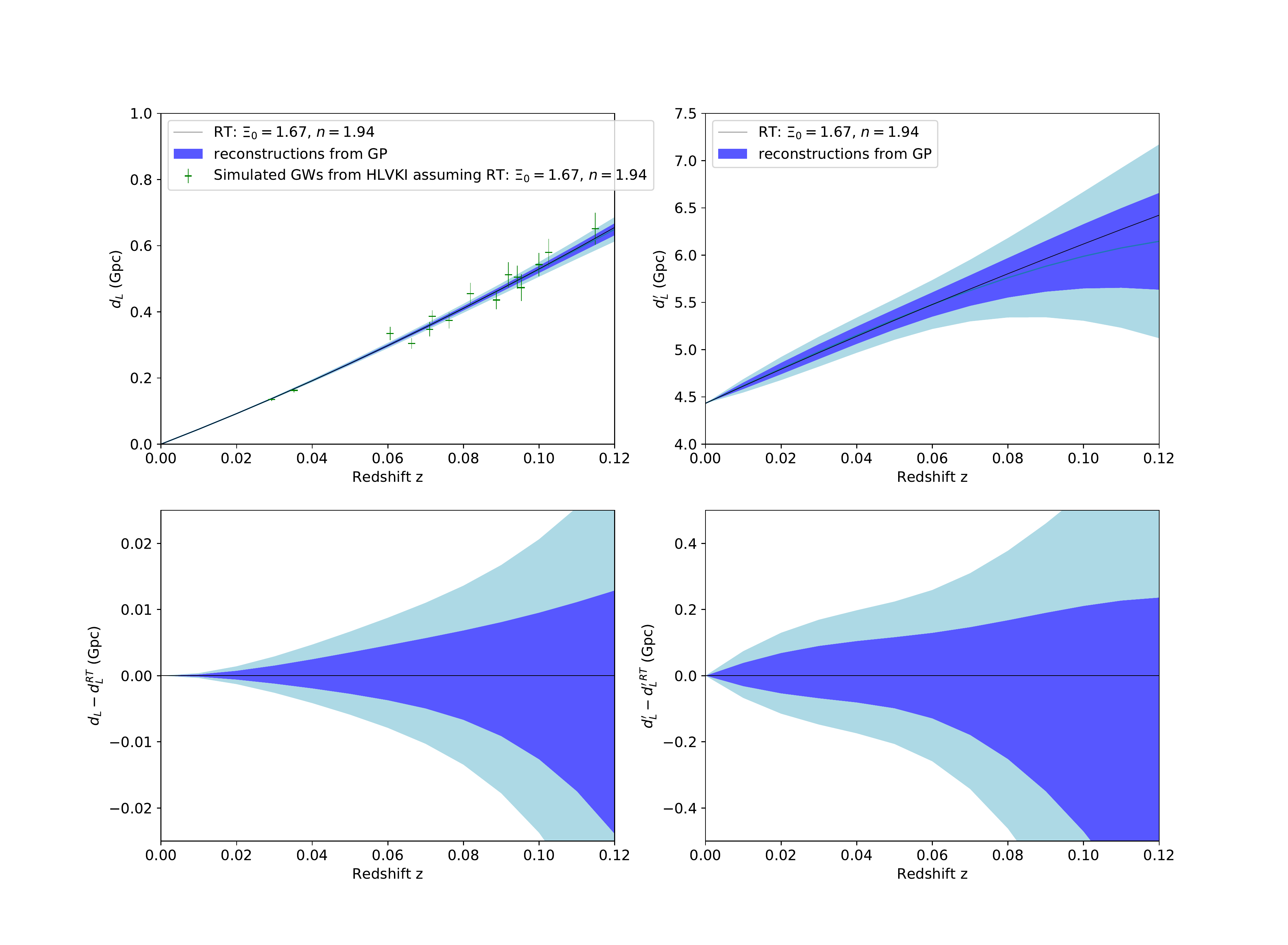}
\caption{Reconstruction of the GW luminosity distance from mock detections at HLVKI with electromagnetic counterpart, for the RT fiducial cosmology. Panels as in Fig.~\ref{fig:dL_HLVKI}.}
\label{fig:dL_MGHLVKI}
\end{figure*}

\begin{figure*}
\centering
\includegraphics[width=0.45\textwidth]{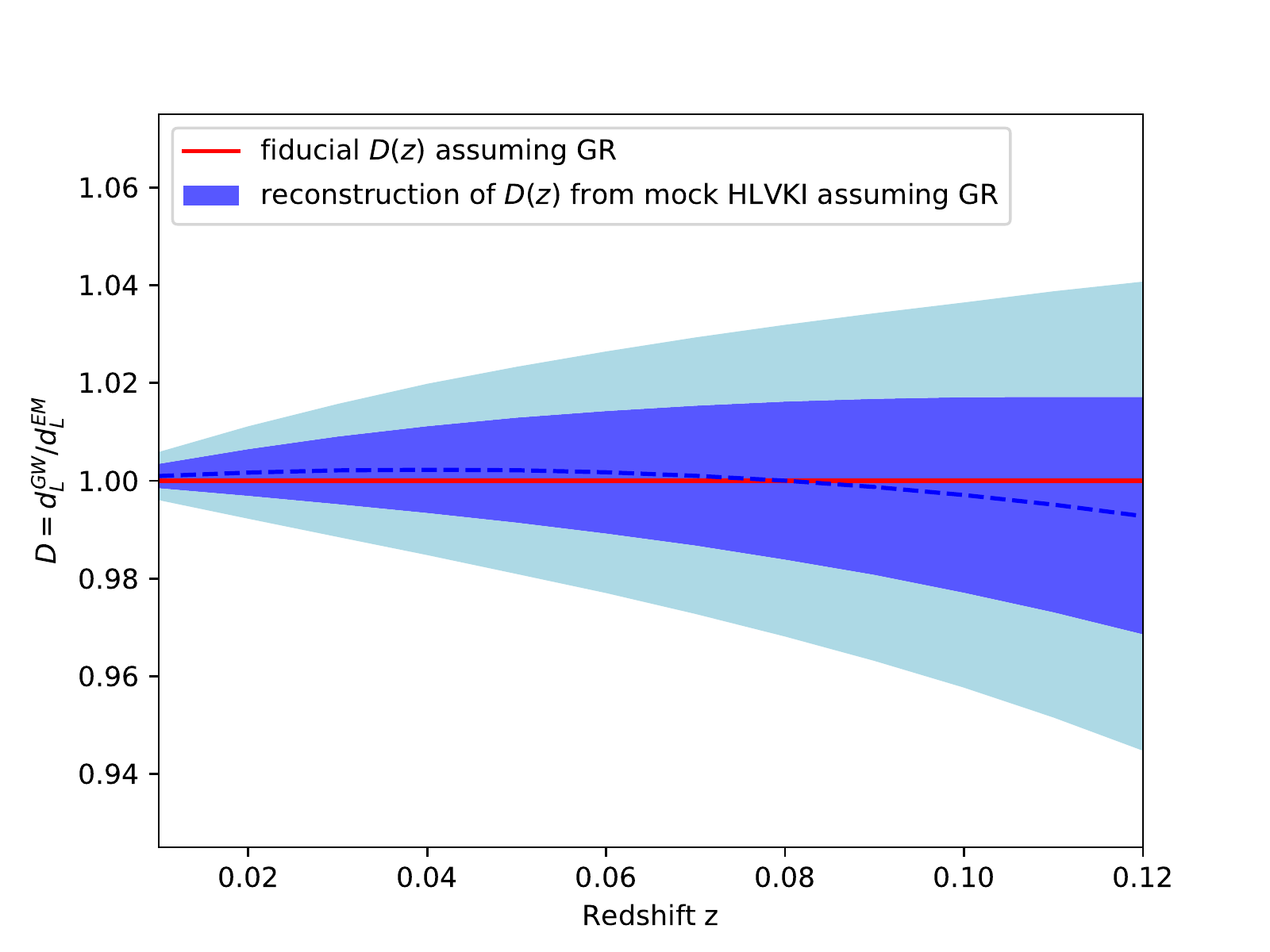}
\includegraphics[width=0.45\textwidth]{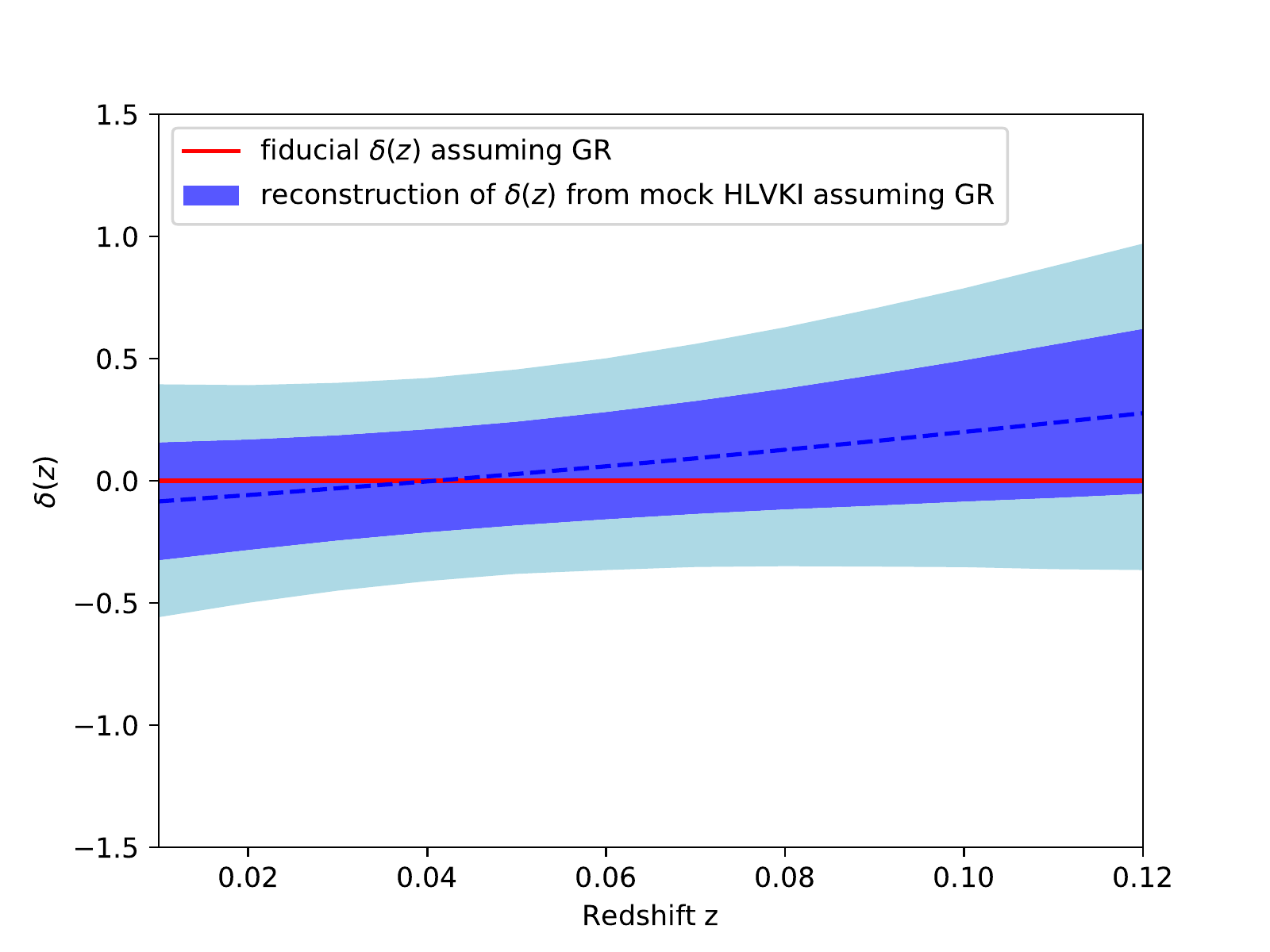}
\caption{Results from the mock HLVKI and DES catalogs, assuming the $\Lambda$CDM fiducial cosmology. Left panel: reconstruction of the ratio $D(z)=\dgw(z)/\dem(z)$. Right panel: reconstruction of $\delta(z)$.}
\label{fig:results_HLVKI}
\end{figure*}

\begin{figure*}
\centering
\includegraphics[width=0.45\textwidth]{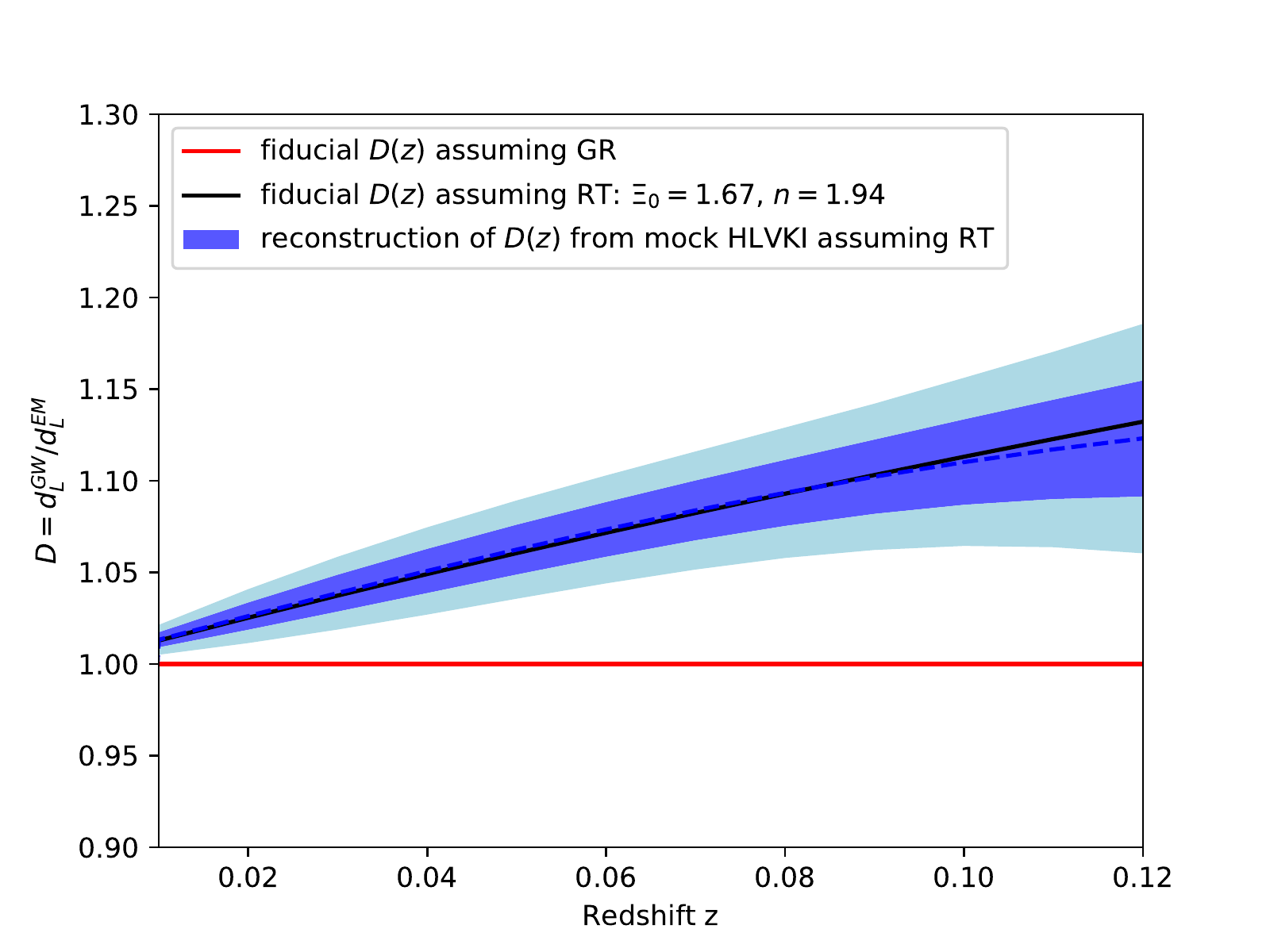}
\includegraphics[width=0.45\textwidth]{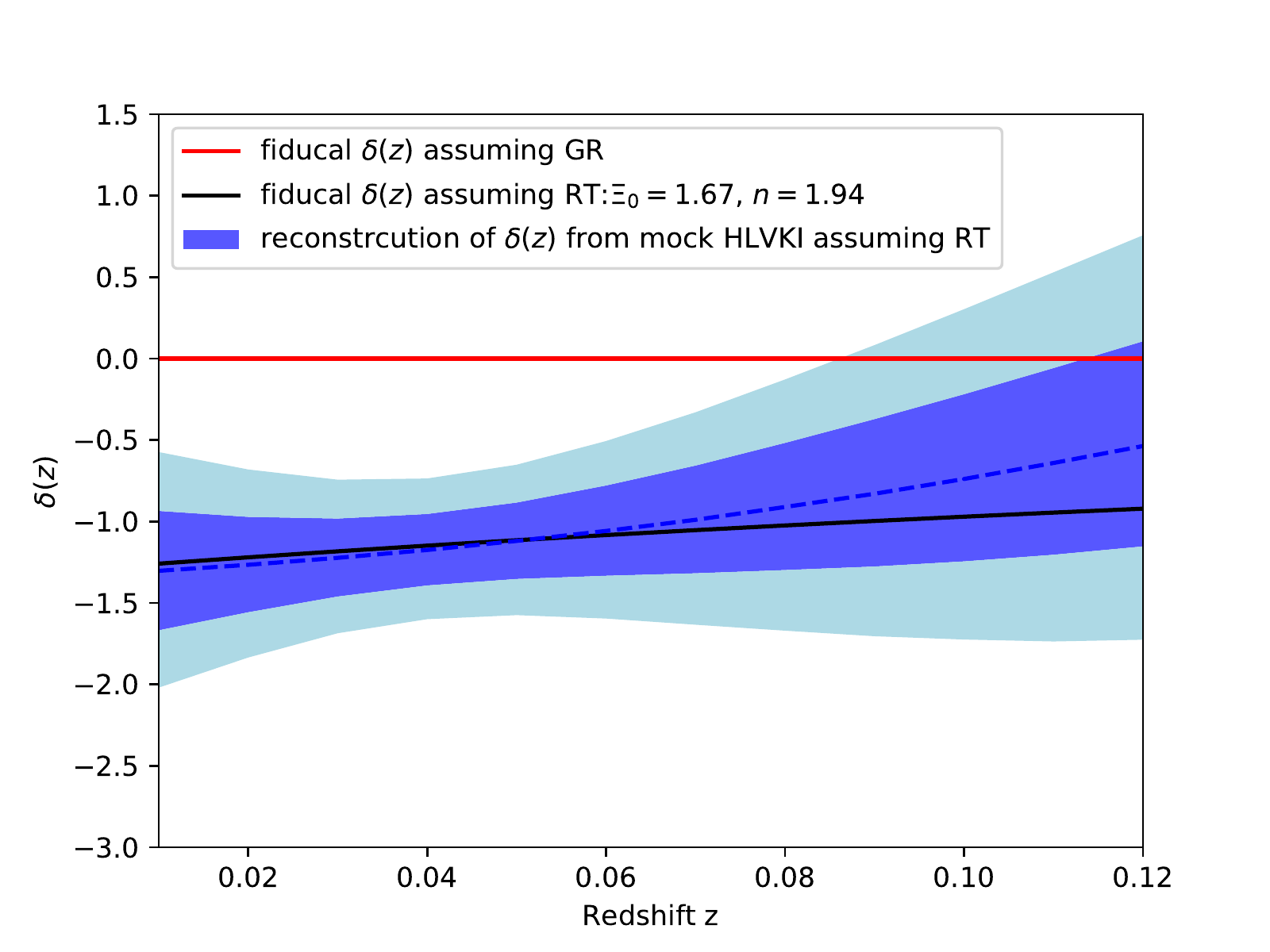}
\caption{As in Fig.~\ref{fig:results_HLVKI}, using the mock HLVKI and DES catalogs for the RT fiducial cosmology.}
\label{fig:results_MGHLVKI}
\end{figure*}

In Fig.~\ref{fig:dL_HLVKI} we plot the reconstructed GW luminosity distance obtained from the mock catalog of standard sirens at the second-generation network HLVKI, assuming $\Lambda$CDM as fiducial cosmology, while Fig.~\ref{fig:dL_MGHLVKI} is the analogous plot in the case of the RT fiducial.
In the case of $\Lambda$CDM fiducial, the reconstructions of the ratio $D(z)=\dgw(z)/\dem(z)$ and of $\delta(z)$ [using \eq{eq:deltaz}] from the HLVKI and DES mock catalogs are plotted in Fig.~\ref{fig:results_HLVKI}.
Similarly, Fig.~\ref{fig:results_MGHLVKI} shows the results assuming the RT fiducial cosmology.
We provide the reconstructions up to the maximum redshift reached in the HLVKI mock catalog $z_{\rm max}^{(\rm HLVKI)}\simeq0.12$ (see Table~\ref{tab:cat2G}), which is smaller than the maximum redshift $z_{\rm max}^{(\rm DES)}$=1.2 in the simulated DES catalog of supernovae.

It is interesting to compare the results on modified GW propagation from Gaussian processes reconstruction with those obtained by using the $(\Xi_0,n)$ parametrization for the function $D(z)=\dgw(z)/\dem(z)$ in \eq{eq:fit}. In this case our strategy is to implement the parametrization in the CLASS Boltzmann code and run a MCMC. The parameter $n$ plays a much less important role than $\Xi_0$ for $D(z)$ because it just regulates the precise interpolation between the value $D(z=0)=1$ and the asymptotic $D(z\gg 1)=\Xi_0$. Furthermore, chains with both $\Xi_0$ and $n$ as free parameters fail to converge and for these reasons we just keep $n$ fixed and look at the precision reached by the MCMC in obtaining the $\Xi_0$ value of the fiducial cosmological model. For the RT fiducial cosmology we set $n$ to the actual fiducial value $n^{\rm (RT)}=1.94$ and we want the MCMC to recover the fiducial $\Xi_0^{\rm (RT)}=1.67$. In the $\Lambda$CDM case, $n^{(\Lambda \rm{CDM})}$ is not determined and when running the MCMC we choose to set it to $n=2.5$, which is in the ballpark of typical values predicted in modified gravity theories (for example in some of the nonlocal gravity models). We then want to recover $\Xi_0^{(\Lambda \rm{CDM})}=1$.

\begin{table}[H]
\centering
\begin{tabular}{|c|c|c|}
 \hline
                               &  DES+HLVKI& CMB+BAO+DES+HLVKI     \\ \hline
$\Delta\Xi_0$ &  0.127 (12.7\%)            & 0.127 (12.7\%)                   \\
$\Delta H_0/H_0$ &  0.38\%                   & 0.21\%                    \\
$\Delta\oma/\oma$ &  3.30\%                  & 0.71\%                       \\
\hline
\end{tabular}
\caption{Accuracy ($1\sigma$ level) in the reconstruction of $\Xi_0$, $H_0$ and $\oma$ with DES+HLVKI and CMB+BAO+DES+HLVKI, assuming $\Lambda$CDM as the fiducial cosmology for the HLVKI dataset. The relative error on $\Xi_0$ is the same as the absolute error, because of the fiducial value $\Xi_0^{(\Lambda \rm{CDM})}=1$.}
\label{tab:HLVKI_LCDM}
\end{table}

\begin{table}[H]
\centering
\begin{tabular}{|c|c|c|}
 \hline
                               &  DES+HLVKI& CMB+BAO+DES+HLVKI     \\ \hline
$\Delta\Xi_0/\Xi_0$ &  8.32\%            & 8.32\%                   \\
$\Delta H_0/H_0$ &  0.38\%                   & 0.21\%                    \\
$\Delta\oma/\oma$ &  3.30\%                  & 0.74\%                       \\
\hline
\end{tabular}
\caption{As in~Table~\ref{tab:HLVKI_LCDM}, assuming the RT model as the fiducial cosmology for the HLVKI dataset. The absolute error on $\Xi_0$ can be easily found recalling the fiducial value $\Xi_0^{\rm (RT)}=1.67$.}
\label{tab:HLVKI_RT}
\end{table}

For each fiducial cosmology we run MCMCs with two choices of combined datasets:
\begin{enumerate}
\item the simulated DES supernovae and the mock catalog of GW detections at HLVKI (with electromagnetic counterpart) described in Section~\ref{sec:data};
\item to reduce the degeneracies between cosmological parameters, in addition to the DES and HLVKI mock catalogs, we also consider the following CMB and BAO data
\begin{itemize}
\item{\em CMB.} We use the 2015 \textit{Planck}  \cite{Adam:2015rua} measurements of the angular (cross-)power spectra, including  full-mission lowTEB data for low multipoles ($\ell \leq 29$) and the high-$\ell$ Plik  TT,TE,EE (cross-half-mission) ones for the high multipoles ($\ell > 29$) of the temperature and polarization auto- and cross- power spectra \cite{Ade:2015rim}.
We also include the temperature$+$polarization (T$+$P) lensing data, using  only the conservative multipole range $\ell =40-400$  \cite{Planck_2015_Lkl,Ade:2015zua}.

\item {\em Baryon Acoustic Oscillations} (BAO). We use the isotropic constraints provided by 6dFGS at $z_{\rm eff}=0.106$ \cite{Beutler:2011hx}, SDSS-MGS DR7 at $z_{\rm eff}=0.15$ \cite{Ross_SDSS_2014} and BOSS LOWZ at $z_{\rm eff}=0.32$  \cite{Anderson:2013zyy}, as well as  the anisotropic constraints from CMASS at $z_{\rm eff}=0.57$ \cite{Anderson:2013zyy}.
\end{itemize}
\end{enumerate}

The two combined datasets will be denoted as ``DES+HLVKI" and  ``CMB+BAO+DES+HLVKI", respectively. In the first case the MonteCarlo is sensitive to the set of parameters $\{H_0,\Omega_M,\Xi_0\}$, while in the second case the parameters that come into play are $\{H_0,\omega_b,\omega_c,A_s,n_s,\tau_{re},\Xi_0\}$ and the total matter fraction $\Omega_M$ is a derived parameter.
The quantities $\omega_b = \Omega_b h^2$ and $\omega_c = \Omega_c h^2$ are the physical baryon and cold dark matter density fractions today, respectively (where $h$ is defined by the relation $H_0 = 100 h \, \rm{km} \, \rm{s}^{-1} \rm{Mpc}^{-1}$). $A_s$ and $n_s$ are the amplitude and tilt of the primordial scalar perturbations,   and $\tau_{\rm re}$ is  the reionization optical depth. We keep the sum of neutrino masses fixed, at the value $\sum_{\nu}m_{\nu}=0.06$~eV, as in the {\em Planck} baseline analysis~\cite{Planck_2015_CP}.

\begin{figure*}
\includegraphics[width=0.35\textwidth]{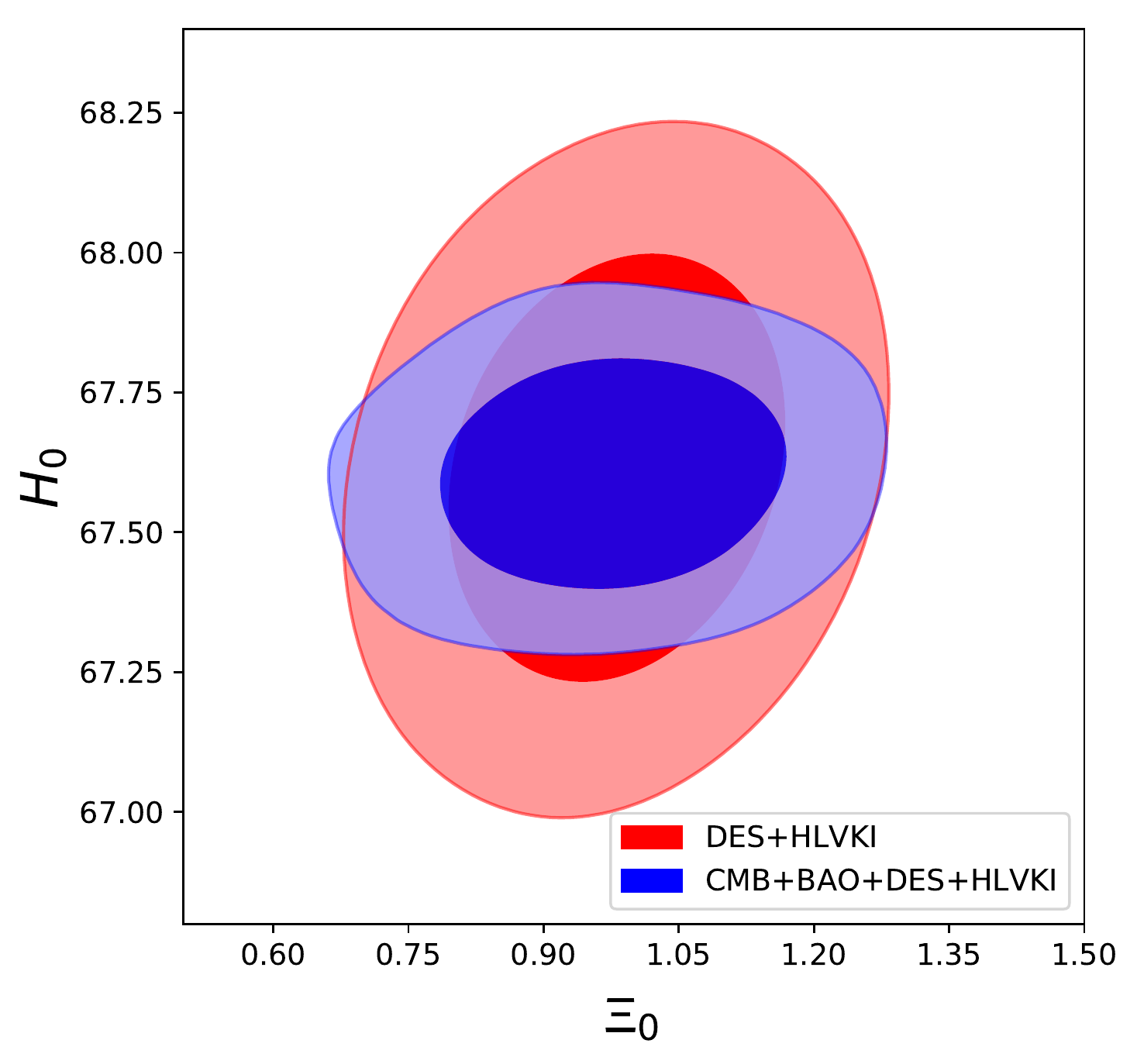}
\includegraphics[width=0.35\textwidth]{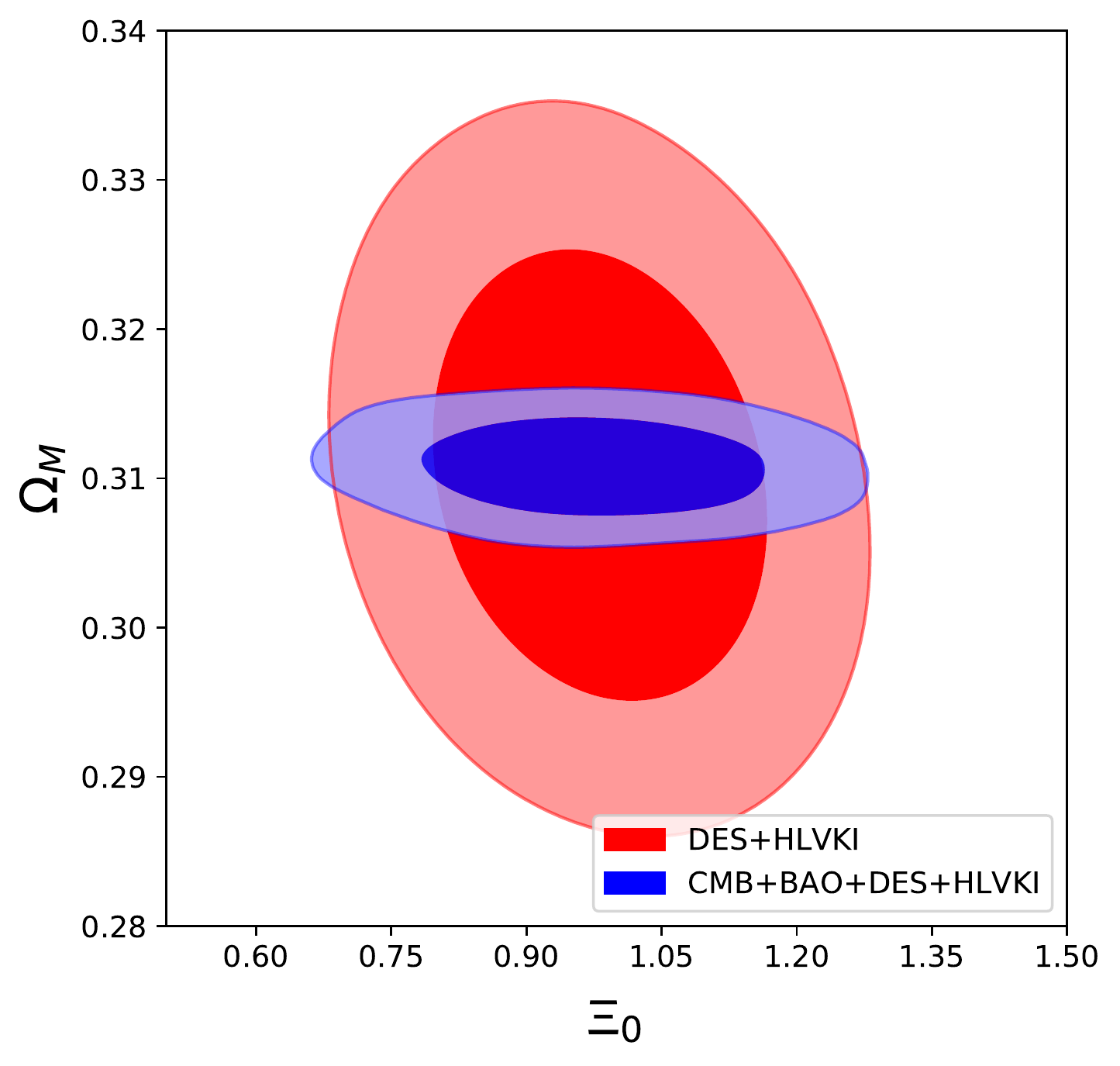}
\caption{The  $1\sigma$ and $2\sigma$
contours  of two-dimensional likelihoods, from DES+HLVKI (red) and CMB+BAO+DES+HLVKI (blue). The fiducial cosmology for the HLVKI dataset is $\Lambda$CDM. Left:  in the $(\Xi_0,H_0)$ plane. Right:  in the $(\Xi_0,\oma)$ plane.}
\label{fig:HLVKI_fidLCDM}
\end{figure*}

\begin{figure*}
\includegraphics[width=0.35\textwidth]{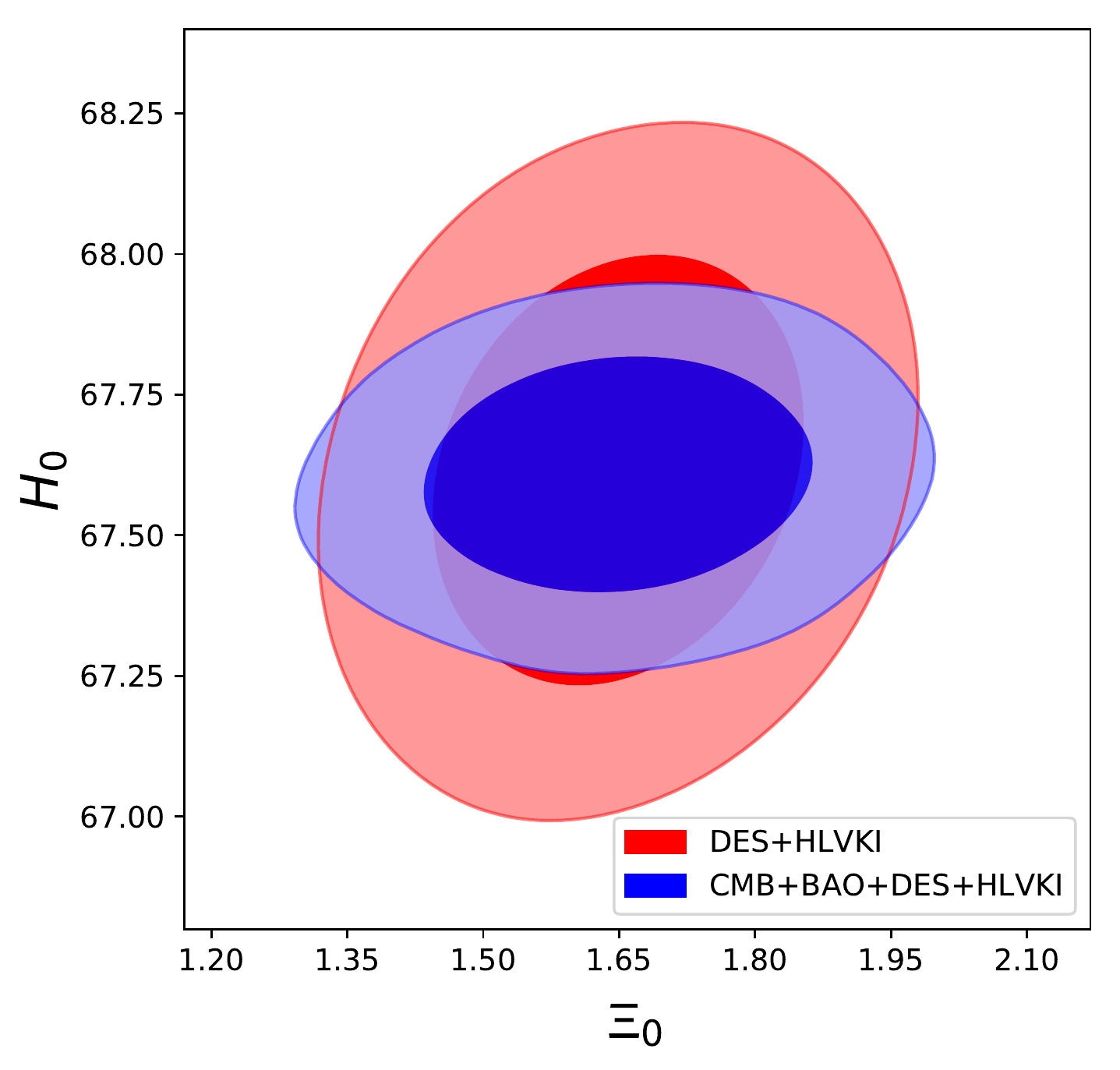}
\includegraphics[width=0.35\textwidth]{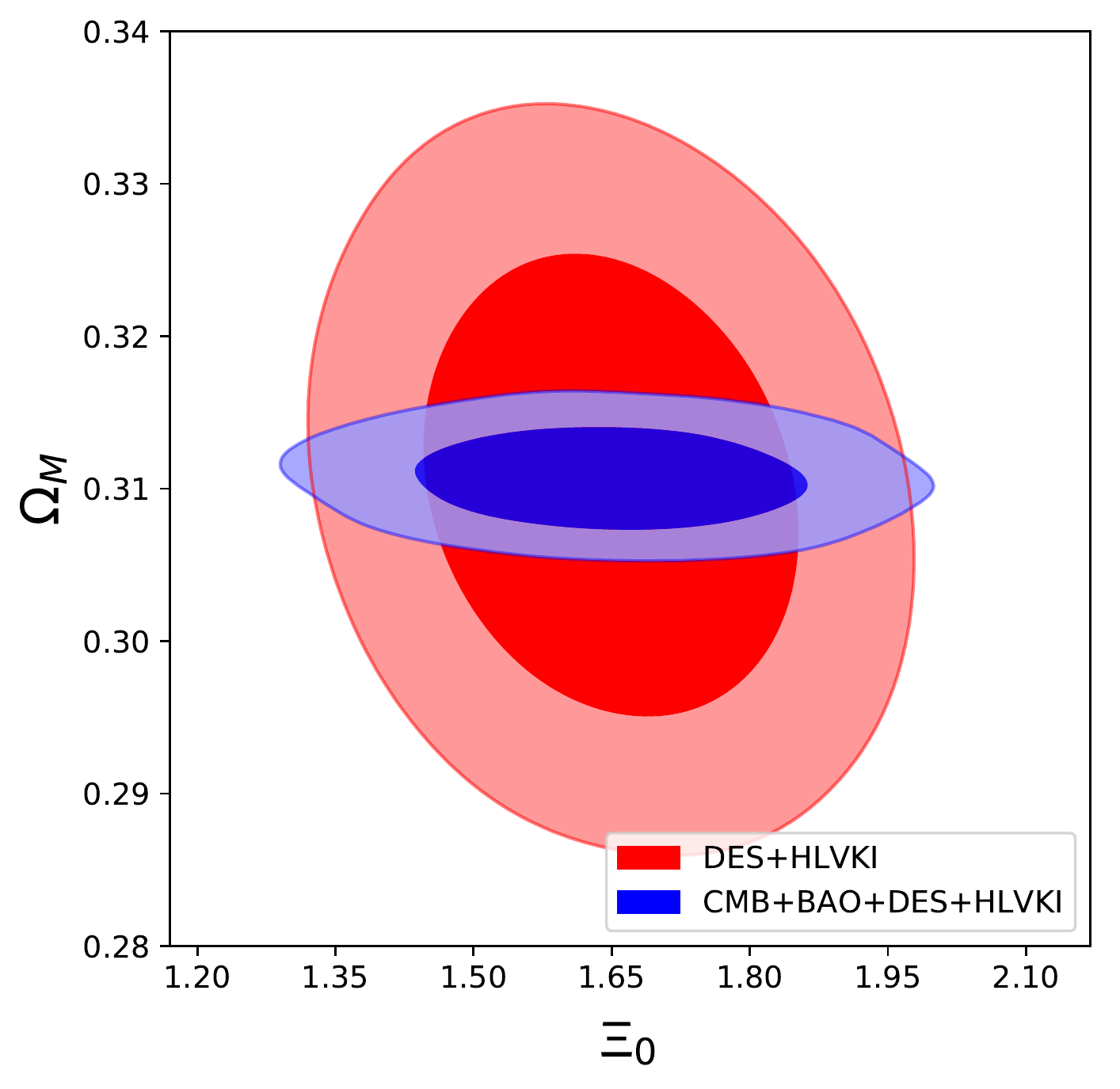}
\caption{As in Fig.~\ref{fig:HLVKI_fidLCDM}, assuming the RT nonlocal model as the fiducial cosmology for the HLVKI dataset.}
\label{fig:HLVKI_fidRT}
\end{figure*}

Fig.~\ref{fig:HLVKI_fidLCDM} and Fig.~\ref{fig:HLVKI_fidRT} show the results for the two-dimensional likelihoods of cosmological parameters, in the case where $\Lambda$CDM or RT are assumed as fiducial cosmologies for the mock catalog of GW detections at the HLVKI network. Tables~\ref{tab:HLVKI_LCDM} and~\ref{tab:HLVKI_RT} contain the errors obtained from the corresponding one-dimensional marginalized likelihoods.

\subsection{Results for the Einstein Telescope}

\begin{figure*}
\includegraphics[width=0.9\textwidth]{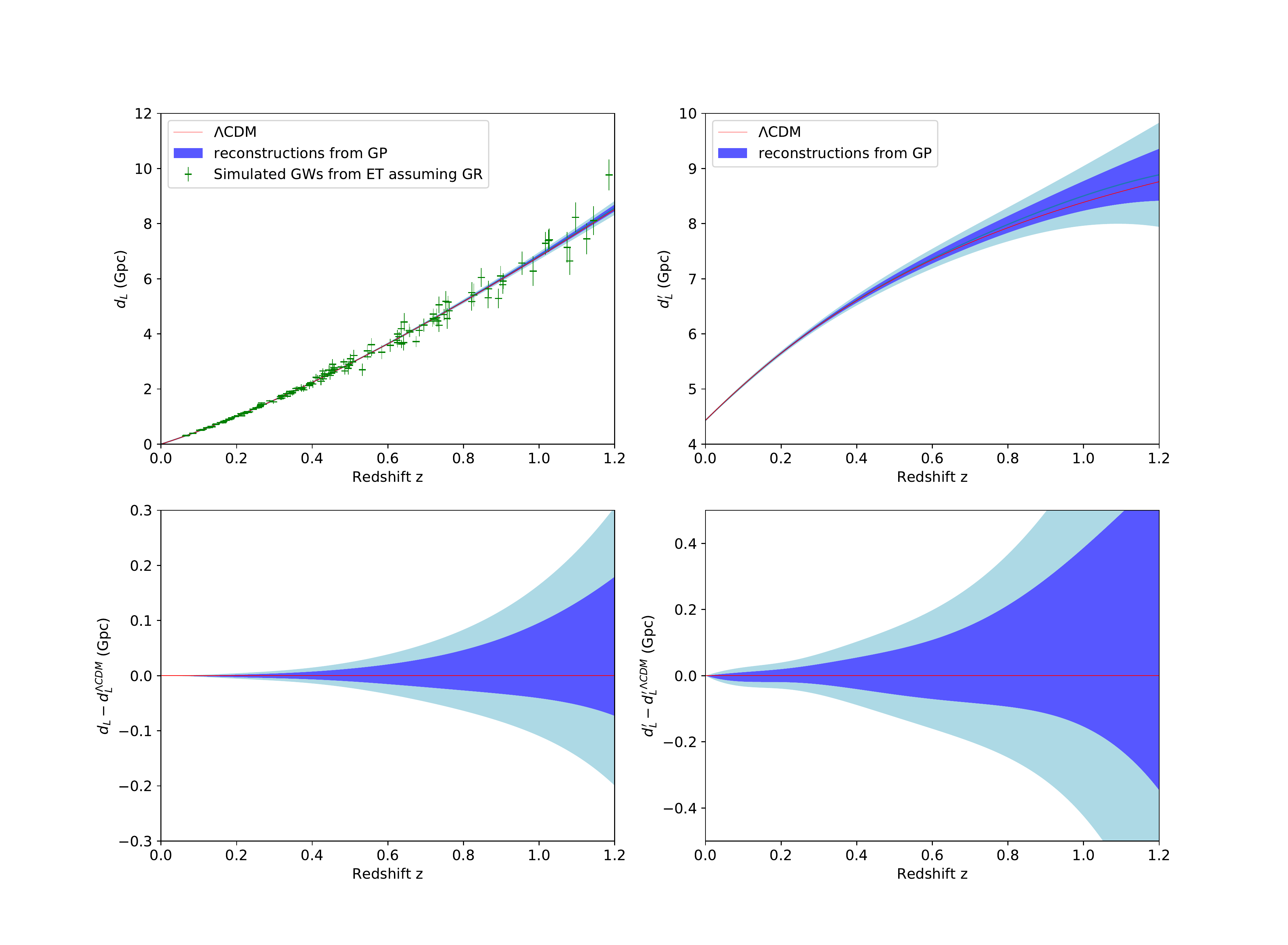}
\caption{Reconstruction of the GW luminosity distance from mock detections at the Einstein Telescope with electromagnetic counterpart, assuming $\Lambda$CDM as the fiducial cosmological model, again with the same meaning of the panels as in Fig.~\ref{fig:dL_DES}. The green points in the upper left panel are the mock data in the ET catalog.}
\label{fig:dL_ET}
\end{figure*}

\begin{figure*}
\includegraphics[width=0.9\textwidth]{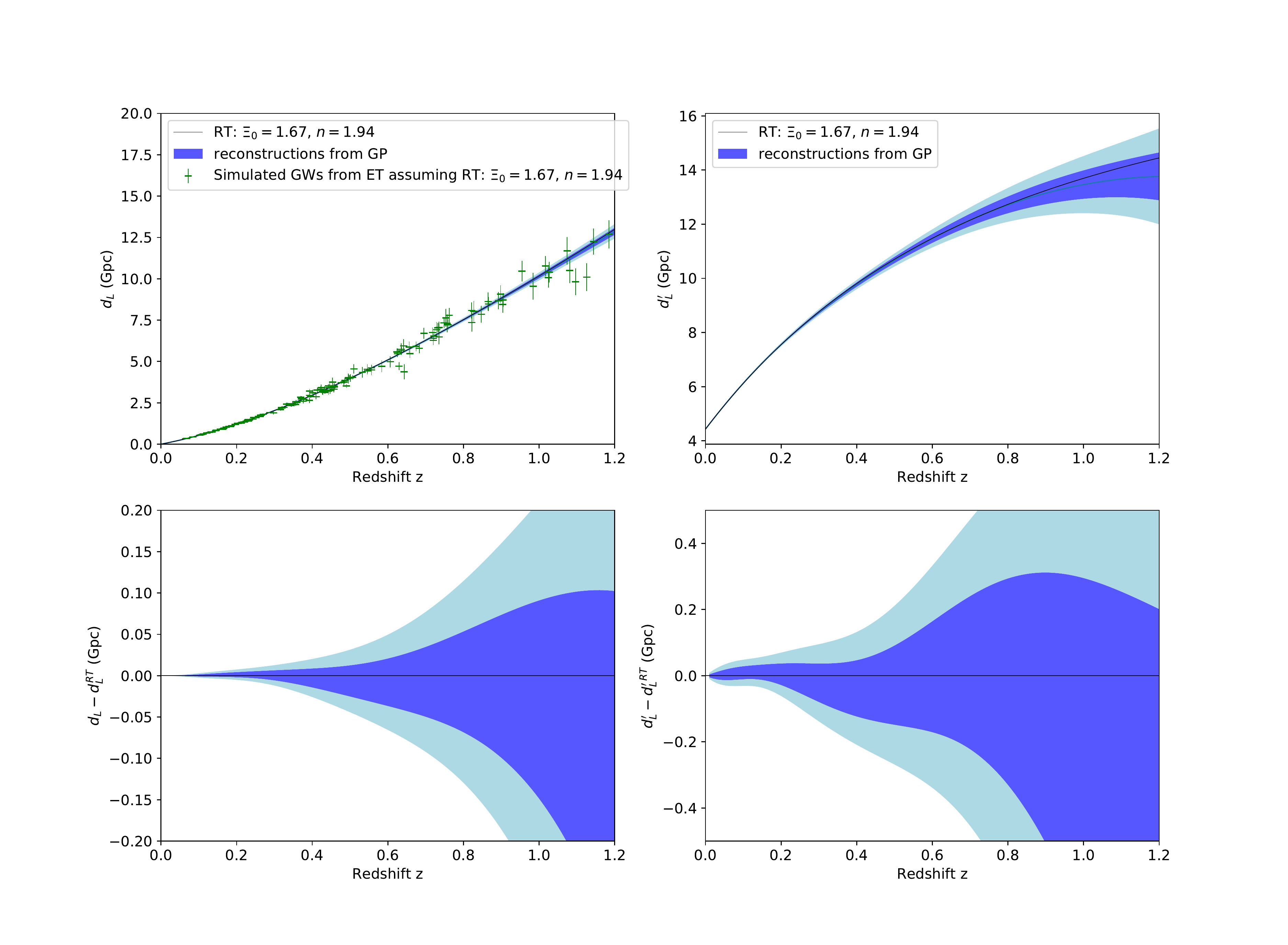}
\caption{As in Fig.~\ref{fig:dL_ET}, assuming the RT fiducial cosmology.}
\label{fig:dL_MGET}
\end{figure*}

\begin{figure*}
\centering
\includegraphics[width=0.45\textwidth]{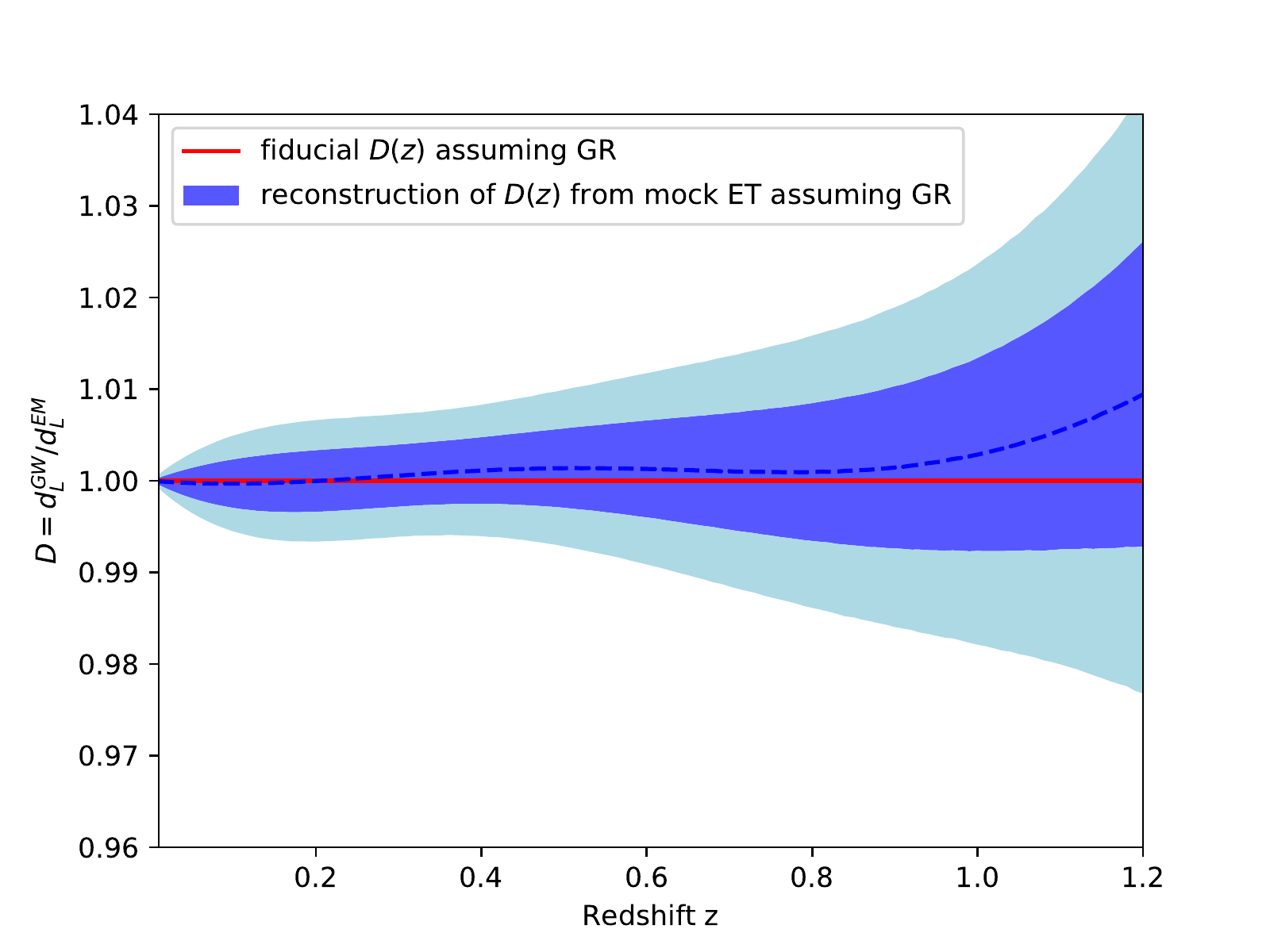}
\includegraphics[width=0.45\textwidth]{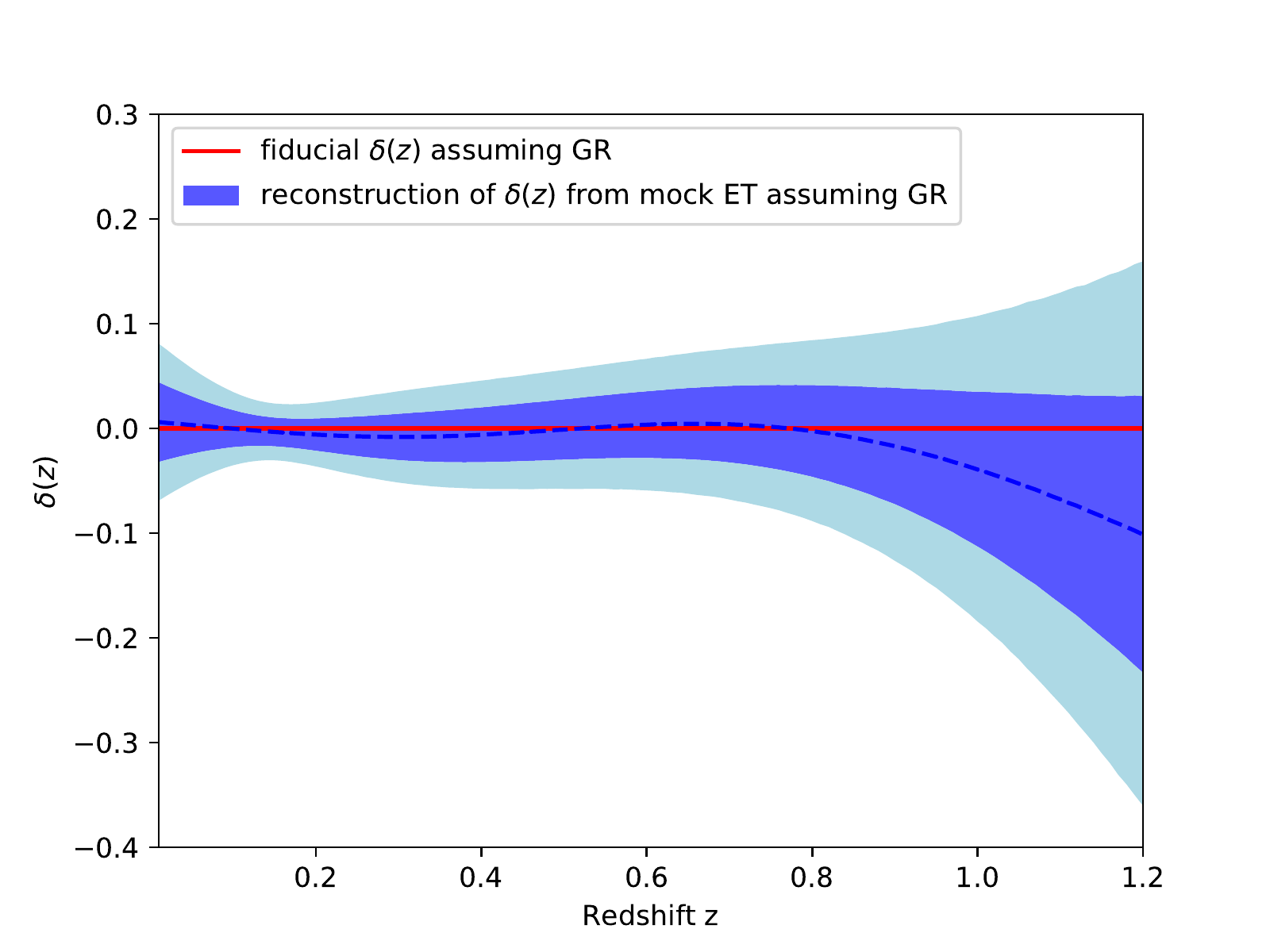}
\caption{Results from the mock ET and DES catalogs, assuming the $\Lambda$CDM fiducial cosmology. Left panel: reconstruction of the ratio $D(z)=\dgw(z)/\dem(z)$. Right panel: reconstruction of $\delta(z)$.}
\label{fig:results_ET}
\end{figure*}

\begin{figure*}
\centering
\includegraphics[width=0.45\textwidth]{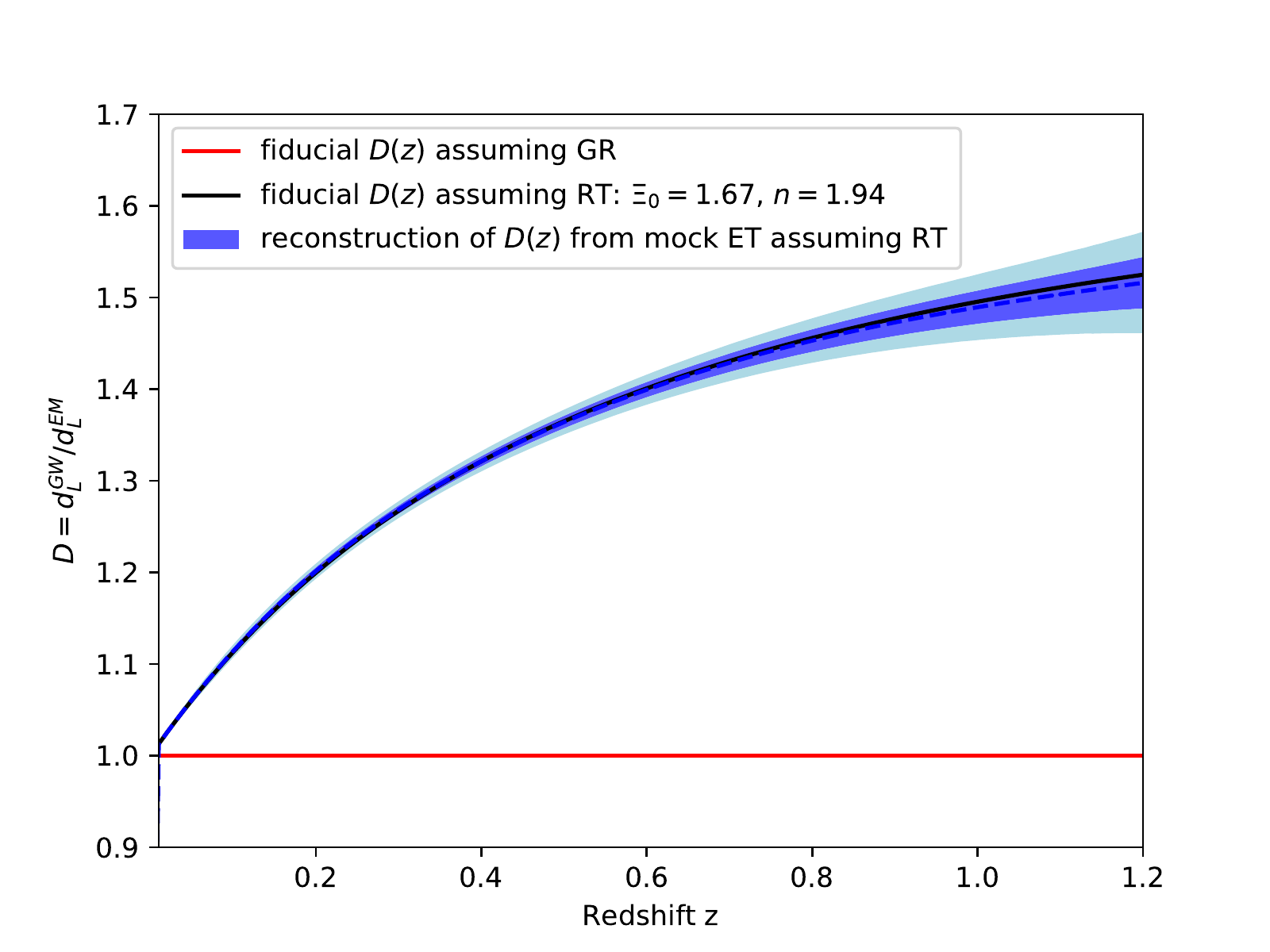}
\includegraphics[width=0.45\textwidth]{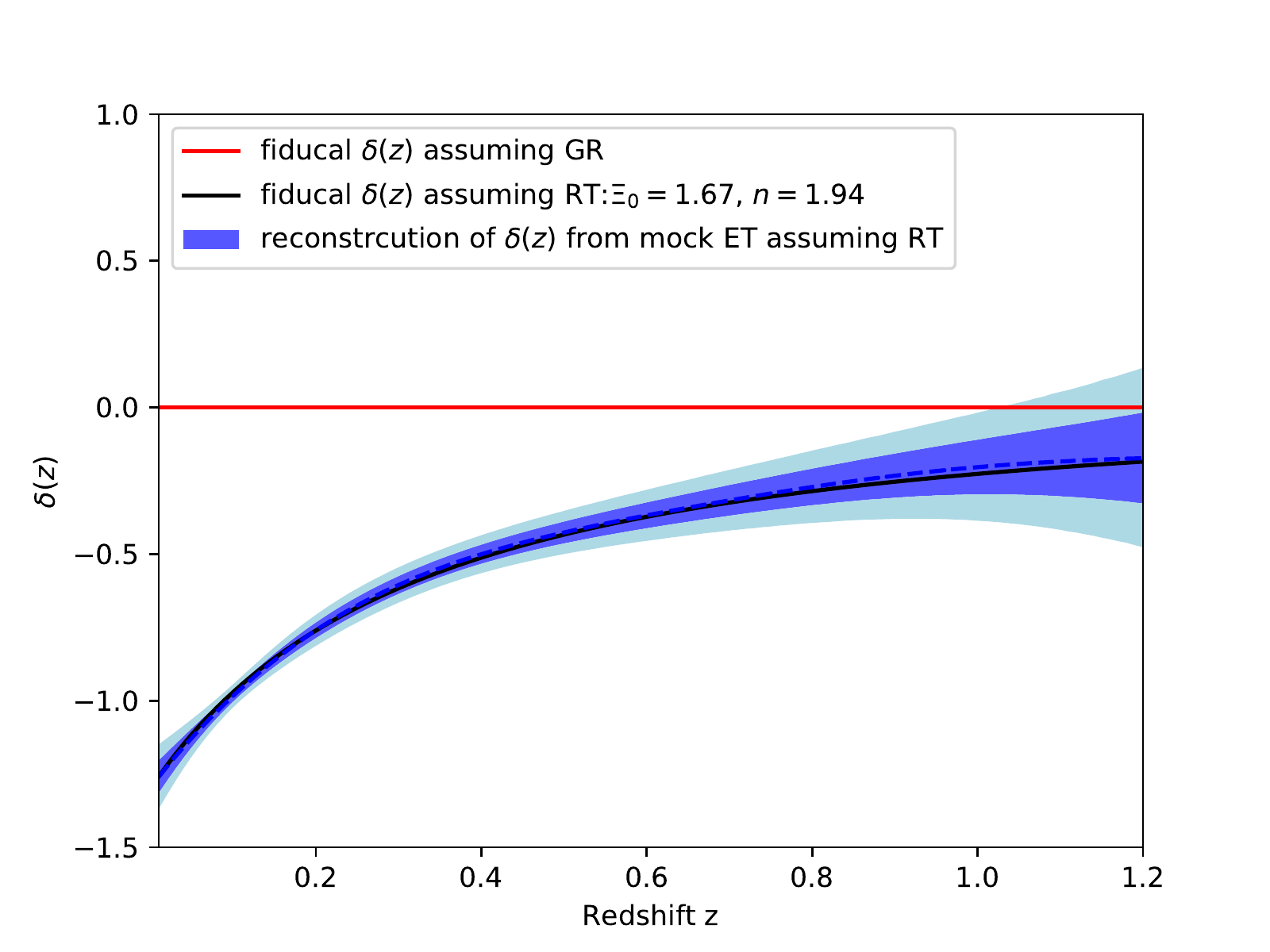}
\caption{As in Fig.~\ref{fig:results_ET}, using the mock ET and DES catalogs for the RT fiducial cosmology.}
\label{fig:results_MGET}
\end{figure*}

Fig.~\ref{fig:dL_ET} and Fig.~\ref{fig:dL_MGET} show the reconstruction of the GW luminosity distance from mock detections at the Einstein Telescope, obtained by Gaussian processes, for the $\Lambda$CDM and the RT fiducial cosmologies respectively. The final results for the luminosity distance ratio $D(z)=\dgw(z)/\dem(z)$ and the funcion $\delta(z)$ are given in Fig.~\ref{fig:results_ET} assuming the $\Lambda$CDM fiducial and in Fig.~\ref{fig:results_MGET} for the RT fiducial. We provide the reconstructions up to the maximum redshift reached in the DES mock catalog of supernovae $z_{\rm max}^{(\rm DES)}$=1.2, which is smaller than the maximum redshift $z_{\rm max}^{(\rm ET)}\simeq1.63$ in the simulated catalog of GWs from binary neutron stars detected at ET (see Table~\ref{tab:relerr_ET}).

\begin{figure*}
\includegraphics[width=0.4\textwidth]{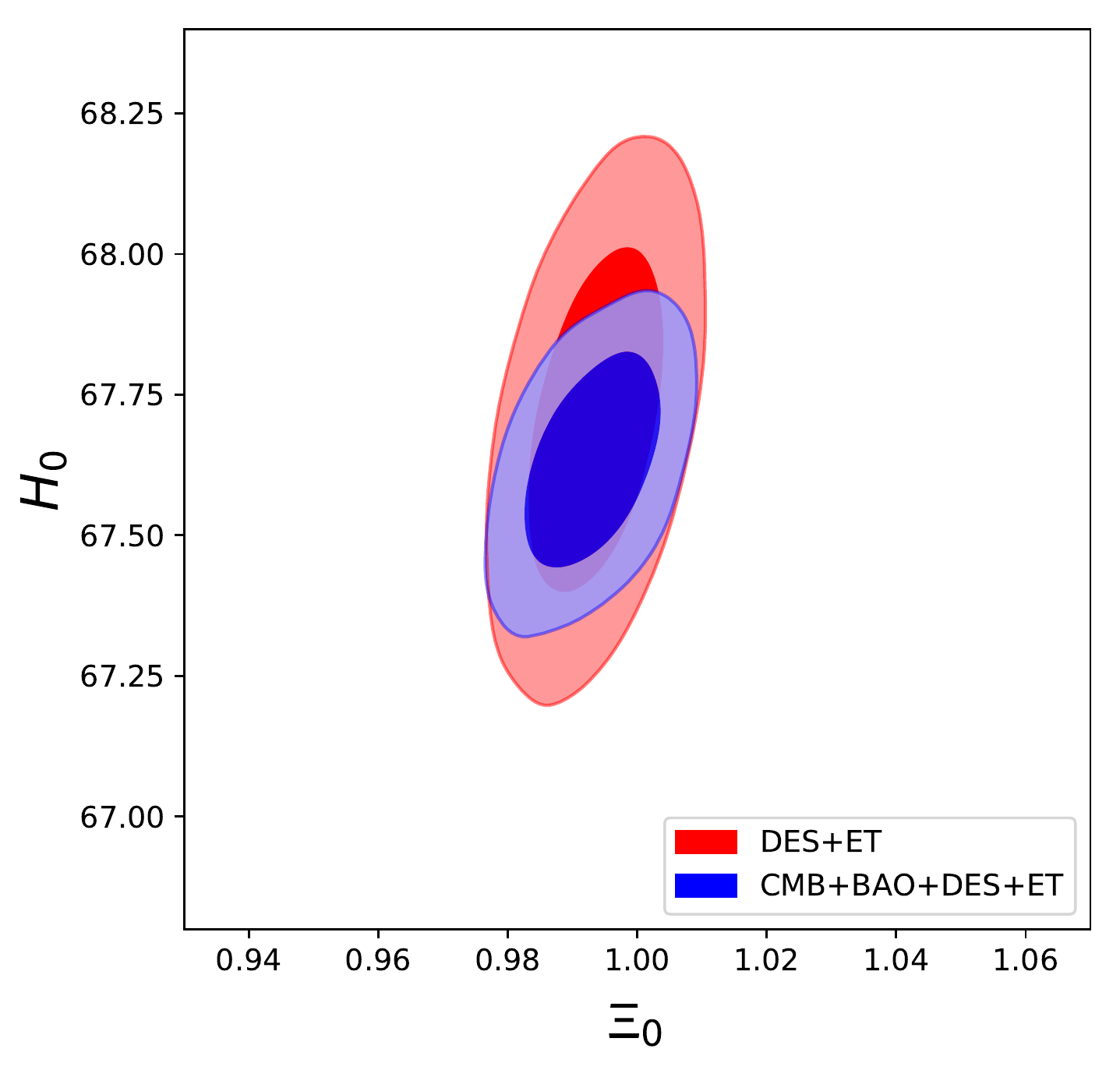}
\includegraphics[width=0.4\textwidth]{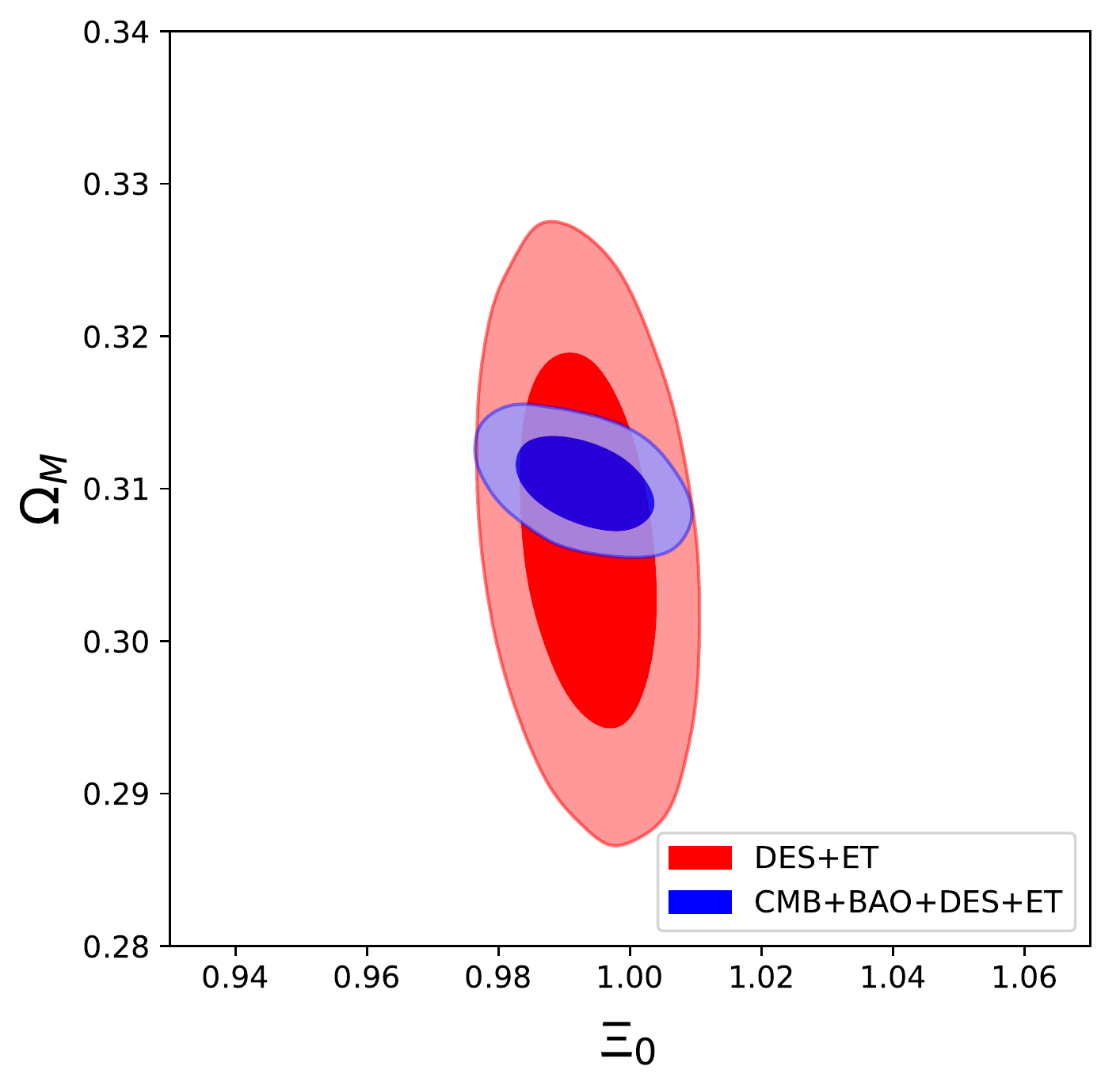}
\caption{The  $1\sigma$ and $2\sigma$
contours  of two-dimensional likelihoods, from DES+ET (red) and CMB+BAO+DES+ET (blue). The fiducial cosmology for the ET dataset is $\Lambda$CDM. Left:  in the $(\Xi_0,H_0)$ plane. Right:  in the $(\Xi_0,\oma)$ plane.}
\label{fig:ET_fidLCDM}
\end{figure*}

\begin{figure*}
\includegraphics[width=0.4\textwidth]{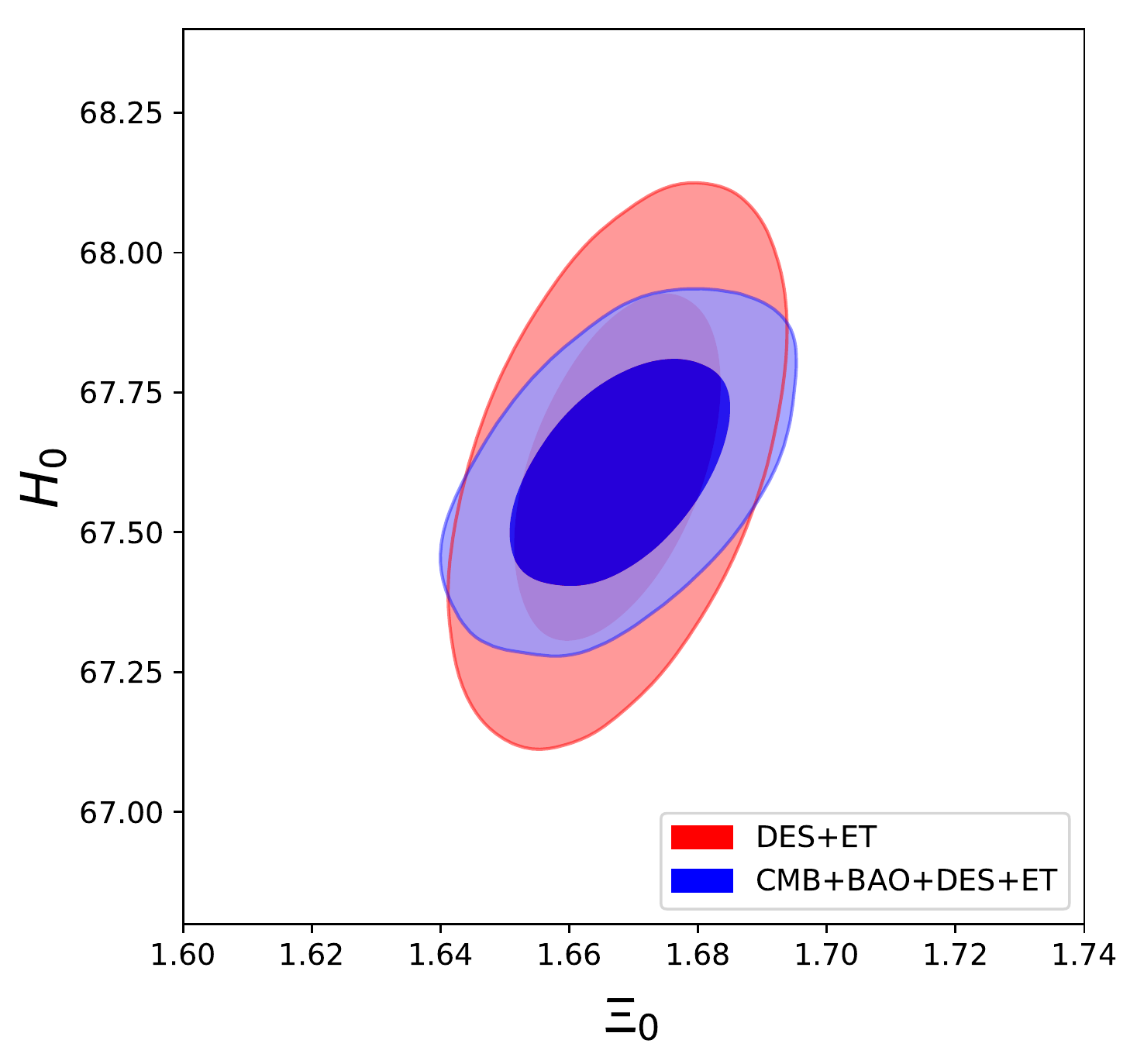}
\includegraphics[width=0.4\textwidth]{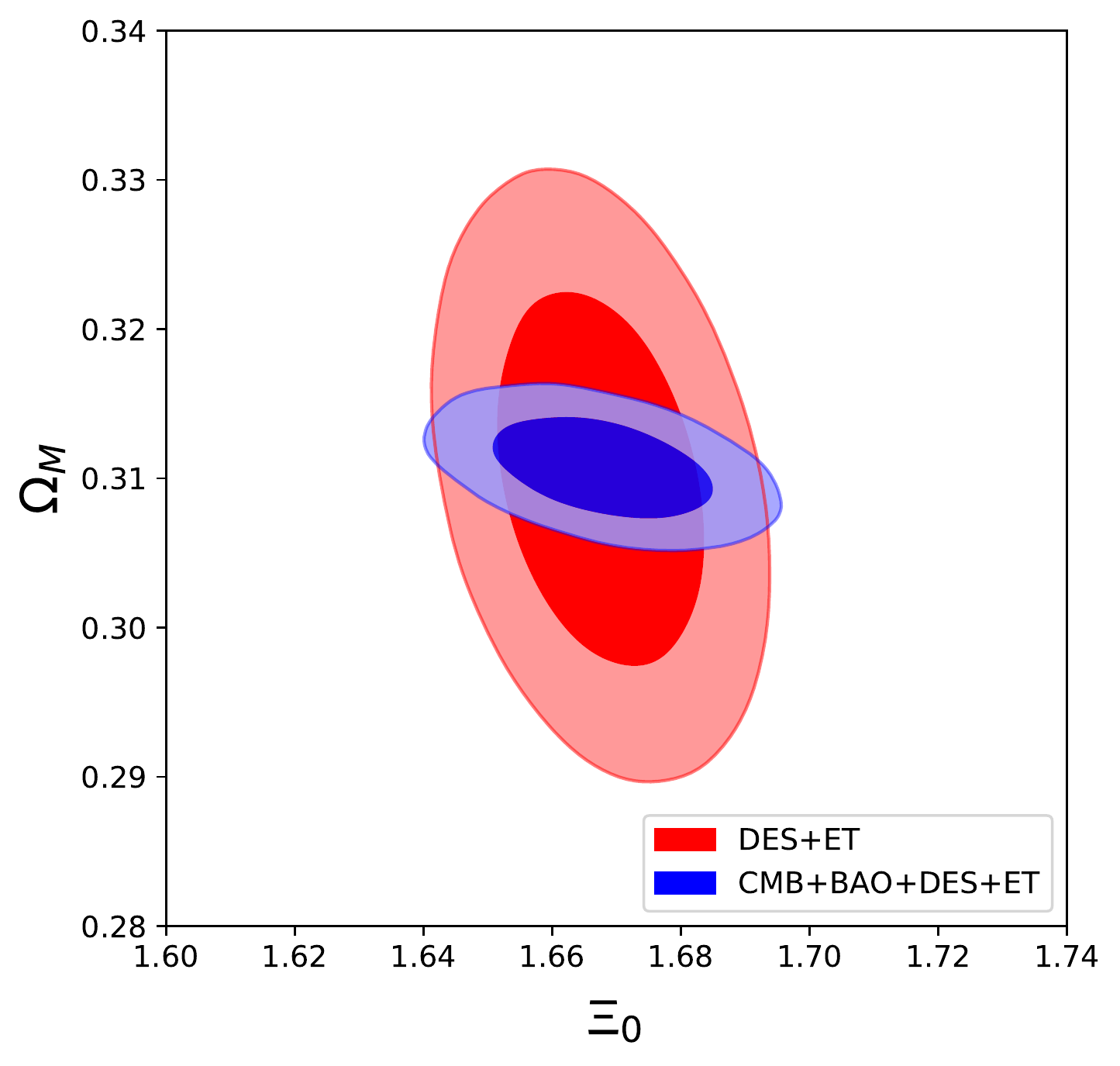}
\caption{As in Fig.~\ref{fig:ET_fidLCDM}, assuming the RT nonlocal model as the fiducial cosmology for the ET dataset.}
\label{fig:ET_fidRT}
\end{figure*}

As we did or the HLVKI network, we also show the results obtained by running MCMCs with the parametrization given by \eq{eq:fit}. The two-dimensional likelihoods for the cosmological parameters are shown in Fig.~\ref{fig:ET_fidLCDM} and Fig. ~\ref{fig:ET_fidRT}, assuming $\Lambda$CDM or RT, respectively, as fiducial cosmologies for the mock catalog of GW detections at the ET network. The errors on cosmological parameters are listed in Tables~\ref{tab:ET_LCDM} and~\ref{tab:ET_RT}.

\begin{table}[H]
\centering
\begin{tabular}{|c|c|c|}
 \hline
                               &  DES+ET& CMB+BAO+DES+ET     \\ \hline
$\Delta\Xi_0$ &  0.008 (0.8\%)            & 0.007 (0.7\%)                   \\
$\Delta H_0/H_0$ &  0.31\%                   & 0.19\%                    \\
$\Delta\oma/\oma$ &  2.78\%                  & 0.68\%                       \\
\hline
\end{tabular}
\caption{Accuracy ($1\sigma$ level) in the reconstruction of $\Xi_0$, $H_0$ and $\oma$ with DES+ET and CMB+BAO+DES+ET, assuming $\Lambda$CDM as the fiducial cosmology for the ET dataset.}
\label{tab:ET_LCDM}
\end{table}

\begin{table}[H]
\centering
\begin{tabular}{|c|c|c|}
 \hline
                               &  DES+ET& CMB+BAO+DES+ET     \\ \hline
$\Delta\Xi_0/\Xi_0$ &  0.66\%            & 0.66\%                   \\
$\Delta H_0/H_0$ &  0.31\%                   & 0.20\%                    \\
$\Delta\oma/\oma$ &  2.76\%                  & 0.74\%                       \\
\hline
\end{tabular}
\caption{As in~Table~\ref{tab:ET_LCDM}, assuming the RT model as the fiducial cosmology for the ET dataset.}
\label{tab:ET_RT}
\end{table}

\section{Conclusions}

Modified GW propagation can play a crucial role in telling apart modified gravity models from $\Lambda$CDM. We have explored two very convenient ways of extracting the information from the observations, either using the parametrization (\ref{eq:fit}), or using a parametrization-independent reconstruction method based on Gaussian processes. The first conclusion is that the two methods are quite complementary. The parametrization (\ref{eq:fit}) of the ratio $\dgw(z)/\dem(z)$
has been shown to fit extremely well the predictions of most of the best studied modified gravity models; this, together with its simplicity, makes it an extremely convenient tool. On the other hand, the study in \cite{Belgacem:2019pkk} has also identified one model, bigravity, where this parametrization is not adequate since it does not catch a series of oscillations that take place in that model; this already makes it worthwhile to test the data also against a parametrization-independent technique. For the function $\delta(z)$ the use of Gaussian processes is even more informative; indeed,  the directly observable quantity is $\dgw(z)/\dem(z)$, which is related to $\delta(z)$ by \eq{eq:dLgwdLem}. Features of the function $\delta(z)$, such as bumps, can be smoothed out by the integration in \eq{eq:dLgwdLem}, so that eventually the parametrization (\ref{eq:fit}) could still  fit the data relatively well even when some features of the function $\delta(z)$ are not correctly reproduced by the corresponding parametrization (\ref{paramdeltaz}). 
As we already mentioned, this happens indeed in DHOST theories~\cite{Belgacem:2019pkk}. In this case, a parametrization-independent reconstruction of the function $\delta(z)$ using Gaussian processes might put in evidence structures in $\delta(z)$, and therefore signatures of the underlying model, that would be lost using 
the parametrization (\ref{eq:fit}).

The complementarity between the two methods can also be seen in the informations that we get from them. In particular, the parametrization (\ref{eq:fit}) allows us to get in a very direct manner the asymptotic value of the ratio $\dgw(z)/\dem(z)$ for large redshift, which is given by the parameter $\Xi_0$. In contrast,
the  Gaussian processes reconstruction can identify the correct functional form at moderate redshift without making any assumptions, but
becomes less and less accurate with increasing redshift, as we see from all figures showing the result of such reconstructions. 

Beside testing the accuracy to which one could confirm the validity of $\Lambda$CDM, assuming it as fiducial model, we have also tested the accuracy to which one could validate a modified gravity model. We have focused in particular on a nonlocal modification of gravity, the so-called RT model~\cite{Maggiore:2013mea}, for which recent work~\cite{Belgacem:2019lwx} has shown that $\Xi_0$ can reach values as large as $1.6$. The model therefore gives a $60\%$ deviation from $\Lambda$CDM in the tensor sector, despite the fact that, both in the background evolution and in scalar perturbations it is very close to $\Lambda$CDM, and indeed it fits the existing cosmological datasets at a level statistically indistinguishable from $\Lambda$CDM. The result are very interesting, since they show that already with 15 binary neutron stars with counterpart, the second-generation detector network LIGO/Virgo/Kagra (HVLKI) could very clearly detect this effect. Indeed,  using the parametrization
(\ref{eq:fit}), the results in Table~\ref{tab:HLVKI_RT} show that $\Xi_0$ can be determined by HVLKI (combining the results with CMB, BAO and DES SNe) to about $8\%$ level accuracy, well below the $60\%$ deviation predicted by the RT model in the more optimistic case. This is fully consistent with the results obtained in \cite{Belgacem:2019tbw} where, using the current JLA SN dataset rather than the mock DES SNe used here, it was found that HVLKI combined with CMB, BAO and JLA can measure $\Xi_0$ to about $10\%$.
In this paper we see that a similar conclusion can be obtained from Gaussian process reconstruction. We see indeed from Fig.~\ref{fig:results_MGHLVKI} that, with 15 binary neutron stars with counterpart at HVLKI,
it is possible to  perform a parametrization-independent reconstruction of the ratio $\dgw(z)/\dem(z)$ to a few percent, across the whole range of redshifts considered. For instance, this implies that
the prediction of the RT model with $\Delta N=64$ and the prediction of $\Lambda$CDM would be very clearly distinguished.

For third-generation detectors, such as the Einstein Telescope, the situation will be even more exciting: in terms of $\Xi_0$, ET combined with the CMB+BAO+DES will reach an accuracy on $\Xi_0$ better than $1\%$, see Table~\ref{tab:ET_LCDM}, while Fig.~\ref{fig:results_MGET} shows the remarkable accuracy of the Gaussian-processes reconstruction. 

\vspace{5mm}\noindent
{\bf Acknowledgments.} 
The work  of EB, SF and MM is supported by the Fonds National Suisse and  by the SwissMap National Center for Competence in Research. The work of TY is supported by an appointment to the YST Program at the APCTP through the Science and Technology Promotion Fund and Lottery Fund of the Korean Government, and the Korean Local Governments - Gyeongsangbuk-do Province and Pohang City.

\bibliography{myrefs}

\begin{thebibliography}{81}%
\makeatletter
\providecommand \@ifxundefined [1]{%
 \@ifx{#1\undefined}
}%
\providecommand \@ifnum [1]{%
 \ifnum #1\expandafter \@firstoftwo
 \else \expandafter \@secondoftwo
 \fi
}%
\providecommand \@ifx [1]{%
 \ifx #1\expandafter \@firstoftwo
 \else \expandafter \@secondoftwo
 \fi
}%
\providecommand \natexlab [1]{#1}%
\providecommand \enquote  [1]{``#1''}%
\providecommand \bibnamefont  [1]{#1}%
\providecommand \bibfnamefont [1]{#1}%
\providecommand \citenamefont [1]{#1}%
\providecommand \href@noop [0]{\@secondoftwo}%
\providecommand \href [0]{\begingroup \@sanitize@url \@href}%
\providecommand \@href[1]{\@@startlink{#1}\@@href}%
\providecommand \@@href[1]{\endgroup#1\@@endlink}%
\providecommand \@sanitize@url [0]{\catcode `\\12\catcode `\$12\catcode
  `\&12\catcode `\#12\catcode `\^12\catcode `\_12\catcode `\%12\relax}%
\providecommand \@@startlink[1]{}%
\providecommand \@@endlink[0]{}%
\providecommand \url  [0]{\begingroup\@sanitize@url \@url }%
\providecommand \@url [1]{\endgroup\@href {#1}{\urlprefix }}%
\providecommand \urlprefix  [0]{URL }%
\providecommand \Eprint [0]{\href }%
\providecommand \doibase [0]{http://dx.doi.org/}%
\providecommand \selectlanguage [0]{\@gobble}%
\providecommand \bibinfo  [0]{\@secondoftwo}%
\providecommand \bibfield  [0]{\@secondoftwo}%
\providecommand \translation [1]{[#1]}%
\providecommand \BibitemOpen [0]{}%
\providecommand \bibitemStop [0]{}%
\providecommand \bibitemNoStop [0]{.\EOS\space}%
\providecommand \EOS [0]{\spacefactor3000\relax}%
\providecommand \BibitemShut  [1]{\csname bibitem#1\endcsname}%
\let\auto@bib@innerbib\@empty
\bibitem [{\citenamefont {Riess}\ \emph {et~al.}(1998)\citenamefont {Riess}
  \emph {et~al.}}]{Riess:1998cb}%
  \BibitemOpen
  \bibfield  {author} {\bibinfo {author} {\bibfnamefont {A.~G.}\ \bibnamefont
  {Riess}} \emph {et~al.} (\bibinfo {collaboration} {Supernova Search Team}),\
  }\href {\doibase 10.1086/300499} {\bibfield  {journal} {\bibinfo  {journal}
  {Astron.J.}\ }\textbf {\bibinfo {volume} {116}},\ \bibinfo {pages} {1009}
  (\bibinfo {year} {1998})},\ \Eprint {http://arxiv.org/abs/astro-ph/9805201}
  {arXiv:astro-ph/9805201 [astro-ph]} \BibitemShut {NoStop}%
\bibitem [{\citenamefont {Perlmutter}\ \emph {et~al.}(1999)\citenamefont
  {Perlmutter} \emph {et~al.}}]{Perlmutter:1998np}%
  \BibitemOpen
  \bibfield  {author} {\bibinfo {author} {\bibfnamefont {S.}~\bibnamefont
  {Perlmutter}} \emph {et~al.} (\bibinfo {collaboration} {Supernova Cosmology
  Project}),\ }\href {\doibase 10.1086/307221} {\bibfield  {journal} {\bibinfo
  {journal} {Astrophys.J.}\ }\textbf {\bibinfo {volume} {517}},\ \bibinfo
  {pages} {565} (\bibinfo {year} {1999})},\ \Eprint
  {http://arxiv.org/abs/astro-ph/9812133} {arXiv:astro-ph/9812133 [astro-ph]}
  \BibitemShut {NoStop}%
\bibitem [{\citenamefont {Schutz}(1986)}]{Schutz:1986gp}%
  \BibitemOpen
  \bibfield  {author} {\bibinfo {author} {\bibfnamefont {B.~F.}\ \bibnamefont
  {Schutz}},\ }\href {\doibase 10.1038/323310a0} {\bibfield  {journal}
  {\bibinfo  {journal} {Nature}\ }\textbf {\bibinfo {volume} {323}},\ \bibinfo
  {pages} {310} (\bibinfo {year} {1986})}\BibitemShut {NoStop}%
\bibitem [{\citenamefont {Holz}\ and\ \citenamefont
  {Hughes}(2005)}]{Holz:2005df}%
  \BibitemOpen
  \bibfield  {author} {\bibinfo {author} {\bibfnamefont {D.~E.}\ \bibnamefont
  {Holz}}\ and\ \bibinfo {author} {\bibfnamefont {S.~A.}\ \bibnamefont
  {Hughes}},\ }\href {\doibase 10.1086/431341} {\bibfield  {journal} {\bibinfo
  {journal} {Astrophys. J.}\ }\textbf {\bibinfo {volume} {629}},\ \bibinfo
  {pages} {15} (\bibinfo {year} {2005})},\ \Eprint
  {http://arxiv.org/abs/astro-ph/0504616} {arXiv:astro-ph/0504616 [astro-ph]}
  \BibitemShut {NoStop}%
\bibitem [{\citenamefont {Dalal}\ \emph {et~al.}(2006)\citenamefont {Dalal},
  \citenamefont {Holz}, \citenamefont {Hughes},\ and\ \citenamefont
  {Jain}}]{Dalal:2006qt}%
  \BibitemOpen
  \bibfield  {author} {\bibinfo {author} {\bibfnamefont {N.}~\bibnamefont
  {Dalal}}, \bibinfo {author} {\bibfnamefont {D.~E.}\ \bibnamefont {Holz}},
  \bibinfo {author} {\bibfnamefont {S.~A.}\ \bibnamefont {Hughes}}, \ and\
  \bibinfo {author} {\bibfnamefont {B.}~\bibnamefont {Jain}},\ }\href {\doibase
  10.1103/PhysRevD.74.063006} {\bibfield  {journal} {\bibinfo  {journal} {Phys.
  Rev.}\ }\textbf {\bibinfo {volume} {D74}},\ \bibinfo {pages} {063006}
  (\bibinfo {year} {2006})},\ \Eprint {http://arxiv.org/abs/astro-ph/0601275}
  {arXiv:astro-ph/0601275 [astro-ph]} \BibitemShut {NoStop}%
\bibitem [{\citenamefont {MacLeod}\ and\ \citenamefont
  {Hogan}(2008)}]{MacLeod:2007jd}%
  \BibitemOpen
  \bibfield  {author} {\bibinfo {author} {\bibfnamefont {C.~L.}\ \bibnamefont
  {MacLeod}}\ and\ \bibinfo {author} {\bibfnamefont {C.~J.}\ \bibnamefont
  {Hogan}},\ }\href {\doibase 10.1103/PhysRevD.77.043512} {\bibfield  {journal}
  {\bibinfo  {journal} {Phys. Rev.}\ }\textbf {\bibinfo {volume} {D77}},\
  \bibinfo {pages} {043512} (\bibinfo {year} {2008})},\ \Eprint
  {http://arxiv.org/abs/0712.0618} {arXiv:0712.0618 [astro-ph]} \BibitemShut
  {NoStop}%
\bibitem [{\citenamefont {Nissanke}\ \emph {et~al.}(2010)\citenamefont
  {Nissanke}, \citenamefont {Holz}, \citenamefont {Hughes}, \citenamefont
  {Dalal},\ and\ \citenamefont {Sievers}}]{Nissanke:2009kt}%
  \BibitemOpen
  \bibfield  {author} {\bibinfo {author} {\bibfnamefont {S.}~\bibnamefont
  {Nissanke}}, \bibinfo {author} {\bibfnamefont {D.~E.}\ \bibnamefont {Holz}},
  \bibinfo {author} {\bibfnamefont {S.}~\bibnamefont {Hughes}}, \bibinfo
  {author} {\bibfnamefont {N.}~\bibnamefont {Dalal}}, \ and\ \bibinfo {author}
  {\bibfnamefont {J.~L.}\ \bibnamefont {Sievers}},\ }\href {\doibase
  10.1088/0004-637X/725/1/496} {\bibfield  {journal} {\bibinfo  {journal}
  {Astrophys. J.}\ }\textbf {\bibinfo {volume} {725}},\ \bibinfo {pages} {496}
  (\bibinfo {year} {2010})},\ \Eprint {http://arxiv.org/abs/0904.1017}
  {arXiv:0904.1017 [astro-ph.CO]} \BibitemShut {NoStop}%
\bibitem [{\citenamefont {Cutler}\ and\ \citenamefont
  {Holz}(2009)}]{Cutler:2009qv}%
  \BibitemOpen
  \bibfield  {author} {\bibinfo {author} {\bibfnamefont {C.}~\bibnamefont
  {Cutler}}\ and\ \bibinfo {author} {\bibfnamefont {D.~E.}\ \bibnamefont
  {Holz}},\ }\href {\doibase 10.1103/PhysRevD.80.104009} {\bibfield  {journal}
  {\bibinfo  {journal} {Phys. Rev.}\ }\textbf {\bibinfo {volume} {D80}},\
  \bibinfo {pages} {104009} (\bibinfo {year} {2009})},\ \Eprint
  {http://arxiv.org/abs/0906.3752} {arXiv:0906.3752 [astro-ph.CO]} \BibitemShut
  {NoStop}%
\bibitem [{\citenamefont {Sathyaprakash}\ \emph {et~al.}(2010)\citenamefont
  {Sathyaprakash}, \citenamefont {Schutz},\ and\ \citenamefont {Van
  Den~Broeck}}]{Sathyaprakash:2009xt}%
  \BibitemOpen
  \bibfield  {author} {\bibinfo {author} {\bibfnamefont {B.~S.}\ \bibnamefont
  {Sathyaprakash}}, \bibinfo {author} {\bibfnamefont {B.~F.}\ \bibnamefont
  {Schutz}}, \ and\ \bibinfo {author} {\bibfnamefont {C.}~\bibnamefont {Van
  Den~Broeck}},\ }\href {\doibase 10.1088/0264-9381/27/21/215006} {\bibfield
  {journal} {\bibinfo  {journal} {Class. Quant. Grav.}\ }\textbf {\bibinfo
  {volume} {27}},\ \bibinfo {pages} {215006} (\bibinfo {year} {2010})},\
  \Eprint {http://arxiv.org/abs/0906.4151} {arXiv:0906.4151 [astro-ph.CO]}
  \BibitemShut {NoStop}%
\bibitem [{\citenamefont {Zhao}\ \emph {et~al.}(2011)\citenamefont {Zhao},
  \citenamefont {Van Den~Broeck}, \citenamefont {Baskaran},\ and\ \citenamefont
  {Li}}]{Zhao:2010sz}%
  \BibitemOpen
  \bibfield  {author} {\bibinfo {author} {\bibfnamefont {W.}~\bibnamefont
  {Zhao}}, \bibinfo {author} {\bibfnamefont {C.}~\bibnamefont {Van
  Den~Broeck}}, \bibinfo {author} {\bibfnamefont {D.}~\bibnamefont {Baskaran}},
  \ and\ \bibinfo {author} {\bibfnamefont {T.~G.~F.}\ \bibnamefont {Li}},\
  }\href {\doibase 10.1103/PhysRevD.83.023005} {\bibfield  {journal} {\bibinfo
  {journal} {Phys. Rev.}\ }\textbf {\bibinfo {volume} {D83}},\ \bibinfo {pages}
  {023005} (\bibinfo {year} {2011})},\ \Eprint {http://arxiv.org/abs/1009.0206}
  {arXiv:1009.0206 [astro-ph.CO]} \BibitemShut {NoStop}%
\bibitem [{\citenamefont {Del~Pozzo}(2012)}]{DelPozzo:2011yh}%
  \BibitemOpen
  \bibfield  {author} {\bibinfo {author} {\bibfnamefont {W.}~\bibnamefont
  {Del~Pozzo}},\ }\href {\doibase 10.1103/PhysRevD.86.043011} {\bibfield
  {journal} {\bibinfo  {journal} {Phys. Rev.}\ }\textbf {\bibinfo {volume}
  {D86}},\ \bibinfo {pages} {043011} (\bibinfo {year} {2012})},\ \Eprint
  {http://arxiv.org/abs/1108.1317} {arXiv:1108.1317 [astro-ph.CO]} \BibitemShut
  {NoStop}%
\bibitem [{\citenamefont {Nishizawa}\ \emph {et~al.}(2012)\citenamefont
  {Nishizawa}, \citenamefont {Yagi}, \citenamefont {Taruya},\ and\
  \citenamefont {Tanaka}}]{Nishizawa:2011eq}%
  \BibitemOpen
  \bibfield  {author} {\bibinfo {author} {\bibfnamefont {A.}~\bibnamefont
  {Nishizawa}}, \bibinfo {author} {\bibfnamefont {K.}~\bibnamefont {Yagi}},
  \bibinfo {author} {\bibfnamefont {A.}~\bibnamefont {Taruya}}, \ and\ \bibinfo
  {author} {\bibfnamefont {T.}~\bibnamefont {Tanaka}},\ }\href {\doibase
  10.1103/PhysRevD.85.044047} {\bibfield  {journal} {\bibinfo  {journal} {Phys.
  Rev.}\ }\textbf {\bibinfo {volume} {D85}},\ \bibinfo {pages} {044047}
  (\bibinfo {year} {2012})},\ \Eprint {http://arxiv.org/abs/1110.2865}
  {arXiv:1110.2865 [astro-ph.CO]} \BibitemShut {NoStop}%
\bibitem [{\citenamefont {Taylor}\ and\ \citenamefont
  {Gair}(2012)}]{Taylor:2012db}%
  \BibitemOpen
  \bibfield  {author} {\bibinfo {author} {\bibfnamefont {S.~R.}\ \bibnamefont
  {Taylor}}\ and\ \bibinfo {author} {\bibfnamefont {J.~R.}\ \bibnamefont
  {Gair}},\ }\href {\doibase 10.1103/PhysRevD.86.023502} {\bibfield  {journal}
  {\bibinfo  {journal} {Phys. Rev.}\ }\textbf {\bibinfo {volume} {D86}},\
  \bibinfo {pages} {023502} (\bibinfo {year} {2012})},\ \Eprint
  {http://arxiv.org/abs/1204.6739} {arXiv:1204.6739 [astro-ph.CO]} \BibitemShut
  {NoStop}%
\bibitem [{\citenamefont {Camera}\ and\ \citenamefont
  {Nishizawa}(2013)}]{Camera:2013xfa}%
  \BibitemOpen
  \bibfield  {author} {\bibinfo {author} {\bibfnamefont {S.}~\bibnamefont
  {Camera}}\ and\ \bibinfo {author} {\bibfnamefont {A.}~\bibnamefont
  {Nishizawa}},\ }\href {\doibase 10.1103/PhysRevLett.110.151103} {\bibfield
  {journal} {\bibinfo  {journal} {Phys. Rev. Lett.}\ }\textbf {\bibinfo
  {volume} {110}},\ \bibinfo {pages} {151103} (\bibinfo {year} {2013})},\
  \Eprint {http://arxiv.org/abs/1303.5446} {arXiv:1303.5446 [astro-ph.CO]}
  \BibitemShut {NoStop}%
\bibitem [{\citenamefont {Tamanini}\ \emph {et~al.}(2016)\citenamefont
  {Tamanini}, \citenamefont {Caprini}, \citenamefont {Barausse}, \citenamefont
  {Sesana}, \citenamefont {Klein},\ and\ \citenamefont
  {Petiteau}}]{Tamanini:2016zlh}%
  \BibitemOpen
  \bibfield  {author} {\bibinfo {author} {\bibfnamefont {N.}~\bibnamefont
  {Tamanini}}, \bibinfo {author} {\bibfnamefont {C.}~\bibnamefont {Caprini}},
  \bibinfo {author} {\bibfnamefont {E.}~\bibnamefont {Barausse}}, \bibinfo
  {author} {\bibfnamefont {A.}~\bibnamefont {Sesana}}, \bibinfo {author}
  {\bibfnamefont {A.}~\bibnamefont {Klein}}, \ and\ \bibinfo {author}
  {\bibfnamefont {A.}~\bibnamefont {Petiteau}},\ }\href {\doibase
  10.1088/1475-7516/2016/04/002} {\bibfield  {journal} {\bibinfo  {journal}
  {JCAP}\ }\textbf {\bibinfo {volume} {1604}},\ \bibinfo {pages} {002}
  (\bibinfo {year} {2016})},\ \Eprint {http://arxiv.org/abs/1601.07112}
  {arXiv:1601.07112 [astro-ph.CO]} \BibitemShut {NoStop}%
\bibitem [{\citenamefont {Caprini}\ and\ \citenamefont
  {Tamanini}(2016)}]{Caprini:2016qxs}%
  \BibitemOpen
  \bibfield  {author} {\bibinfo {author} {\bibfnamefont {C.}~\bibnamefont
  {Caprini}}\ and\ \bibinfo {author} {\bibfnamefont {N.}~\bibnamefont
  {Tamanini}},\ }\href {\doibase 10.1088/1475-7516/2016/10/006} {\bibfield
  {journal} {\bibinfo  {journal} {JCAP}\ }\textbf {\bibinfo {volume} {1610}},\
  \bibinfo {pages} {006} (\bibinfo {year} {2016})},\ \Eprint
  {http://arxiv.org/abs/1607.08755} {arXiv:1607.08755 [astro-ph.CO]}
  \BibitemShut {NoStop}%
\bibitem [{\citenamefont {Cai}\ and\ \citenamefont {Yang}(2017)}]{Cai:2016sby}%
  \BibitemOpen
  \bibfield  {author} {\bibinfo {author} {\bibfnamefont {R.-G.}\ \bibnamefont
  {Cai}}\ and\ \bibinfo {author} {\bibfnamefont {T.}~\bibnamefont {Yang}},\
  }\href {\doibase 10.1103/PhysRevD.95.044024} {\bibfield  {journal} {\bibinfo
  {journal} {Phys. Rev.}\ }\textbf {\bibinfo {volume} {D95}},\ \bibinfo {pages}
  {044024} (\bibinfo {year} {2017})},\ \Eprint
  {http://arxiv.org/abs/1608.08008} {arXiv:1608.08008 [astro-ph.CO]}
  \BibitemShut {NoStop}%
\bibitem [{\citenamefont {Del~Pozzo}\ \emph {et~al.}(2018)\citenamefont
  {Del~Pozzo}, \citenamefont {Sesana},\ and\ \citenamefont
  {Klein}}]{DelPozzo:2017kme}%
  \BibitemOpen
  \bibfield  {author} {\bibinfo {author} {\bibfnamefont {W.}~\bibnamefont
  {Del~Pozzo}}, \bibinfo {author} {\bibfnamefont {A.}~\bibnamefont {Sesana}}, \
  and\ \bibinfo {author} {\bibfnamefont {A.}~\bibnamefont {Klein}},\ }\href
  {\doibase 10.1093/mnras/sty057} {\bibfield  {journal} {\bibinfo  {journal}
  {Mon. Not. Roy. Astron. Soc.}\ }\textbf {\bibinfo {volume} {475}},\ \bibinfo
  {pages} {3485} (\bibinfo {year} {2018})},\ \Eprint
  {http://arxiv.org/abs/1703.01300} {arXiv:1703.01300 [astro-ph.CO]}
  \BibitemShut {NoStop}%
\bibitem [{\citenamefont {Belgacem}\ \emph
  {et~al.}(2018{\natexlab{a}})\citenamefont {Belgacem}, \citenamefont {Dirian},
  \citenamefont {Foffa},\ and\ \citenamefont {Maggiore}}]{Belgacem:2017ihm}%
  \BibitemOpen
  \bibfield  {author} {\bibinfo {author} {\bibfnamefont {E.}~\bibnamefont
  {Belgacem}}, \bibinfo {author} {\bibfnamefont {Y.}~\bibnamefont {Dirian}},
  \bibinfo {author} {\bibfnamefont {S.}~\bibnamefont {Foffa}}, \ and\ \bibinfo
  {author} {\bibfnamefont {M.}~\bibnamefont {Maggiore}},\ }\href@noop {}
  {\bibfield  {journal} {\bibinfo  {journal} {Phys. Rev.}\ }\textbf {\bibinfo
  {volume} {D97}},\ \bibinfo {pages} {104066} (\bibinfo {year}
  {2018}{\natexlab{a}})},\ \Eprint {http://arxiv.org/abs/1712.08108}
  {arXiv:1712.08108 [astro-ph.CO]} \BibitemShut {NoStop}%
\bibitem [{\citenamefont {Belgacem}\ \emph
  {et~al.}(2018{\natexlab{b}})\citenamefont {Belgacem}, \citenamefont {Dirian},
  \citenamefont {Foffa},\ and\ \citenamefont {Maggiore}}]{Belgacem:2018lbp}%
  \BibitemOpen
  \bibfield  {author} {\bibinfo {author} {\bibfnamefont {E.}~\bibnamefont
  {Belgacem}}, \bibinfo {author} {\bibfnamefont {Y.}~\bibnamefont {Dirian}},
  \bibinfo {author} {\bibfnamefont {S.}~\bibnamefont {Foffa}}, \ and\ \bibinfo
  {author} {\bibfnamefont {M.}~\bibnamefont {Maggiore}},\ }\href {\doibase
  10.1103/PhysRevD.98.023510} {\bibfield  {journal} {\bibinfo  {journal} {Phys.
  Rev.}\ }\textbf {\bibinfo {volume} {D98}},\ \bibinfo {pages} {023510}
  (\bibinfo {year} {2018}{\natexlab{b}})},\ \Eprint
  {http://arxiv.org/abs/1805.08731} {arXiv:1805.08731 [gr-qc]} \BibitemShut
  {NoStop}%
\bibitem [{\citenamefont {Mendon\c{c}a}\ and\ \citenamefont
  {Sturani}(2019)}]{Mendonca:2019yfo}%
  \BibitemOpen
  \bibfield  {author} {\bibinfo {author} {\bibfnamefont {J.}~\bibnamefont
  {Mendon\c{c}a}}\ and\ \bibinfo {author} {\bibfnamefont {R.}~\bibnamefont
  {Sturani}},\ }\href@noop {} {\  (\bibinfo {year} {2019})},\ \Eprint
  {http://arxiv.org/abs/1905.03848} {arXiv:1905.03848 [gr-qc]} \BibitemShut
  {NoStop}%
\bibitem [{\citenamefont {Abbott}\ \emph
  {et~al.}(2017{\natexlab{a}})\citenamefont {Abbott} \emph
  {et~al.}}]{TheLIGOScientific:2017qsa}%
  \BibitemOpen
  \bibfield  {author} {\bibinfo {author} {\bibfnamefont {B.}~\bibnamefont
  {Abbott}} \emph {et~al.},\ }\href {\doibase 10.1103/PhysRevLett.119.161101}
  {\bibfield  {journal} {\bibinfo  {journal} {Phys. Rev. Lett.}\ }\textbf
  {\bibinfo {volume} {119}},\ \bibinfo {pages} {161101} (\bibinfo {year}
  {2017}{\natexlab{a}})},\ \Eprint {http://arxiv.org/abs/1710.05832}
  {arXiv:1710.05832 [gr-qc]} \BibitemShut {NoStop}%
\bibitem [{\citenamefont {Abbott}\ \emph
  {et~al.}(2017{\natexlab{b}})\citenamefont {Abbott} \emph
  {et~al.}}]{GBM:2017lvd}%
  \BibitemOpen
  \bibfield  {author} {\bibinfo {author} {\bibfnamefont {B.~P.}\ \bibnamefont
  {Abbott}} \emph {et~al.},\ }\href {\doibase 10.3847/2041-8213/aa91c9}
  {\bibfield  {journal} {\bibinfo  {journal} {Astrophys. J.}\ }\textbf
  {\bibinfo {volume} {848}},\ \bibinfo {pages} {L12} (\bibinfo {year}
  {2017}{\natexlab{b}})},\ \Eprint {http://arxiv.org/abs/1710.05833}
  {arXiv:1710.05833 [astro-ph.HE]} \BibitemShut {NoStop}%
\bibitem [{\citenamefont {Abbott}\ \emph
  {et~al.}(2017{\natexlab{c}})\citenamefont {Abbott} \emph
  {et~al.}}]{Monitor:2017mdv}%
  \BibitemOpen
  \bibfield  {author} {\bibinfo {author} {\bibfnamefont {B.~P.}\ \bibnamefont
  {Abbott}} \emph {et~al.},\ }\href {\doibase 10.3847/2041-8213/aa920c}
  {\bibfield  {journal} {\bibinfo  {journal} {Astrophys. J.}\ }\textbf
  {\bibinfo {volume} {848}},\ \bibinfo {pages} {L13} (\bibinfo {year}
  {2017}{\natexlab{c}})},\ \Eprint {http://arxiv.org/abs/1710.05834}
  {arXiv:1710.05834 [astro-ph.HE]} \BibitemShut {NoStop}%
\bibitem [{\citenamefont {Abbott}\ \emph
  {et~al.}(2017{\natexlab{d}})\citenamefont {Abbott} \emph
  {et~al.}}]{Abbott:2017xzu}%
  \BibitemOpen
  \bibfield  {author} {\bibinfo {author} {\bibfnamefont {B.~P.}\ \bibnamefont
  {Abbott}} \emph {et~al.},\ }\href {\doibase 10.1038/nature24471} {\bibfield
  {journal} {\bibinfo  {journal} {Nature}\ }\textbf {\bibinfo {volume} {551}},\
  \bibinfo {pages} {85} (\bibinfo {year} {2017}{\natexlab{d}})},\ \Eprint
  {http://arxiv.org/abs/1710.05835} {arXiv:1710.05835 [astro-ph.CO]}
  \BibitemShut {NoStop}%
\bibitem [{\citenamefont {Punturo}\ \emph {et~al.}(2010)\citenamefont {Punturo}
  \emph {et~al.}}]{Punturo:2010zz}%
  \BibitemOpen
  \bibfield  {author} {\bibinfo {author} {\bibfnamefont {M.}~\bibnamefont
  {Punturo}} \emph {et~al.},\ }\href {\doibase 10.1088/0264-9381/27/19/194002}
  {\bibfield  {journal} {\bibinfo  {journal} {Class. Quant. Grav.}\ }\textbf
  {\bibinfo {volume} {27}},\ \bibinfo {pages} {194002} (\bibinfo {year}
  {2010})}\BibitemShut {NoStop}%
\bibitem [{\citenamefont {Dwyer}\ \emph {et~al.}(2015)\citenamefont {Dwyer},
  \citenamefont {Sigg}, \citenamefont {Ballmer}, \citenamefont {Barsotti},
  \citenamefont {Mavalvala},\ and\ \citenamefont {Evans}}]{Dwyer:2014fpa}%
  \BibitemOpen
  \bibfield  {author} {\bibinfo {author} {\bibfnamefont {S.}~\bibnamefont
  {Dwyer}}, \bibinfo {author} {\bibfnamefont {D.}~\bibnamefont {Sigg}},
  \bibinfo {author} {\bibfnamefont {S.~W.}\ \bibnamefont {Ballmer}}, \bibinfo
  {author} {\bibfnamefont {L.}~\bibnamefont {Barsotti}}, \bibinfo {author}
  {\bibfnamefont {N.}~\bibnamefont {Mavalvala}}, \ and\ \bibinfo {author}
  {\bibfnamefont {M.}~\bibnamefont {Evans}},\ }\href {\doibase
  10.1103/PhysRevD.91.082001} {\bibfield  {journal} {\bibinfo  {journal} {Phys.
  Rev.}\ }\textbf {\bibinfo {volume} {D91}},\ \bibinfo {pages} {082001}
  (\bibinfo {year} {2015})},\ \Eprint {http://arxiv.org/abs/1410.0612}
  {arXiv:1410.0612 [astro-ph.IM]} \BibitemShut {NoStop}%
\bibitem [{\citenamefont {Sathyaprakash}\ \emph {et~al.}(2019)\citenamefont
  {Sathyaprakash} \emph {et~al.}}]{Sathyaprakash:2019nnu}%
  \BibitemOpen
  \bibfield  {author} {\bibinfo {author} {\bibfnamefont {B.~S.}\ \bibnamefont
  {Sathyaprakash}} \emph {et~al.},\ }\href@noop {} {\  (\bibinfo {year}
  {2019})},\ \Eprint {http://arxiv.org/abs/1903.09260} {arXiv:1903.09260
  [astro-ph.HE]} \BibitemShut {NoStop}%
\bibitem [{\citenamefont {Belgacem}\ \emph
  {et~al.}(2019{\natexlab{a}})\citenamefont {Belgacem}, \citenamefont {Dirian},
  \citenamefont {Foffa}, \citenamefont {Howell}, \citenamefont {Maggiore},\
  and\ \citenamefont {Regimbau}}]{Belgacem:2019tbw}%
  \BibitemOpen
  \bibfield  {author} {\bibinfo {author} {\bibfnamefont {E.}~\bibnamefont
  {Belgacem}}, \bibinfo {author} {\bibfnamefont {Y.}~\bibnamefont {Dirian}},
  \bibinfo {author} {\bibfnamefont {S.}~\bibnamefont {Foffa}}, \bibinfo
  {author} {\bibfnamefont {E.~J.}\ \bibnamefont {Howell}}, \bibinfo {author}
  {\bibfnamefont {M.}~\bibnamefont {Maggiore}}, \ and\ \bibinfo {author}
  {\bibfnamefont {T.}~\bibnamefont {Regimbau}},\ }\href {\doibase
  10.1088/1475-7516/2019/08/015} {\bibfield  {journal} {\bibinfo  {journal}
  {JCAP}\ }\textbf {\bibinfo {volume} {1908}},\ \bibinfo {pages} {015}
  (\bibinfo {year} {2019}{\natexlab{a}})},\ \Eprint
  {http://arxiv.org/abs/1907.01487} {arXiv:1907.01487 [astro-ph.CO]}
  \BibitemShut {NoStop}%
\bibitem [{\citenamefont {Stratta}\ \emph {et~al.}(2018)\citenamefont {Stratta}
  \emph {et~al.}}]{Stratta:2017bwq}%
  \BibitemOpen
  \bibfield  {author} {\bibinfo {author} {\bibfnamefont {G.}~\bibnamefont
  {Stratta}} \emph {et~al.},\ }\bibfield  {booktitle} {\emph {\bibinfo
  {booktitle} {{THESEUS Workshop 2017 Naples, Italy, October 5-6, 2017}}},\
  }\href {\doibase 10.1016/j.asr.2018.04.013} {\bibfield  {journal} {\bibinfo
  {journal} {Adv. Space Res.}\ }\textbf {\bibinfo {volume} {62}},\ \bibinfo
  {pages} {662} (\bibinfo {year} {2018})},\ \Eprint
  {http://arxiv.org/abs/1712.08153} {arXiv:1712.08153 [astro-ph.HE]}
  \BibitemShut {NoStop}%
\bibitem [{\citenamefont {Saltas}\ \emph {et~al.}(2014)\citenamefont {Saltas},
  \citenamefont {Sawicki}, \citenamefont {Amendola},\ and\ \citenamefont
  {Kunz}}]{Saltas:2014dha}%
  \BibitemOpen
  \bibfield  {author} {\bibinfo {author} {\bibfnamefont {I.~D.}\ \bibnamefont
  {Saltas}}, \bibinfo {author} {\bibfnamefont {I.}~\bibnamefont {Sawicki}},
  \bibinfo {author} {\bibfnamefont {L.}~\bibnamefont {Amendola}}, \ and\
  \bibinfo {author} {\bibfnamefont {M.}~\bibnamefont {Kunz}},\ }\href {\doibase
  10.1103/PhysRevLett.113.191101} {\bibfield  {journal} {\bibinfo  {journal}
  {Phys. Rev. Lett.}\ }\textbf {\bibinfo {volume} {113}},\ \bibinfo {pages}
  {191101} (\bibinfo {year} {2014})},\ \Eprint {http://arxiv.org/abs/1406.7139}
  {arXiv:1406.7139 [astro-ph.CO]} \BibitemShut {NoStop}%
\bibitem [{\citenamefont {Gleyzes}\ \emph {et~al.}(2014)\citenamefont
  {Gleyzes}, \citenamefont {Langlois},\ and\ \citenamefont
  {Vernizzi}}]{Gleyzes:2014rba}%
  \BibitemOpen
  \bibfield  {author} {\bibinfo {author} {\bibfnamefont {J.}~\bibnamefont
  {Gleyzes}}, \bibinfo {author} {\bibfnamefont {D.}~\bibnamefont {Langlois}}, \
  and\ \bibinfo {author} {\bibfnamefont {F.}~\bibnamefont {Vernizzi}},\ }\href
  {\doibase 10.1142/S021827181443010X} {\bibfield  {journal} {\bibinfo
  {journal} {Int. J. Mod. Phys.}\ }\textbf {\bibinfo {volume} {D23}},\ \bibinfo
  {pages} {1443010} (\bibinfo {year} {2014})},\ \Eprint
  {http://arxiv.org/abs/1411.3712} {arXiv:1411.3712 [hep-th]} \BibitemShut
  {NoStop}%
\bibitem [{\citenamefont {Lombriser}\ and\ \citenamefont
  {Taylor}(2016)}]{Lombriser:2015sxa}%
  \BibitemOpen
  \bibfield  {author} {\bibinfo {author} {\bibfnamefont {L.}~\bibnamefont
  {Lombriser}}\ and\ \bibinfo {author} {\bibfnamefont {A.}~\bibnamefont
  {Taylor}},\ }\href {\doibase 10.1088/1475-7516/2016/03/031} {\bibfield
  {journal} {\bibinfo  {journal} {JCAP}\ }\textbf {\bibinfo {volume} {1603}},\
  \bibinfo {pages} {031} (\bibinfo {year} {2016})},\ \Eprint
  {http://arxiv.org/abs/1509.08458} {arXiv:1509.08458 [astro-ph.CO]}
  \BibitemShut {NoStop}%
\bibitem [{\citenamefont {Nishizawa}(2018)}]{Nishizawa:2017nef}%
  \BibitemOpen
  \bibfield  {author} {\bibinfo {author} {\bibfnamefont {A.}~\bibnamefont
  {Nishizawa}},\ }\href {\doibase 10.1103/PhysRevD.97.104037} {\bibfield
  {journal} {\bibinfo  {journal} {Phys. Rev.}\ }\textbf {\bibinfo {volume}
  {D97}},\ \bibinfo {pages} {104037} (\bibinfo {year} {2018})},\ \Eprint
  {http://arxiv.org/abs/1710.04825} {arXiv:1710.04825 [gr-qc]} \BibitemShut
  {NoStop}%
\bibitem [{\citenamefont {Arai}\ and\ \citenamefont
  {Nishizawa}(2018)}]{Arai:2017hxj}%
  \BibitemOpen
  \bibfield  {author} {\bibinfo {author} {\bibfnamefont {S.}~\bibnamefont
  {Arai}}\ and\ \bibinfo {author} {\bibfnamefont {A.}~\bibnamefont
  {Nishizawa}},\ }\href {\doibase 10.1103/PhysRevD.97.104038} {\bibfield
  {journal} {\bibinfo  {journal} {Phys. Rev.}\ }\textbf {\bibinfo {volume}
  {D97}},\ \bibinfo {pages} {104038} (\bibinfo {year} {2018})},\ \Eprint
  {http://arxiv.org/abs/1711.03776} {arXiv:1711.03776 [gr-qc]} \BibitemShut
  {NoStop}%
\bibitem [{\citenamefont {Amendola}\ \emph {et~al.}(2018)\citenamefont
  {Amendola}, \citenamefont {Sawicki}, \citenamefont {Kunz},\ and\
  \citenamefont {Saltas}}]{Amendola:2017ovw}%
  \BibitemOpen
  \bibfield  {author} {\bibinfo {author} {\bibfnamefont {L.}~\bibnamefont
  {Amendola}}, \bibinfo {author} {\bibfnamefont {I.}~\bibnamefont {Sawicki}},
  \bibinfo {author} {\bibfnamefont {M.}~\bibnamefont {Kunz}}, \ and\ \bibinfo
  {author} {\bibfnamefont {I.~D.}\ \bibnamefont {Saltas}},\ }\href {\doibase
  10.1088/1475-7516/2018/08/030} {\bibfield  {journal} {\bibinfo  {journal}
  {JCAP}\ }\textbf {\bibinfo {volume} {1808}},\ \bibinfo {pages} {030}
  (\bibinfo {year} {2018})},\ \Eprint {http://arxiv.org/abs/1712.08623}
  {arXiv:1712.08623 [astro-ph.CO]} \BibitemShut {NoStop}%
\bibitem [{\citenamefont {Linder}(2018)}]{Linder:2018jil}%
  \BibitemOpen
  \bibfield  {author} {\bibinfo {author} {\bibfnamefont {E.~V.}\ \bibnamefont
  {Linder}},\ }\href {\doibase 10.1088/1475-7516/2018/03/005} {\bibfield
  {journal} {\bibinfo  {journal} {JCAP}\ }\textbf {\bibinfo {volume} {1803}},\
  \bibinfo {pages} {005} (\bibinfo {year} {2018})},\ \Eprint
  {http://arxiv.org/abs/1801.01503} {arXiv:1801.01503 [astro-ph.CO]}
  \BibitemShut {NoStop}%
\bibitem [{\citenamefont {Lagos}\ \emph {et~al.}(2019)\citenamefont {Lagos},
  \citenamefont {Fishbach}, \citenamefont {Landry},\ and\ \citenamefont
  {Holz}}]{Lagos:2019kds}%
  \BibitemOpen
  \bibfield  {author} {\bibinfo {author} {\bibfnamefont {M.}~\bibnamefont
  {Lagos}}, \bibinfo {author} {\bibfnamefont {M.}~\bibnamefont {Fishbach}},
  \bibinfo {author} {\bibfnamefont {P.}~\bibnamefont {Landry}}, \ and\ \bibinfo
  {author} {\bibfnamefont {D.~E.}\ \bibnamefont {Holz}},\ }\href {\doibase
  10.1103/PhysRevD.99.083504} {\bibfield  {journal} {\bibinfo  {journal} {Phys.
  Rev.}\ }\textbf {\bibinfo {volume} {D99}},\ \bibinfo {pages} {083504}
  (\bibinfo {year} {2019})},\ \Eprint {http://arxiv.org/abs/1901.03321}
  {arXiv:1901.03321 [astro-ph.CO]} \BibitemShut {NoStop}%
\bibitem [{\citenamefont {Nishizawa}\ and\ \citenamefont
  {Arai}(2019)}]{Nishizawa:2019rra}%
  \BibitemOpen
  \bibfield  {author} {\bibinfo {author} {\bibfnamefont {A.}~\bibnamefont
  {Nishizawa}}\ and\ \bibinfo {author} {\bibfnamefont {S.}~\bibnamefont
  {Arai}},\ }\href {\doibase 10.1103/PhysRevD.99.104038} {\bibfield  {journal}
  {\bibinfo  {journal} {Phys. Rev.}\ }\textbf {\bibinfo {volume} {D99}},\
  \bibinfo {pages} {104038} (\bibinfo {year} {2019})},\ \Eprint
  {http://arxiv.org/abs/1901.08249} {arXiv:1901.08249 [gr-qc]} \BibitemShut
  {NoStop}%
\bibitem [{\citenamefont {Maggiore}(2007)}]{Maggiore:1900zz}%
  \BibitemOpen
  \bibfield  {author} {\bibinfo {author} {\bibfnamefont {M.}~\bibnamefont
  {Maggiore}},\ }\href@noop {} {\emph {\bibinfo {title} {{Gravitational Waves.
  Vol. 1. Theory and Experiments}}}}\ (\bibinfo  {publisher} {Oxford University
  Press, 574 p},\ \bibinfo {year} {2007})\BibitemShut {NoStop}%
\bibitem [{\citenamefont {Belgacem}\ \emph
  {et~al.}(2019{\natexlab{b}})\citenamefont {Belgacem} \emph
  {et~al.}}]{Belgacem:2019pkk}%
  \BibitemOpen
  \bibfield  {author} {\bibinfo {author} {\bibfnamefont {E.}~\bibnamefont
  {Belgacem}} \emph {et~al.} (\bibinfo {collaboration} {LISA Cosmology Working
  Group}),\ }\href {\doibase 10.1088/1475-7516/2019/07/024} {\bibfield
  {journal} {\bibinfo  {journal} {JCAP}\ }\textbf {\bibinfo {volume} {1907}},\
  \bibinfo {pages} {024} (\bibinfo {year} {2019}{\natexlab{b}})},\ \Eprint
  {http://arxiv.org/abs/1906.01593} {arXiv:1906.01593 [astro-ph.CO]}
  \BibitemShut {NoStop}%
\bibitem [{\citenamefont {Deffayet}\ and\ \citenamefont
  {Menou}(2007)}]{Deffayet:2007kf}%
  \BibitemOpen
  \bibfield  {author} {\bibinfo {author} {\bibfnamefont {C.}~\bibnamefont
  {Deffayet}}\ and\ \bibinfo {author} {\bibfnamefont {K.}~\bibnamefont
  {Menou}},\ }\href {\doibase 10.1086/522931} {\bibfield  {journal} {\bibinfo
  {journal} {Astrophys. J.}\ }\textbf {\bibinfo {volume} {668}},\ \bibinfo
  {pages} {L143} (\bibinfo {year} {2007})},\ \Eprint
  {http://arxiv.org/abs/0709.0003} {arXiv:0709.0003 [astro-ph]} \BibitemShut
  {NoStop}%
\bibitem [{\citenamefont {Pardo}\ \emph {et~al.}(2018)\citenamefont {Pardo},
  \citenamefont {Fishbach}, \citenamefont {Holz},\ and\ \citenamefont
  {Spergel}}]{Pardo:2018ipy}%
  \BibitemOpen
  \bibfield  {author} {\bibinfo {author} {\bibfnamefont {K.}~\bibnamefont
  {Pardo}}, \bibinfo {author} {\bibfnamefont {M.}~\bibnamefont {Fishbach}},
  \bibinfo {author} {\bibfnamefont {D.~E.}\ \bibnamefont {Holz}}, \ and\
  \bibinfo {author} {\bibfnamefont {D.~N.}\ \bibnamefont {Spergel}},\ }\href
  {\doibase 10.1088/1475-7516/2018/07/048} {\bibfield  {journal} {\bibinfo
  {journal} {JCAP}\ }\textbf {\bibinfo {volume} {1807}},\ \bibinfo {pages}
  {048} (\bibinfo {year} {2018})},\ \Eprint {http://arxiv.org/abs/1801.08160}
  {arXiv:1801.08160 [gr-qc]} \BibitemShut {NoStop}%
\bibitem [{\citenamefont {Maggiore}(2018)}]{Maggiore:2018zz}%
  \BibitemOpen
  \bibfield  {author} {\bibinfo {author} {\bibfnamefont {M.}~\bibnamefont
  {Maggiore}},\ }\href@noop {} {\emph {\bibinfo {title} {{Gravitational Waves.
  Vol. 2. Astrophysics and Cosmology}}}}\ (\bibinfo  {publisher} {Oxford
  University Press, 848 p},\ \bibinfo {year} {2018})\BibitemShut {NoStop}%
\bibitem [{\citenamefont {Ade}\ \emph {et~al.}(2016{\natexlab{a}})\citenamefont
  {Ade} \emph {et~al.}}]{Planck_2015_CP}%
  \BibitemOpen
  \bibfield  {author} {\bibinfo {author} {\bibfnamefont {P.~A.~R.}\
  \bibnamefont {Ade}} \emph {et~al.} (\bibinfo {collaboration} {Planck}),\
  }\href {\doibase 10.1051/0004-6361/201525830} {\bibfield  {journal} {\bibinfo
   {journal} {Astron. Astrophys.}\ }\textbf {\bibinfo {volume} {594}},\
  \bibinfo {pages} {A13} (\bibinfo {year} {2016}{\natexlab{a}})},\ \Eprint
  {http://arxiv.org/abs/1502.01589} {arXiv:1502.01589 [astro-ph.CO]}
  \BibitemShut {NoStop}%
\bibitem [{\citenamefont {Abbott}\ \emph
  {et~al.}(2019{\natexlab{a}})\citenamefont {Abbott} \emph
  {et~al.}}]{Abbott:2018xao}%
  \BibitemOpen
  \bibfield  {author} {\bibinfo {author} {\bibfnamefont {T.~M.~C.}\
  \bibnamefont {Abbott}} \emph {et~al.} (\bibinfo {collaboration} {DES}),\
  }\href {\doibase 10.1103/PhysRevD.99.123505} {\bibfield  {journal} {\bibinfo
  {journal} {Phys. Rev.}\ }\textbf {\bibinfo {volume} {D99}},\ \bibinfo {pages}
  {123505} (\bibinfo {year} {2019}{\natexlab{a}})},\ \Eprint
  {http://arxiv.org/abs/1810.02499} {arXiv:1810.02499 [astro-ph.CO]}
  \BibitemShut {NoStop}%
\bibitem [{\citenamefont {Belgacem}\ \emph
  {et~al.}(2019{\natexlab{c}})\citenamefont {Belgacem}, \citenamefont {Dirian},
  \citenamefont {Finke}, \citenamefont {Foffa},\ and\ \citenamefont
  {Maggiore}}]{Belgacem:2019lwx}%
  \BibitemOpen
  \bibfield  {author} {\bibinfo {author} {\bibfnamefont {E.}~\bibnamefont
  {Belgacem}}, \bibinfo {author} {\bibfnamefont {Y.}~\bibnamefont {Dirian}},
  \bibinfo {author} {\bibfnamefont {A.}~\bibnamefont {Finke}}, \bibinfo
  {author} {\bibfnamefont {S.}~\bibnamefont {Foffa}}, \ and\ \bibinfo {author}
  {\bibfnamefont {M.}~\bibnamefont {Maggiore}},\ }\href {\doibase
  10.1088/1475-7516/2019/11/022} {\bibfield  {journal} {\bibinfo  {journal}
  {JCAP}\ }\textbf {\bibinfo {volume} {1911}},\ \bibinfo {pages} {022}
  (\bibinfo {year} {2019}{\natexlab{c}})},\ \Eprint
  {http://arxiv.org/abs/1907.02047} {arXiv:1907.02047 [astro-ph.CO]}
  \BibitemShut {NoStop}%
\bibitem [{\citenamefont {Maggiore}(2014)}]{Maggiore:2013mea}%
  \BibitemOpen
  \bibfield  {author} {\bibinfo {author} {\bibfnamefont {M.}~\bibnamefont
  {Maggiore}},\ }\href@noop {} {\bibfield  {journal} {\bibinfo  {journal}
  {Phys.Rev.}\ }\textbf {\bibinfo {volume} {D89}},\ \bibinfo {pages} {043008}
  (\bibinfo {year} {2014})},\ \Eprint {http://arxiv.org/abs/1307.3898}
  {arXiv:1307.3898 [hep-th]} \BibitemShut {NoStop}%
\bibitem [{\citenamefont {Maggiore}(2017)}]{Maggiore:2016gpx}%
  \BibitemOpen
  \bibfield  {author} {\bibinfo {author} {\bibfnamefont {M.}~\bibnamefont
  {Maggiore}},\ }\href {\doibase 10.1007/978-3-319-51700-1_16} {\bibfield
  {journal} {\bibinfo  {journal} {Fundam. Theor. Phys.}\ }\textbf {\bibinfo
  {volume} {187}},\ \bibinfo {pages} {221} (\bibinfo {year} {2017})},\ \Eprint
  {http://arxiv.org/abs/1606.08784} {arXiv:1606.08784 [hep-th]} \BibitemShut
  {NoStop}%
\bibitem [{\citenamefont {Chevallier}\ and\ \citenamefont
  {Polarski}(2001)}]{Chevallier:2000qy}%
  \BibitemOpen
  \bibfield  {author} {\bibinfo {author} {\bibfnamefont {M.}~\bibnamefont
  {Chevallier}}\ and\ \bibinfo {author} {\bibfnamefont {D.}~\bibnamefont
  {Polarski}},\ }\href {\doibase 10.1142/S0218271801000822} {\bibfield
  {journal} {\bibinfo  {journal} {Int.J.Mod.Phys.}\ }\textbf {\bibinfo {volume}
  {D10}},\ \bibinfo {pages} {213} (\bibinfo {year} {2001})},\ \Eprint
  {http://arxiv.org/abs/gr-qc/0009008} {arXiv:gr-qc/0009008 [gr-qc]}
  \BibitemShut {NoStop}%
\bibitem [{\citenamefont {Linder}(2003)}]{Linder:2002et}%
  \BibitemOpen
  \bibfield  {author} {\bibinfo {author} {\bibfnamefont {E.~V.}\ \bibnamefont
  {Linder}},\ }\href {\doibase 10.1103/PhysRevLett.90.091301} {\bibfield
  {journal} {\bibinfo  {journal} {Phys.Rev.Lett.}\ }\textbf {\bibinfo {volume}
  {90}},\ \bibinfo {pages} {091301} (\bibinfo {year} {2003})},\ \Eprint
  {http://arxiv.org/abs/astro-ph/0208512} {arXiv:astro-ph/0208512 [astro-ph]}
  \BibitemShut {NoStop}%
\bibitem [{\citenamefont {Belgacem}\ \emph
  {et~al.}(2018{\natexlab{c}})\citenamefont {Belgacem}, \citenamefont {Dirian},
  \citenamefont {Foffa},\ and\ \citenamefont {Maggiore}}]{Belgacem:2017cqo}%
  \BibitemOpen
  \bibfield  {author} {\bibinfo {author} {\bibfnamefont {E.}~\bibnamefont
  {Belgacem}}, \bibinfo {author} {\bibfnamefont {Y.}~\bibnamefont {Dirian}},
  \bibinfo {author} {\bibfnamefont {S.}~\bibnamefont {Foffa}}, \ and\ \bibinfo
  {author} {\bibfnamefont {M.}~\bibnamefont {Maggiore}},\ }\href {\doibase
  10.1088/1475-7516/2018/03/002} {\bibfield  {journal} {\bibinfo  {journal}
  {JCAP}\ }\textbf {\bibinfo {volume} {1803}},\ \bibinfo {pages} {002}
  (\bibinfo {year} {2018}{\natexlab{c}})},\ \Eprint
  {http://arxiv.org/abs/1712.07066} {arXiv:1712.07066 [hep-th]} \BibitemShut
  {NoStop}%
\bibitem [{\citenamefont {Foffa}\ \emph {et~al.}(2014)\citenamefont {Foffa},
  \citenamefont {Maggiore},\ and\ \citenamefont {Mitsou}}]{Foffa:2013vma}%
  \BibitemOpen
  \bibfield  {author} {\bibinfo {author} {\bibfnamefont {S.}~\bibnamefont
  {Foffa}}, \bibinfo {author} {\bibfnamefont {M.}~\bibnamefont {Maggiore}}, \
  and\ \bibinfo {author} {\bibfnamefont {E.}~\bibnamefont {Mitsou}},\
  }\href@noop {} {\bibfield  {journal} {\bibinfo  {journal} {Int.J.Mod.Phys.}\
  }\textbf {\bibinfo {volume} {A29}},\ \bibinfo {pages} {1450116} (\bibinfo
  {year} {2014})},\ \Eprint {http://arxiv.org/abs/1311.3435} {arXiv:1311.3435
  [hep-th]} \BibitemShut {NoStop}%
\bibitem [{\citenamefont {Dirian}\ \emph {et~al.}(2014)\citenamefont {Dirian},
  \citenamefont {Foffa}, \citenamefont {Khosravi}, \citenamefont {Kunz},\ and\
  \citenamefont {Maggiore}}]{Dirian:2014ara}%
  \BibitemOpen
  \bibfield  {author} {\bibinfo {author} {\bibfnamefont {Y.}~\bibnamefont
  {Dirian}}, \bibinfo {author} {\bibfnamefont {S.}~\bibnamefont {Foffa}},
  \bibinfo {author} {\bibfnamefont {N.}~\bibnamefont {Khosravi}}, \bibinfo
  {author} {\bibfnamefont {M.}~\bibnamefont {Kunz}}, \ and\ \bibinfo {author}
  {\bibfnamefont {M.}~\bibnamefont {Maggiore}},\ }\href {\doibase
  10.1088/1475-7516/2014/06/033} {\bibfield  {journal} {\bibinfo  {journal}
  {JCAP}\ }\textbf {\bibinfo {volume} {1406}},\ \bibinfo {pages} {033}
  (\bibinfo {year} {2014})},\ \Eprint {http://arxiv.org/abs/1403.6068}
  {arXiv:1403.6068 [astro-ph.CO]} \BibitemShut {NoStop}%
\bibitem [{\citenamefont {Dirian}\ \emph {et~al.}(2015)\citenamefont {Dirian},
  \citenamefont {Foffa}, \citenamefont {Kunz}, \citenamefont {Maggiore},\ and\
  \citenamefont {Pettorino}}]{Dirian:2014bma}%
  \BibitemOpen
  \bibfield  {author} {\bibinfo {author} {\bibfnamefont {Y.}~\bibnamefont
  {Dirian}}, \bibinfo {author} {\bibfnamefont {S.}~\bibnamefont {Foffa}},
  \bibinfo {author} {\bibfnamefont {M.}~\bibnamefont {Kunz}}, \bibinfo {author}
  {\bibfnamefont {M.}~\bibnamefont {Maggiore}}, \ and\ \bibinfo {author}
  {\bibfnamefont {V.}~\bibnamefont {Pettorino}},\ }\href {\doibase
  10.1088/1475-7516/2015/04/044} {\bibfield  {journal} {\bibinfo  {journal}
  {JCAP}\ }\textbf {\bibinfo {volume} {1504}},\ \bibinfo {pages} {044}
  (\bibinfo {year} {2015})},\ \Eprint {http://arxiv.org/abs/1411.7692}
  {arXiv:1411.7692 [astro-ph.CO]} \BibitemShut {NoStop}%
\bibitem [{\citenamefont {Dirian}\ \emph {et~al.}(2016)\citenamefont {Dirian},
  \citenamefont {Foffa}, \citenamefont {Kunz}, \citenamefont {Maggiore},\ and\
  \citenamefont {Pettorino}}]{Dirian:2016puz}%
  \BibitemOpen
  \bibfield  {author} {\bibinfo {author} {\bibfnamefont {Y.}~\bibnamefont
  {Dirian}}, \bibinfo {author} {\bibfnamefont {S.}~\bibnamefont {Foffa}},
  \bibinfo {author} {\bibfnamefont {M.}~\bibnamefont {Kunz}}, \bibinfo {author}
  {\bibfnamefont {M.}~\bibnamefont {Maggiore}}, \ and\ \bibinfo {author}
  {\bibfnamefont {V.}~\bibnamefont {Pettorino}},\ }\href {\doibase
  10.1088/1475-7516/2016/05/068} {\bibfield  {journal} {\bibinfo  {journal}
  {JCAP}\ }\textbf {\bibinfo {volume} {1605}},\ \bibinfo {pages} {068}
  (\bibinfo {year} {2016})},\ \Eprint {http://arxiv.org/abs/1602.03558}
  {arXiv:1602.03558 [astro-ph.CO]} \BibitemShut {NoStop}%
\bibitem [{\citenamefont {Dirian}(2017)}]{Dirian:2017pwp}%
  \BibitemOpen
  \bibfield  {author} {\bibinfo {author} {\bibfnamefont {Y.}~\bibnamefont
  {Dirian}},\ }\href {\doibase 10.1103/PhysRevD.96.083513} {\bibfield
  {journal} {\bibinfo  {journal} {Phys. Rev.}\ }\textbf {\bibinfo {volume}
  {D96}},\ \bibinfo {pages} {083513} (\bibinfo {year} {2017})},\ \Eprint
  {http://arxiv.org/abs/1704.04075} {arXiv:1704.04075 [astro-ph.CO]}
  \BibitemShut {NoStop}%
\bibitem [{\citenamefont {Belgacem}\ \emph
  {et~al.}(2019{\natexlab{d}})\citenamefont {Belgacem}, \citenamefont {Finke},
  \citenamefont {Frassino},\ and\ \citenamefont {Maggiore}}]{Belgacem:2018wtb}%
  \BibitemOpen
  \bibfield  {author} {\bibinfo {author} {\bibfnamefont {E.}~\bibnamefont
  {Belgacem}}, \bibinfo {author} {\bibfnamefont {A.}~\bibnamefont {Finke}},
  \bibinfo {author} {\bibfnamefont {A.}~\bibnamefont {Frassino}}, \ and\
  \bibinfo {author} {\bibfnamefont {M.}~\bibnamefont {Maggiore}},\ }\href
  {\doibase 10.1088/1475-7516/2019/02/035} {\bibfield  {journal} {\bibinfo
  {journal} {JCAP}\ }\textbf {\bibinfo {volume} {1902}},\ \bibinfo {pages}
  {035} (\bibinfo {year} {2019}{\natexlab{d}})},\ \Eprint
  {http://arxiv.org/abs/1812.11181} {arXiv:1812.11181 [gr-qc]} \BibitemShut
  {NoStop}%
\bibitem [{\citenamefont {Seikel}\ \emph
  {et~al.}(2012{\natexlab{a}})\citenamefont {Seikel}, \citenamefont
  {Clarkson},\ and\ \citenamefont {Smith}}]{Seikel:2012uu}%
  \BibitemOpen
  \bibfield  {author} {\bibinfo {author} {\bibfnamefont {M.}~\bibnamefont
  {Seikel}}, \bibinfo {author} {\bibfnamefont {C.}~\bibnamefont {Clarkson}}, \
  and\ \bibinfo {author} {\bibfnamefont {M.}~\bibnamefont {Smith}},\ }\href
  {\doibase 10.1088/1475-7516/2012/06/036} {\bibfield  {journal} {\bibinfo
  {journal} {JCAP}\ }\textbf {\bibinfo {volume} {1206}},\ \bibinfo {pages}
  {036} (\bibinfo {year} {2012}{\natexlab{a}})},\ \Eprint
  {http://arxiv.org/abs/1204.2832} {arXiv:1204.2832 [astro-ph.CO]} \BibitemShut
  {NoStop}%
\bibitem [{\citenamefont {Seikel}\ \emph
  {et~al.}(2012{\natexlab{b}})\citenamefont {Seikel}, \citenamefont {Yahya},
  \citenamefont {Maartens},\ and\ \citenamefont {Clarkson}}]{Seikel:2012cs}%
  \BibitemOpen
  \bibfield  {author} {\bibinfo {author} {\bibfnamefont {M.}~\bibnamefont
  {Seikel}}, \bibinfo {author} {\bibfnamefont {S.}~\bibnamefont {Yahya}},
  \bibinfo {author} {\bibfnamefont {R.}~\bibnamefont {Maartens}}, \ and\
  \bibinfo {author} {\bibfnamefont {C.}~\bibnamefont {Clarkson}},\ }\href
  {\doibase 10.1103/PhysRevD.86.083001} {\bibfield  {journal} {\bibinfo
  {journal} {Phys. Rev.}\ }\textbf {\bibinfo {volume} {D86}},\ \bibinfo {pages}
  {083001} (\bibinfo {year} {2012}{\natexlab{b}})},\ \Eprint
  {http://arxiv.org/abs/1205.3431} {arXiv:1205.3431 [astro-ph.CO]} \BibitemShut
  {NoStop}%
\bibitem [{\citenamefont {Yahya}\ \emph {et~al.}(2014)\citenamefont {Yahya},
  \citenamefont {Seikel}, \citenamefont {Clarkson}, \citenamefont {Maartens},\
  and\ \citenamefont {Smith}}]{Yahya:2013xma}%
  \BibitemOpen
  \bibfield  {author} {\bibinfo {author} {\bibfnamefont {S.}~\bibnamefont
  {Yahya}}, \bibinfo {author} {\bibfnamefont {M.}~\bibnamefont {Seikel}},
  \bibinfo {author} {\bibfnamefont {C.}~\bibnamefont {Clarkson}}, \bibinfo
  {author} {\bibfnamefont {R.}~\bibnamefont {Maartens}}, \ and\ \bibinfo
  {author} {\bibfnamefont {M.}~\bibnamefont {Smith}},\ }\href {\doibase
  10.1103/PhysRevD.89.023503} {\bibfield  {journal} {\bibinfo  {journal} {Phys.
  Rev.}\ }\textbf {\bibinfo {volume} {D89}},\ \bibinfo {pages} {023503}
  (\bibinfo {year} {2014})},\ \Eprint {http://arxiv.org/abs/1308.4099}
  {arXiv:1308.4099 [astro-ph.CO]} \BibitemShut {NoStop}%
\bibitem [{\citenamefont {Busti}\ \emph {et~al.}(2014)\citenamefont {Busti},
  \citenamefont {Clarkson},\ and\ \citenamefont {Seikel}}]{Busti:2014dua}%
  \BibitemOpen
  \bibfield  {author} {\bibinfo {author} {\bibfnamefont {V.~C.}\ \bibnamefont
  {Busti}}, \bibinfo {author} {\bibfnamefont {C.}~\bibnamefont {Clarkson}}, \
  and\ \bibinfo {author} {\bibfnamefont {M.}~\bibnamefont {Seikel}},\ }\href
  {\doibase 10.1093/mnrasl/slu035} {\bibfield  {journal} {\bibinfo  {journal}
  {Mon. Not. Roy. Astron. Soc.}\ }\textbf {\bibinfo {volume} {441}},\ \bibinfo
  {pages} {11} (\bibinfo {year} {2014})},\ \Eprint
  {http://arxiv.org/abs/1402.5429} {arXiv:1402.5429 [astro-ph.CO]} \BibitemShut
  {NoStop}%
\bibitem [{\citenamefont {Busti}\ and\ \citenamefont
  {Clarkson}(2016)}]{Busti:2015aqa}%
  \BibitemOpen
  \bibfield  {author} {\bibinfo {author} {\bibfnamefont {V.~C.}\ \bibnamefont
  {Busti}}\ and\ \bibinfo {author} {\bibfnamefont {C.}~\bibnamefont
  {Clarkson}},\ }\href {\doibase 10.1088/1475-7516/2016/05/008} {\bibfield
  {journal} {\bibinfo  {journal} {JCAP}\ }\textbf {\bibinfo {volume} {1605}},\
  \bibinfo {pages} {008} (\bibinfo {year} {2016})},\ \Eprint
  {http://arxiv.org/abs/1505.01821} {arXiv:1505.01821 [astro-ph.CO]}
  \BibitemShut {NoStop}%
\bibitem [{\citenamefont {Yang}\ \emph {et~al.}(2015)\citenamefont {Yang},
  \citenamefont {Guo},\ and\ \citenamefont {Cai}}]{Cai:2015zoa}%
  \BibitemOpen
  \bibfield  {author} {\bibinfo {author} {\bibfnamefont {T.}~\bibnamefont
  {Yang}}, \bibinfo {author} {\bibfnamefont {Z.-K.}\ \bibnamefont {Guo}}, \
  and\ \bibinfo {author} {\bibfnamefont {R.-G.}\ \bibnamefont {Cai}},\ }\href
  {\doibase 10.1103/PhysRevD.91.123533} {\bibfield  {journal} {\bibinfo
  {journal} {Phys. Rev.}\ }\textbf {\bibinfo {volume} {D91}},\ \bibinfo {pages}
  {123533} (\bibinfo {year} {2015})},\ \Eprint
  {http://arxiv.org/abs/1505.04443} {arXiv:1505.04443 [astro-ph.CO]}
  \BibitemShut {NoStop}%
\bibitem [{\citenamefont {Cai}\ \emph {et~al.}(2016{\natexlab{a}})\citenamefont
  {Cai}, \citenamefont {Guo},\ and\ \citenamefont {Yang}}]{Cai:2015pia}%
  \BibitemOpen
  \bibfield  {author} {\bibinfo {author} {\bibfnamefont {R.-G.}\ \bibnamefont
  {Cai}}, \bibinfo {author} {\bibfnamefont {Z.-K.}\ \bibnamefont {Guo}}, \ and\
  \bibinfo {author} {\bibfnamefont {T.}~\bibnamefont {Yang}},\ }\href {\doibase
  10.1103/PhysRevD.93.043517} {\bibfield  {journal} {\bibinfo  {journal} {Phys.
  Rev.}\ }\textbf {\bibinfo {volume} {D93}},\ \bibinfo {pages} {043517}
  (\bibinfo {year} {2016}{\natexlab{a}})},\ \Eprint
  {http://arxiv.org/abs/1509.06283} {arXiv:1509.06283 [astro-ph.CO]}
  \BibitemShut {NoStop}%
\bibitem [{\citenamefont {Cai}\ \emph {et~al.}(2016{\natexlab{b}})\citenamefont
  {Cai}, \citenamefont {Guo},\ and\ \citenamefont {Yang}}]{Cai:2016vmn}%
  \BibitemOpen
  \bibfield  {author} {\bibinfo {author} {\bibfnamefont {R.-G.}\ \bibnamefont
  {Cai}}, \bibinfo {author} {\bibfnamefont {Z.-K.}\ \bibnamefont {Guo}}, \ and\
  \bibinfo {author} {\bibfnamefont {T.}~\bibnamefont {Yang}},\ }\href {\doibase
  10.1088/1475-7516/2016/08/016} {\bibfield  {journal} {\bibinfo  {journal}
  {JCAP}\ }\textbf {\bibinfo {volume} {1608}},\ \bibinfo {pages} {016}
  (\bibinfo {year} {2016}{\natexlab{b}})},\ \Eprint
  {http://arxiv.org/abs/1601.05497} {arXiv:1601.05497 [astro-ph.CO]}
  \BibitemShut {NoStop}%
\bibitem [{\citenamefont {Cai}\ \emph {et~al.}(2017)\citenamefont {Cai},
  \citenamefont {Tamanini},\ and\ \citenamefont {Yang}}]{Cai:2017yww}%
  \BibitemOpen
  \bibfield  {author} {\bibinfo {author} {\bibfnamefont {R.-G.}\ \bibnamefont
  {Cai}}, \bibinfo {author} {\bibfnamefont {N.}~\bibnamefont {Tamanini}}, \
  and\ \bibinfo {author} {\bibfnamefont {T.}~\bibnamefont {Yang}},\ }\href
  {\doibase 10.1088/1475-7516/2017/05/031} {\bibfield  {journal} {\bibinfo
  {journal} {JCAP}\ }\textbf {\bibinfo {volume} {1705}},\ \bibinfo {pages}
  {031} (\bibinfo {year} {2017})},\ \Eprint {http://arxiv.org/abs/1703.07323}
  {arXiv:1703.07323 [astro-ph.CO]} \BibitemShut {NoStop}%
\bibitem [{\citenamefont {Bernstein}\ \emph {et~al.}(2012)\citenamefont
  {Bernstein} \emph {et~al.}}]{Bernstein:2011zf}%
  \BibitemOpen
  \bibfield  {author} {\bibinfo {author} {\bibfnamefont {J.~P.}\ \bibnamefont
  {Bernstein}} \emph {et~al.},\ }\href {\doibase 10.1088/0004-637X/753/2/152}
  {\bibfield  {journal} {\bibinfo  {journal} {Astrophys. J.}\ }\textbf
  {\bibinfo {volume} {753}},\ \bibinfo {pages} {152} (\bibinfo {year}
  {2012})},\ \Eprint {http://arxiv.org/abs/1111.1969} {arXiv:1111.1969
  [astro-ph.CO]} \BibitemShut {NoStop}%
\bibitem [{\citenamefont {Vangioni}\ \emph {et~al.}(2015)\citenamefont
  {Vangioni}, \citenamefont {Olive}, \citenamefont {Prestegard}, \citenamefont
  {Silk}, \citenamefont {Petitjean},\ and\ \citenamefont
  {Mandic}}]{Vangioni:2014axa}%
  \BibitemOpen
  \bibfield  {author} {\bibinfo {author} {\bibfnamefont {E.}~\bibnamefont
  {Vangioni}}, \bibinfo {author} {\bibfnamefont {K.~A.}\ \bibnamefont {Olive}},
  \bibinfo {author} {\bibfnamefont {T.}~\bibnamefont {Prestegard}}, \bibinfo
  {author} {\bibfnamefont {J.}~\bibnamefont {Silk}}, \bibinfo {author}
  {\bibfnamefont {P.}~\bibnamefont {Petitjean}}, \ and\ \bibinfo {author}
  {\bibfnamefont {V.}~\bibnamefont {Mandic}},\ }\href {\doibase
  10.1093/mnras/stu2600} {\bibfield  {journal} {\bibinfo  {journal} {Mon. Not.
  Roy. Astron. Soc.}\ }\textbf {\bibinfo {volume} {447}},\ \bibinfo {pages}
  {2575} (\bibinfo {year} {2015})},\ \Eprint {http://arxiv.org/abs/1409.2462}
  {arXiv:1409.2462 [astro-ph.GA]} \BibitemShut {NoStop}%
\bibitem [{\citenamefont {Abbott}\ \emph
  {et~al.}(2019{\natexlab{b}})\citenamefont {Abbott} \emph
  {et~al.}}]{LIGOScientific:2018mvr}%
  \BibitemOpen
  \bibfield  {author} {\bibinfo {author} {\bibfnamefont {B.~P.}\ \bibnamefont
  {Abbott}} \emph {et~al.} (\bibinfo {collaboration} {LIGO Scientific,
  Virgo}),\ }\href {\doibase 10.1103/PhysRevX.9.031040} {\bibfield  {journal}
  {\bibinfo  {journal} {Phys. Rev.}\ }\textbf {\bibinfo {volume} {X9}},\
  \bibinfo {pages} {031040} (\bibinfo {year} {2019}{\natexlab{b}})},\ \Eprint
  {http://arxiv.org/abs/1811.12907} {arXiv:1811.12907 [astro-ph.HE]}
  \BibitemShut {NoStop}%
\bibitem [{\citenamefont {Bertacca}\ \emph {et~al.}(2018)\citenamefont
  {Bertacca}, \citenamefont {Raccanelli}, \citenamefont {Bartolo},\ and\
  \citenamefont {Matarrese}}]{Bertacca:2017vod}%
  \BibitemOpen
  \bibfield  {author} {\bibinfo {author} {\bibfnamefont {D.}~\bibnamefont
  {Bertacca}}, \bibinfo {author} {\bibfnamefont {A.}~\bibnamefont
  {Raccanelli}}, \bibinfo {author} {\bibfnamefont {N.}~\bibnamefont {Bartolo}},
  \ and\ \bibinfo {author} {\bibfnamefont {S.}~\bibnamefont {Matarrese}},\
  }\href {\doibase 10.1016/j.dark.2018.03.001} {\bibfield  {journal} {\bibinfo
  {journal} {Phys. Dark Univ.}\ }\textbf {\bibinfo {volume} {20}},\ \bibinfo
  {pages} {32} (\bibinfo {year} {2018})},\ \Eprint
  {http://arxiv.org/abs/1702.01750} {arXiv:1702.01750 [gr-qc]} \BibitemShut
  {NoStop}%
\bibitem [{\citenamefont {Chen}\ \emph {et~al.}(2018)\citenamefont {Chen},
  \citenamefont {Fishbach},\ and\ \citenamefont {Holz}}]{Chen:2017rfc}%
  \BibitemOpen
  \bibfield  {author} {\bibinfo {author} {\bibfnamefont {H.-Y.}\ \bibnamefont
  {Chen}}, \bibinfo {author} {\bibfnamefont {M.}~\bibnamefont {Fishbach}}, \
  and\ \bibinfo {author} {\bibfnamefont {D.~E.}\ \bibnamefont {Holz}},\ }\href
  {\doibase 10.1038/s41586-018-0606-0} {\bibfield  {journal} {\bibinfo
  {journal} {Nature}\ }\textbf {\bibinfo {volume} {562}},\ \bibinfo {pages}
  {545} (\bibinfo {year} {2018})},\ \Eprint {http://arxiv.org/abs/1712.06531}
  {arXiv:1712.06531 [astro-ph.CO]} \BibitemShut {NoStop}%
\bibitem [{\citenamefont {Mukherjee}\ \emph {et~al.}(2019)\citenamefont
  {Mukherjee}, \citenamefont {Lavaux}, \citenamefont {Bouchet}, \citenamefont
  {Jasche}, \citenamefont {Wandelt}, \citenamefont {Nissanke}, \citenamefont
  {Leclercq},\ and\ \citenamefont {Hotokezaka}}]{Mukherjee:2019qmm}%
  \BibitemOpen
  \bibfield  {author} {\bibinfo {author} {\bibfnamefont {S.}~\bibnamefont
  {Mukherjee}}, \bibinfo {author} {\bibfnamefont {G.}~\bibnamefont {Lavaux}},
  \bibinfo {author} {\bibfnamefont {F.~R.}\ \bibnamefont {Bouchet}}, \bibinfo
  {author} {\bibfnamefont {J.}~\bibnamefont {Jasche}}, \bibinfo {author}
  {\bibfnamefont {B.~D.}\ \bibnamefont {Wandelt}}, \bibinfo {author}
  {\bibfnamefont {S.~M.}\ \bibnamefont {Nissanke}}, \bibinfo {author}
  {\bibfnamefont {F.}~\bibnamefont {Leclercq}}, \ and\ \bibinfo {author}
  {\bibfnamefont {K.}~\bibnamefont {Hotokezaka}},\ }\href@noop {} {\  (\bibinfo
  {year} {2019})},\ \Eprint {http://arxiv.org/abs/1909.08627} {arXiv:1909.08627
  [astro-ph.CO]} \BibitemShut {NoStop}%
\bibitem [{\citenamefont {Amati}\ \emph {et~al.}(2018)\citenamefont {Amati}
  \emph {et~al.}}]{Amati:2017npy}%
  \BibitemOpen
  \bibfield  {author} {\bibinfo {author} {\bibfnamefont {L.}~\bibnamefont
  {Amati}} \emph {et~al.},\ }\href {\doibase 10.1016/j.asr.2018.03.010}
  {\bibfield  {journal} {\bibinfo  {journal} {Adv. Space Res.}\ }\textbf
  {\bibinfo {volume} {62}},\ \bibinfo {pages} {191} (\bibinfo {year} {2018})},\
  \Eprint {http://arxiv.org/abs/1710.04638} {arXiv:1710.04638 [astro-ph.IM]}
  \BibitemShut {NoStop}%
\bibitem [{\citenamefont {Adam}\ \emph {et~al.}(2016)\citenamefont {Adam} \emph
  {et~al.}}]{Adam:2015rua}%
  \BibitemOpen
  \bibfield  {author} {\bibinfo {author} {\bibfnamefont {R.}~\bibnamefont
  {Adam}} \emph {et~al.} (\bibinfo {collaboration} {Planck}),\ }\href {\doibase
  10.1051/0004-6361/201527101} {\bibfield  {journal} {\bibinfo  {journal}
  {Astron. Astrophys.}\ }\textbf {\bibinfo {volume} {594}},\ \bibinfo {pages}
  {A1} (\bibinfo {year} {2016})},\ \Eprint {http://arxiv.org/abs/1502.01582}
  {arXiv:1502.01582 [astro-ph.CO]} \BibitemShut {NoStop}%
\bibitem [{\citenamefont {Ade}\ \emph {et~al.}(2016{\natexlab{b}})\citenamefont
  {Ade} \emph {et~al.}}]{Ade:2015rim}%
  \BibitemOpen
  \bibfield  {author} {\bibinfo {author} {\bibfnamefont {P.~A.~R.}\
  \bibnamefont {Ade}} \emph {et~al.} (\bibinfo {collaboration} {Planck}),\
  }\href {\doibase 10.1051/0004-6361/201525814} {\bibfield  {journal} {\bibinfo
   {journal} {Astron. Astrophys.}\ }\textbf {\bibinfo {volume} {594}},\
  \bibinfo {pages} {A14} (\bibinfo {year} {2016}{\natexlab{b}})},\ \Eprint
  {http://arxiv.org/abs/1502.01590} {arXiv:1502.01590 [astro-ph.CO]}
  \BibitemShut {NoStop}%
\bibitem [{\citenamefont {Aghanim}\ \emph {et~al.}(2016)\citenamefont {Aghanim}
  \emph {et~al.}}]{Planck_2015_Lkl}%
  \BibitemOpen
  \bibfield  {author} {\bibinfo {author} {\bibfnamefont {N.}~\bibnamefont
  {Aghanim}} \emph {et~al.} (\bibinfo {collaboration} {Planck}),\ }\href
  {\doibase 10.1051/0004-6361/201526926} {\bibfield  {journal} {\bibinfo
  {journal} {Astron. Astrophys.}\ }\textbf {\bibinfo {volume} {594}},\ \bibinfo
  {pages} {A11} (\bibinfo {year} {2016})},\ \Eprint
  {http://arxiv.org/abs/1507.02704} {arXiv:1507.02704 [astro-ph.CO]}
  \BibitemShut {NoStop}%
\bibitem [{\citenamefont {Ade}\ \emph {et~al.}(2016{\natexlab{c}})\citenamefont
  {Ade} \emph {et~al.}}]{Ade:2015zua}%
  \BibitemOpen
  \bibfield  {author} {\bibinfo {author} {\bibfnamefont {P.~A.~R.}\
  \bibnamefont {Ade}} \emph {et~al.} (\bibinfo {collaboration} {Planck}),\
  }\href {\doibase 10.1051/0004-6361/201525941} {\bibfield  {journal} {\bibinfo
   {journal} {Astron. Astrophys.}\ }\textbf {\bibinfo {volume} {594}},\
  \bibinfo {pages} {A15} (\bibinfo {year} {2016}{\natexlab{c}})},\ \Eprint
  {http://arxiv.org/abs/1502.01591} {arXiv:1502.01591 [astro-ph.CO]}
  \BibitemShut {NoStop}%
\bibitem [{\citenamefont {Beutler}\ \emph {et~al.}(2011)\citenamefont
  {Beutler}, \citenamefont {Blake}, \citenamefont {Colless}, \citenamefont
  {Jones}, \citenamefont {Staveley-Smith} \emph {et~al.}}]{Beutler:2011hx}%
  \BibitemOpen
  \bibfield  {author} {\bibinfo {author} {\bibfnamefont {F.}~\bibnamefont
  {Beutler}}, \bibinfo {author} {\bibfnamefont {C.}~\bibnamefont {Blake}},
  \bibinfo {author} {\bibfnamefont {M.}~\bibnamefont {Colless}}, \bibinfo
  {author} {\bibfnamefont {D.~H.}\ \bibnamefont {Jones}}, \bibinfo {author}
  {\bibfnamefont {L.}~\bibnamefont {Staveley-Smith}},  \emph {et~al.},\ }\href
  {\doibase 10.1111/j.1365-2966.2011.19250.x} {\bibfield  {journal} {\bibinfo
  {journal} {Mon.Not.Roy.Astron.Soc.}\ }\textbf {\bibinfo {volume} {416}},\
  \bibinfo {pages} {3017} (\bibinfo {year} {2011})},\ \Eprint
  {http://arxiv.org/abs/1106.3366} {arXiv:1106.3366 [astro-ph.CO]} \BibitemShut
  {NoStop}%
\bibitem [{\citenamefont {Ross}\ \emph {et~al.}(2015)\citenamefont {Ross},
  \citenamefont {Samushia}, \citenamefont {Howlett}, \citenamefont {Percival},
  \citenamefont {Burden},\ and\ \citenamefont {Manera}}]{Ross_SDSS_2014}%
  \BibitemOpen
  \bibfield  {author} {\bibinfo {author} {\bibfnamefont {A.~J.}\ \bibnamefont
  {Ross}}, \bibinfo {author} {\bibfnamefont {L.}~\bibnamefont {Samushia}},
  \bibinfo {author} {\bibfnamefont {C.}~\bibnamefont {Howlett}}, \bibinfo
  {author} {\bibfnamefont {W.~J.}\ \bibnamefont {Percival}}, \bibinfo {author}
  {\bibfnamefont {A.}~\bibnamefont {Burden}}, \ and\ \bibinfo {author}
  {\bibfnamefont {M.}~\bibnamefont {Manera}},\ }\href {\doibase
  10.1093/mnras/stv154} {\bibfield  {journal} {\bibinfo  {journal} {Mon. Not.
  Roy. Astron. Soc.}\ }\textbf {\bibinfo {volume} {449}},\ \bibinfo {pages}
  {835} (\bibinfo {year} {2015})},\ \Eprint {http://arxiv.org/abs/1409.3242}
  {arXiv:1409.3242 [astro-ph.CO]} \BibitemShut {NoStop}%
\bibitem [{\citenamefont {Anderson}\ \emph {et~al.}(2014)\citenamefont
  {Anderson} \emph {et~al.}}]{Anderson:2013zyy}%
  \BibitemOpen
  \bibfield  {author} {\bibinfo {author} {\bibfnamefont {L.}~\bibnamefont
  {Anderson}} \emph {et~al.} (\bibinfo {collaboration} {BOSS}),\ }\href
  {\doibase 10.1093/mnras/stu523} {\bibfield  {journal} {\bibinfo  {journal}
  {Mon. Not. Roy. Astron. Soc.}\ }\textbf {\bibinfo {volume} {441}},\ \bibinfo
  {pages} {24} (\bibinfo {year} {2014})},\ \Eprint
  {http://arxiv.org/abs/1312.4877} {arXiv:1312.4877 [astro-ph.CO]} \BibitemShut
  {NoStop}%
\end{thebibliography}%

\end{document}